\documentclass[usenatbib,a4paper,times,fleqn]{mnras}
\usepackage[T1]{fontenc}
\usepackage{graphicx,natbib,times,amssymb,amsmath,pstricks,mathtools,enumitem,array}
\usepackage{aas_macros}
\usepackage{bm,url}
\usepackage{textcomp}
\usepackage{breakurl}
\usepackage[latin1]{inputenc}

\usepackage{nicefrac,xfrac}
\usepackage{mathtools}
\usepackage{cancel}
\usepackage{listings}
\usepackage{pifont}
\usepackage{parskip}
\usepackage{siunitx}
\usepackage{booktabs}
\usepackage[font=Large]{subfig}
\usepackage[normalem]{ulem}
\usepackage{bbold}

\usepackage{cuted}

\newcommand\redsout{\bgroup\markoverwith{\textcolor{red}{\rule[0.5ex]{2pt}{1pt}}}\ULon}
\newcommand {\apgt} {\ {\raise-.5ex\hbox{$\buildrel>\over\sim$}}\ }
\newcommand {\aplt} {\ {\raise-.5ex\hbox{$\buildrel<\over\sim$}}\ }

\title[Dust fragmentation with high-order schemes]{General non-linear fragmentation with discontinuous Galerkin methods}

\author[Lombart et al.]{Maxime Lombart$^{1}$\thanks{maxime.lombart@gapps.ntnu.edu.tw/maxime.lombart@gmail.com}, Charles-Edouard Br\'ehier$^{2}$, Mark Hutchison$^{3,4}$, Yueh-Ning Lee$^{1}$  \\
$^{1}$Department of Earth Sciences, National Taiwan Normal University, 88, Sec.4, Ting-Chou Road, Taipei 11677, Taiwan\\
$^{2}$Universit\'e de Pau et des Pays de l'Adour, E2S UPPA, CNRS, LMAP, Pau, France\\
$^{3}$Universit{\"a}ts-Sternwarte, Ludwig-Maximilians-Universit{\"a}t  M{\"u}nchen, Scheinerstr. 1, 81679 M{\"u}nchen, Germany\\
$^{4}$Hochschule f\"ur angewandte Wissenschaften M\"unchen, Lothstra{\ss}e 34, 80335 M\"unchen, Germany
}
\date{}
\begin{document}
%\label{firstpage}
\bibliographystyle{mnras}
\maketitle

\begin{abstract}
Dust grains play a significant role in several astrophysical processes, including gas/dust dynamics, chemical reactions, and radiative transfer. Replenishment of small-grain populations is mainly governed by fragmentation during pair-wise collisions between grains. The wide spectrum of fragmentation outcomes, from complete disruption to erosion and/or mass transfer, can be modelled by the general non-linear fragmentation equation. Efficiently solving this equation is crucial for an accurate treatment of the dust fragmentation in numerical modelling. However, similar to dust coagulation, numerical errors in current fragmentation algorithms employed in astrophysics are dominated by the numerical over-diffusion problem -- particularly in 3D hydrodynamic simulations where the discrete resolution of the mass density distribution tends to be highly limited. With this in mind, we have derived the first conservative form of the general non-linear fragmentation with a mass flux highlighting the mass transfer phenomenon. Then, to address cases of limited mass density resolution, we applied a high-order discontinuous Galerkin scheme to efficiently solve the conservative fragmentation equation with a reduced number of dust bins. An accuracy of $0.1 -1 \%$ is reached with $20$ dust bins spanning a mass range of 9 orders of magnitude. \\
\end{abstract}


%----------------------------------------------------------------------------------------------------------------
\section{Introduction}
\label{sec:introduction}
Dust grains play a fundamental role at all scales in astrophysics. Dust surfaces act as an efficient catalyst for the formation of $\mathrm{H}_2$ \citep{Cazaux2004}, which impacts the star formation rate in galaxies \citep{Yamasawa2011,Chen2018}. Dust grains absorb, scatter and reemit stellar light, thereby governing the thermal balance between heating and cooling in star forming regions and protoplanetary discs \citep{Mckee2007,Andrews2020}. Gas and dust dynamics are closely linked through drag forces \citep{Testi2014,Lesur2023b}. During star formation, large grains decouple dynamically from the gas and concentrate in regions of high gas density \citep{Lebreuilly2021}. In protoplanetary discs, the decoupling of large grains from the gas leads to momentum transfer and radial drift towards pressure maxima \citep{Weidenschilling1977,Lesur2023b}. When dust becomes sufficiently concentrated in discs, it produces a "backreaction" on the gas which is central for the development of the streaming instability and the eventual formation of planetesimals \citep[][and references therein]{Youdin2005,Gonzalez2017,Squire2020,Lesur2023b}. Importantly, the efficiency of all these physical processes depends on the grain size distribution and how it evolves in time. Thus, accurate dust modelling is a much needed feature of modern astrophysical simulations.

Because the grain size distribution plays such a key role in the above phenomenon, many studies have been devoted to understanding how poly-disperse distributions evolve in time. Evolutionary changes can be dynamical (e.g. advection and diffusion) or collisional (e.g. coagulation and fragmentation) in nature. For purposes of this study, we will focus exclusively on fragmentation. Fragmentation is typically modelled in one of three ways: i) spontaneous breaking driven by an external force, such as radiative force \citep{Hoang2019,Hirashita2020}, modelled by the \textit{linear fragmentation} equation; ii) collision between two grains where only one grain fragments, modelled by the \textit{non-linear fragmentation} equation \citep{Kostoglou2000,Banasiak2019,Lombart2022}; iii) collision between two grains where both grains can fragment, modelled by the \textit{general non-linear fragmentation} equation \citep{Safronov1972,Blum2006,Hirashita2009}. Note the third model is a generalisation of the second one, which produces a size-distribution of solids that can be generically parametrised from experiments. Two main outcomes observed in laboratory experiments are the complete destruction of the two grains or a partial destruction with mass transfer \citep{Guttler2010,Bukhari_Syed2017,Blum2018}. The phenomenon of mass transfer is important since, during a fragmentation event, a grain can increase in mass -- even at high impact velocities. For example, several studies have demonstrated that mass transfer during fragmentation can  overcome the so-called bouncing and fragmentation barriers \citep{Windmark2012,Garaud2013}. Only the general non-linear fragmentation equation can effectively model fragmentation with mass transfer, making it essential for a full treatment of the problem. The general non-linear fragmentation equation is formalised within the framework of the Smoluchowski-like equation by a mean-field approach \citep{Safronov1972,Gillespie1978,Feingold1988,Blum2006,Hirashita2009,Jacobson2011,Banasiak2019}. The model considers spherical grains of the same composition. Because the general non-linear fragmentation equation does not have a generic analytical solution, astrophysical problems inevitably require numerical solutions.

Unfortunately, incorporating accurate numerical solutions to the fragmentation (and/or coagulation), together with tracking the dust size distribution in 3D multi-physics hydrodynamics simulations, is currently out of reach. So far only 3D hydrodynamics simulations with a mono-disperse model of dust coagulation and fragmentation have been performed for protoplanetary discs \citep{Vericel2021}. Meanwhile, poly-disperse models of dust coagulation and fragmentation have been confined to 1D or 2D hydrodynamic simulations \citep{Suttner2001,Brauer2008,Drazkowska2019,Kobayashi2021,Tu2022,Stammler2022,Lebreuilly2022,Robinson2024} primarily due to the near ubiquitous use of piecewise constant functions to model the dust size distribution. Piecewise constant functions are known to suffer from a numerical over-diffusion for insufficient mass grid resolution \citep{Grabowski2022,Birnstiel2023}. This over-diffusion stems from the difficulty to handle the complexity of the integro-differential and the non-linear properties of the coagulation and the fragmentation equations with only a low-order approximation of the continuous dust size distribution. Moreover, these algorithms account for each pair-wise collision by redistributing the mass over the mass grid to conserve the total mass. The combinatorics treatment of the collisions requires the need of a large number of mass bins to keep high accuracy. Therefore, these algorithms need a high resolution of the mass grid (more than 100 mass bins) to accurately follow the evolution of the dust size distribution. However, 3D hydrodynamics codes, such as \texttt{RAMSES} \citep{Teyssier2002} or \texttt{PHANTOM} \citep{Price2018}, can only handle a few tens of dust mass bins for a multi-physics gas and dust simulation. Further generation of Exascale code will reach further performance, but the energy, the carbon impact and the computational costs of the simulation should remain acceptable. Current algorithms for poly-disperse dust evolution and 3D hydrodynamics codes are simply incompatible. In contrast, numerical schemes using high-order approximations of the continuous dust size distribution are able to efficiently solve the coagulation and the fragmentation equations with a reduced number of mass bins. This study contains two main parts: i) the derivation of the flux of fragmentation in mass space, which encompasses all the combinatorics of the pair-wise collisions, in order to obtain the general non-linear fragmentation equation in conservative form and ii) the application of a high-order solver based on the discontinuous Galerkin method, inspired by the recent works of \citet{Liu2019} and \citet{Lombart2022}. The discontinuous Galerkin method is shown to efficiently solve the general non-linear fragmentation equation with a reduced number of mass bins while still maintaining high accuracy.

The paper is structured as follows. Properties of the general non-linear fragmentation equation and derivation of its conservative form are presented in Section~\ref{sec:frag}. The discontinuous Galerkin method applied to the conservative form is presented in Section~\ref{sec:DG}. In Section~\ref{sec:num}, we analyse the numerical performance of our algorithm on some test cases, including how it copes with the numerical diffusion problem. Section~\ref{sec:discussions} presents the applicability of the discontinuous Galerkin method to treat dust fragmentation in astrophysics.

%-----------------------------------------------------------------------------------------------------------------
\section{General non-linear fragmentation}
\label{sec:frag}

The fragmentation process resulting from the collision of two grains of arbitrary size is described by the general non-linear fragmentation equation, also known as the "collision-induced breakup" or the "stochastic breakage" equation in atmospheric science and mathematics communities \citep{Safronov1972,Gillespie1978,Feingold1988,Blum2006,Giri2021a}. The general non-linear fragmentation model is a natural extension of the non-linear fragmentation model which describes how a small grain fragments after collision with a large grain \citep{Kostoglou2000,Cheng1990,Ernst2007,Lombart2022}. This fragmentation model is described by a non-linear partial integro-differential hyperbolic equation which depends on two functions: i) the fragmentation kernel, and ii) the distribution function of fragments. This general non-linear fragmentation model was initially formalised in astrophysics by \citet{Safronov1972} and in atmospheric science by \citet{List1976} to study the evolution of the drop size distribution in clouds. Currently in the literature, only one analytical solution exists for a constant fragmentation kernel and a specific form of the distribution of fragments \citep{Feingold1988}. Several recent mathematical works have been dedicated to the study of the general non-linear fragmentation equation, including proving the existence and uniqueness of mass-conserving solutions for a large class of collision kernels and fragment distribution functions \citep{Giri2021a,Giri2021b}. However, no exact solutions exist for the ballistic collision kernel studied in astrophysics \citep{Safronov1972,Tanaka1996,Dullemond2005,Kobayashi2010a,Stammler2022} and in atmospheric science \citep{Hu1995,McFarquhar2004,Prat2007,Jacobson2011,Khain2018,Grabowski2022}. Therefore, numerical solutions are required for the ballistic collision kernel. The detailed expression of the ballistic kernel is given in Sect.~\ref{sec:kernels}. In this work, numerical results are compared to the exact solution obtained for the constant collision kernel and a distribution of fragments taken from \citet{Feingold1988} (Sect.~\ref{subsec:kconst_tests}) or a multiplicative collision kernel with a power-law distribution of fragments (Sect.~\ref{sec:power_law_distrib}).

\subsection{Collision outcomes for fragmentation}
\label{sec:mt}

\begin{figure}
\centering
\includegraphics[width=\columnwidth,trim=10 100 10 80, clip]{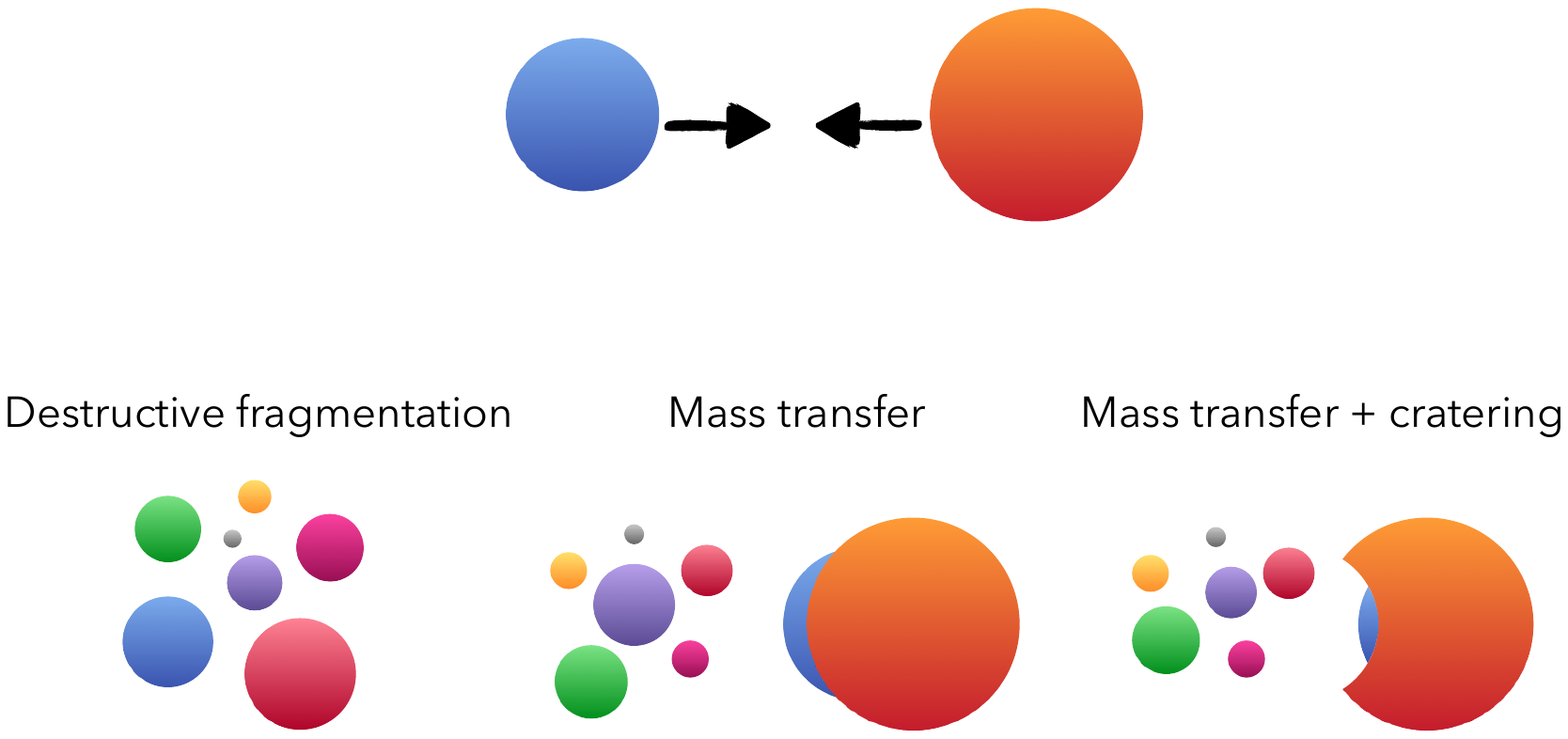}
\caption{Illustration of the collision outcomes leading to the formation of fragments. Depending on the differential velocity of the two colliding grains, three outcomes are possible: destructive fragmentation, mass transfer, and mass transfer with cratering.}
\label{fig:collision_outcome}
\end{figure}

In astrophysics, the outcome of two colliding dust grains leading to the formation of fragments is described by one of three scenarios, illustrated in Fig~\ref{fig:collision_outcome} \citep{Guttler2010,Windmark2012,Blum2018,Birnstiel2023}. First, destructive fragmentation accounts for cases where both grains fragment totally or partially. The second scenario describes mass transfer events where the resulting fragments come from only one body. This scenario has been observed in experiments when a small grain collides with a larger one \citep{Bukhari_Syed2017}. During the collision, the smaller grain fragments and leaves a portion of its mass stuck to the larger (intact) grain, causing the mass of the larger grain to increase. The third scenario is the same as the second, but now with cratering of the larger grain. For sufficiently high-velocities, the small grain both transfers mass to and excavates mass from the large grain. Since the non-linear fragmentation equation \citep{Kostoglou2000,Banasiak2019,Lombart2022} cannot account for mass transfer or the fragmentation of both grains, the above scenarios can only be modelled by the general non-linear fragmentation equation. This is of particular interest to the planetary science community because mass transfer may be an avenue to circumvent the fragmentation barrier \citep{Windmark2012,Garaud2013} and provide larger seeds for planetesimal formation.

\subsection{Original rate equation}
We begin by formulating a rate equation for fragmentation that describes the gains and losses in number density as a function of grain mass. An alternative form is given later in Section~\ref{subsec:alternative} that has improved mass conservation properties. Hereafter, we will distinguish between these two formulations using the labels `original' and `alternative', respectively.

We consider a volume composed of a large number of uniformly-distributed grains undergoing fragmentation through collisions. The number of grains is considered sufficiently large to be treated in a statistical sense, while the even spatial distribution of the grains permits the exclusion of their motion from consideration. Moreover, colliding grains and fragments are considered to be spherical.\\
We denote $K(m',m'')$ to be the fragmentation kernel that encodes the collision frequency of grains leading to fragmentation with dimension $[\mathrm{length}]^3 [\mathrm{time}]^{-1}$. The function $K$ is symmetric in variables $m$ and $m'$. Therefore, the mean number of collisions per unit time and unit volume between the grains of mass in the ranges $[m',m' + \mathrm{d}m']$ and $[m'',m''+\mathrm{d}m'']$ is given by 
\begin{equation}
K(m',m'') n(m',t) n(m'',t) \mathrm{d}m' \mathrm{d}m'' \; (\text{collision rate} ),
\label{eq:collision_rate}
\end{equation}
where $m'$ and $m''$ are the initial masses of the colliding grains and $n(m',t)$ and  $n(m'',t)$ are the number densities function per unit mass of grains in mass ranges $[m',m' + \mathrm{d}m']$ and $[m'',m''+\mathrm{d}m'']$, respectively. We denote by $\tilde{b}(m;m',m'')$ the distribution of fragments of mass $m$ produced by the collision of grains of mass $m'$ and $m''$. The only physical constraint on the function $\tilde{b}$ is that the mass of a fragment cannot exceed the total mass of the colliding grains, therefore one has $b(m;m',m'') = 0$ if $m > m'+m''$. In the derivations that follow, this condition is represented by the operator $\mathbb{1}$, defined as
\begin{equation}
\mathbb{1}_{m'+m'' \geq m} = 
\left\{ 
\begin{aligned}
&1\; \text{if}\; m'+m'' \geq m, \\
& 0\; \text{otherwise}.
\end{aligned}
\right.
\label{eq:operator}
\end{equation}
Therefore, the distribution of fragments writes $\mathbb{1}_{m'+m'' \geq m} \tilde{b}(m;m',m'')$. The term $\tilde{b}(m;m',m'')\mathrm{d}m$ is the mean number of fragments of mass whose masses reside in the range $[m,m+\mathrm{d}m]$ produced in a collision of one pair of grains of mass $m'$ and $m''$. The dimension of $b$ is $[\mathrm{mass}]^{-1}$. The formation rate of grains of mass $[m,m+\mathrm{d}m]$ by fragmentation of larger grains of masses $[m',m' + \mathrm{d}m']$ and $[m'',m''+\mathrm{d}m'']$ is equal to the product of the collision rate (Eq.~\ref{eq:collision_rate}) with the mean number of fragments produced by the collision, $\tilde{b}(m;m',m'')\mathrm{d}m$. The formation rate of grains within mass range $[m,m+\mathrm{d}m]$ is obtained by considering all collisions, and is expressed as
\begin{equation}
\begin{aligned}
& \left[ \frac{1}{2}\int\limits_{0}^{\infty} \int\limits_{0}^{\infty} \mathbb{1}_{m'+m'' \geq m} \tilde{b}(m;m',m'')  K(m',m'') \right.\\
& \qquad \qquad  \left. \vphantom{\int\limits_{0}^{\infty}} \times n(m',t) n(m'',t) \mathrm{d}m' \mathrm{d}m'' \right] \mathrm{d}m \; (\text{formation rate}),
\end{aligned}
\end{equation}
where the factor $1/2$ prevents grain pairs from being double counted. The loss rate of grains of mass $[m,m+\mathrm{d}m]$ by collision with all other grains is given by 
\begin{equation}
\begin{aligned}
&  \left[ \int\limits_{0}^{\infty} K(m,m') n(m,t) n(m',t) \mathrm{d}m' \right] \mathrm{d}m \; (\text{loss rate}).
\end{aligned}
\end{equation}
Therefore, the rate equation for the number density of grains of mass $[m,m+\mathrm{d}m]$ at time $t$ is obtained by balancing the formation and loss rates
\begin{equation}
\begin{aligned}
& \frac{\partial ( n (m,t) \mathrm{d}m)}{\partial t } = \\
&  \quad \left[\frac{1}{2}\int\limits_{0}^{\infty} \int\limits_{0}^{\infty} \mathbb{1}_{m'+m" \geq m} \tilde{b}(m;m',m'') K(m',m'') \right. \\
& \qquad \qquad  \left. \vphantom{\int\limits_{0}^{\infty}} \times  n(m',t) n(m'',t) \mathrm{d}m' \mathrm{d}m'' \right] \mathrm{d}m \\
& \qquad -  \left[ \int\limits_{0}^{\infty} K(m,m') n(m,t) n(m',t) \mathrm{d}m' \right] \mathrm{d}m.
\end{aligned}
\end{equation}
By dividing by $\mathrm{d}m$, we obtain the continuous general non-linear fragmentation equation originally formalised in a mean-field approach by \citet{List1976} and \citet{Gillespie1978} 
\begin{equation}
\begin{aligned}
& \frac{\partial n (m,t)}{\partial t } = \\
& \quad \frac{1}{2}\int\limits_{0}^{\infty} \int\limits_{0}^{\infty} \mathbb{1}_{m'+m'' \geq m}  \tilde{b}(m;m',m'') K(m',m'')  \\
& \qquad \qquad  \times n(m',t) n(m'',t) \mathrm{d}m' \mathrm{d}m'' \\
& \qquad - n(m,t) \int\limits_{0}^{\infty} K(m,m') n(m',t) \mathrm{d}m'.
\end{aligned}
\label{eq:frag_cont}
\end{equation}
\\
To obtain a dimensionless general non-linear fragmentation equation, we first define $n_0(m)=n(m,0)$ to be the initial number density per mass unit. The total mass density $M$, the total number density of particles $N_0$ and the mean mass of the initial distribution $m_0$ can then be written as
\begin{equation}
M \equiv \int\limits_0^{\infty} mn_0(m) \mathrm{d}m, \, N_0 \equiv \int\limits_0^{\infty} n_0(m) \mathrm{d}m,\, m_0\equiv\frac{M}{N_0}.
\end{equation}
Then, we define the dimensionless variables and functions as follows
\begin{equation}
  \left\{ 
  \begin{aligned}
    & x \equiv \frac{m}{m_0},\,y \equiv \frac{m'}{m_0}, \,z \equiv \frac{m''}{m_0},\\
    & \mathcal{K}(x,y) = \frac{K(m,m')}{K(m_0,m_0)},\,  \tau = K(m_0,m_0) N_0 t, \\
    & f(x,\tau) = \frac{m_0}{N_0} n(m,t),\,b(x;y,z) = m_0 \tilde{b}(m;m',m''),
  \end{aligned}
  \right.
\end{equation}
where $K(m_0,m_0)$ is a constant with dimensions of the kernel $[\mathrm{length}]^3 [\mathrm{time}]^{-1}$. To be consistent with the existing literature in mathematics \citep{Banasiak2019,Giri2021a}, we use the variables $x,\tau$, and $f$ for the dimensionless mass, time and number density, respectively, to write the general non-linear fragmentation equation in the following dimensionless form:
\begin{equation}
\begin{aligned}
& \frac{\partial f (x,\tau)}{\partial \tau } = \\
& \qquad \frac{1}{2}\int\limits_{0}^{\infty} \int\limits_{0}^{\infty} \mathbb{1}_{y+z \geq x} b(x;y,z) \mathcal{K}(y,z) f(y,\tau) f(z,\tau) \mathrm{d}y \mathrm{d}z \\
& \qquad \qquad - f(x,\tau) \int\limits_{0}^{\infty} \mathcal{K}(x,y) f(y,\tau) \mathrm{d}y.
\end{aligned}
\label{eq:frag_cont_DL}
\end{equation}
Unless otherwise noted, we will use those dimensionless variables for the remainder of the paper.

\subsection{Fragmentation kernels}
\label{sec:kernels}
The fragmentation kernel describes the collision rate per unit volume of two grain masses. The expression of the kernel is determined by the mechanism driving the collisions between grains (e.g. turbulence, radial drift, vertical settling). The kernel depends on the mass of the grains, and on the properties of any surrounding gas, such as temperature, pressure and the characteristics of the flow field. In astrophysics, the collision between grains is modelled by the ballistic kernel,
\begin{equation}
\mathcal{K}(x,y) = E_{\mathrm{coll}}(x,y) P_{\mathrm{frag}}(x,y,\Delta v) \sigma(x,y)\Delta v(x,y),
\end{equation}
where $ \sigma(x,y)$ is the geometric cross-section, $\Delta v$ is the mean relative velocity between two grains of mass $x$ and $y$, while $P_{\mathrm{frag}}$ denotes the probability that the two colliding grains fragment. The collision efficiency, $E_{\mathrm{coll}}$, describes the probability of two grains embedded in a flow field colliding. For instance, a large grain surrounded by gas in the Stokes regime has a small probability of collision with a small grain dragged along the gas stream lines. The limiting case for impact between large and small grains is referred to as the grazing collision trajectory in atmospheric science for droplets \citep{Pruppacher2010,Wang2013,Khain2018} or grazing impact in astrophysics \citep{Paszun2009,Wada2009}. The standard definition of the collision efficiency is \citep{Paszun2009,Wada2009,Pruppacher2010}
\begin{equation}
 E_{\mathrm{coll}}(s_x,s_y) = \frac{p_{\mathrm{impact}}(s_x,s_y)^2}{(s_x+s_y)^2},
\end{equation}
where $s_x$ and $s_y$ are the radius of the grains of masses $x$ and $y$. $p_{\mathrm{impact}}(s_x,s_y)$ is the impact parameter and depends on the size of the two colliding grains. For a collision event, the impact parameter is defined as the projected distance between the centers of mass of the grains in the perpendicular direction to the collision velocity \citep{Wada2009}. By considering spherical grains, we can directly link $E_{\mathrm{coll}}(s_x,s_y)$ and $E_{\mathrm{coll}}(x,y)$, as was done by \citet{Pinsky2001} by solving the equations of motion of small grains around large grains. In \citet{Paszun2009} and \citet{Wada2009}, the impact parameter is evaluated as a percentage of the maximum impact parameter $p_{\mathrm{impact,max} }= s_x+s_y$. A value of $E_{\mathrm{coll}}(x,y)$ close to zero means that the small grain follows the gas stream lines. This formalism has been adapted to study the collision between planetesimals and small grains \citep{Guillot2014,Visser2016}. However, for purposes of this study, we will consider only head-on collisions, $E_{\mathrm{coll}}(x,y) = 1$. 

The fragmentation probability $P_{\mathrm{frag}}$ is usually defined as 
\begin{equation}
P_{\mathrm{frag}} = 
\left\{
\begin{aligned}
&1\; \text{if}\; \Delta v > \Delta v_{\mathrm{th}},\\
&0\; \text{if}\; \Delta v < \Delta v_{\mathrm{th}},
\end{aligned}
\right.
\end{equation}
where $\Delta v$ is the differential velocity between the two colliding grains and $ \Delta v_{\mathrm{th}}$ is the threshold differential velocity for which fragmentation occurs. The value of the threshold velocity is determined from experiments \citep{Guttler2010,Blum2018} or from theoretical works \citep{Jones1996,Ormel2009}. In this study, because we only consider fragmentation, we set $P_{\mathrm{frag}}=1$.

\subsection{Distribution of fragments}
\label{sec:dist_frag}
The evolution of the number density $f(x,\tau)$ in Eq.~\ref{eq:frag_cont} depends on two physical parameters, the collision kernel $\mathcal{K}$ and the distribution function of fragments $b(x;y,z)$ resulting from a collision between two grains of mass $y$ and $z$ \citep{Gillespie1978,Feingold1988}. The function $b$ is symmetric in the mass variables of the two colliding grains
\begin{equation}
b(x;y,z) = b(x;z,y).
\end{equation}
Because we do not consider sublimation, the fragmentation process must satisfy the following two mass conservation constraints. First, the mass of a fragment cannot exceed the total mass of the colliding grains
\begin{equation}
b(x;y,z)=0\; \text{if}\; x > y+z.
\label{eq:b_constrain_1}
\end{equation}
Secondly, the total mass of the fragments must be equal to the total mass of the colliding grains
\begin{equation}
\int_0^{y+z} x b(x;y,z) \mathrm{d}x = y+z.
\label{eq:b_mass_cons}
\end{equation}
Note that these constraints on the distribution of fragments still allows for the mass transfer phenomenon to occur in Eq.~\ref{eq:frag_cont}. The number of fragments produced for each collision is defined by
\begin{equation}
N_{\mathrm{frag}}(y,z) \equiv \int_0^{y+z} b(x;y,z) \mathrm{d}x.
\label{eq:frag_number}
\end{equation}
By definition, a fragmentation event produces at least two fragments, therefore $N_{\mathrm{frag}}  \geq 2$. The extreme case of $N_{\mathrm{frag}}  = 2$ occurs when a small grain breaks in two and one piece is absorbed by the larger grain.

\subsection{Alternative rate equation}
\label{subsec:alternative}
In atmospheric science, the distribution of fragments from droplets collisions is obtained by experiment \citep{Low1982}. However, the usual formulation of the distribution of fragment does not strictly satisfy the local mass conservation constraint in Eq.~\ref{eq:b_mass_cons} \citep{Brown1986,Feingold1988}. To overcome this problem, an alternative rate equation has been formalised by \citet{List1976} and \citet{Gillespie1978} where the loss rate of droplets (or, in astrophysics, dust grains) is written with the distribution function of fragments. The loss term in Eq.~\ref{eq:frag_cont} is derived by counting all the collisions between droplets of mass $[x,x+\mathrm{d}x]$ and any other droplet. In \citet{Gillespie1978}, the loss rate is defined as the destruction rate of droplets of mass $x$ by fragmentation. We consider the fragmentation of a pair of droplets of masses $(x,y)$. The total mass of fragments created by the broken pair is given by
\begin{equation}
\int\limits_0^{x+y} z b(z;x,y) \mathrm{d}z.
\label{eq:mass_frag}
\end{equation}
Therefore, the mass of fragments created by all the broken pairs of droplets of masses $(x,y)$ per unit time is obtained by multiplying Eq.~\ref{eq:mass_frag} with the mean number of collisions per unit time and per unit volume between the droplets of mass $[x,x+\mathrm{d}x]$ and $[y,y+\mathrm{d}y]$ in Eq.~\ref{eq:collision_rate}, i.e.
\begin{equation}
  \mathcal{K}(x,y) f(x,\tau) f(y,\tau)\mathrm{d}y \mathrm{d}x \, \times \, \int\limits_0^{x+y} z b(z;x,y) \mathrm{d}z.
 \label{eq:mass_all_fragments}
\end{equation}
By the mass conservation, Eq.~\ref{eq:mass_all_fragments} equals the mass of all the broken pairs of droplets per unit time. Then, dividing by the mass of one pair $x+y$, we obtain the number of all broken pairs of masses $[x,x+\mathrm{d}x]$ and $[y,y+\mathrm{d}y]$ per unit time, i.e.
\begin{equation}
 \frac{1}{x+y}  \left[ \mathcal{K}(x,y) f(x,\tau) f(y,\tau) \mathrm{d}y \mathrm{d}x \, \times \, \int\limits_0^{x+y} z b(z;x,y) \mathrm{d}z \right] .
\label{eq:rate_all_broken_pairs}
\end{equation}
Equivalently, Eq.~\ref{eq:rate_all_broken_pairs} gives the number of droplets of mass $[x,x+\mathrm{d}x]$ which break per unit time.
We obtain the total rate of loss of droplets of mass $[x,x+\mathrm{d}x]$  by integrating over $y$,
\begin{equation}
\left[ \int_0^{\infty} \mathcal{K}(x,y) f(x,\tau) f(y,\tau) \frac{1}{x+y} \int\limits_0^{x+y} z b(z;x,y)  \mathrm{d}z \mathrm{d}y \right] \mathrm{d}x,
\end{equation}

and the alternative rate equation writes
\begin{equation}
\begin{aligned}
& \frac{\partial f (x,\tau)}{\partial t } = \\
& \qquad \frac{1}{2}\int\limits_{0}^{\infty} \int\limits_{0}^{\infty} \mathbb{1}_{y+z \geq x} b(x;y,z) \mathcal{K}(y,z) f(y,\tau) f(z,\tau) \mathrm{d}y \mathrm{d}z \\
& \qquad \qquad - f(x,\tau) \int\limits_{0}^{\infty} \frac{\mathcal{K}(x,y)   f(y,\tau)}{x+y} \int\limits_0^{x+y} z b(z;x,y) \mathrm{d}z  \mathrm{d}y.
\end{aligned}
\label{eq:alt_frag_cont_DL}
\end{equation}
The mass conservation with the original equation Eq.~\ref{eq:frag_cont_DL} is ensured by the local mass conservation Eq.~\ref{eq:b_mass_cons}. However, the alternative rate equation is of particular interest because \citet{Feingold1988} proved that the mass is conserved for any choice of collision kernel $\mathcal{K}$ and distribution of fragments $b$, because the function $b$ appears in both term on the right-hand side of Eq.~\ref{eq:alt_frag_cont_DL}. It means that the total mass is conserved even if locally the mass is not strictly conserved for each breakup event \citep{Hu1995}. Note that Eq.~\ref{eq:frag_cont_DL} and Eq.~\ref{eq:alt_frag_cont_DL} are two different models to describe the general non-linear fragmentation equation. In the case where the distribution of fragments respects exactly the local mass conservation, we obtain Eq.~\ref{eq:frag_cont_DL} from Eq.~\ref{eq:alt_frag_cont_DL}.

\subsection{Conservative form}
\label{sec:cons_form}

\subsubsection{Original equation}
\label{sec:cons_form_origin_eq}
The fragmentation process is mass conserving, meaning no mass is lost during the process. In order to solve the general non-linear fragmentation equation by using robust numerical schemes, such as finite volume methods or discontinuous Galerkin methods, which conserve the total mass at machine precision, it is required to derive the conservative form of Eq.~\ref{eq:frag_cont_DL}, as an hyperbolic conservation law. To our knowledge, there have been no studies that have derived the conservative form of the general non-linear fragmentation equation (Eq.~\ref{eq:frag_cont_DL}), which we do now. Note that the calculations are formal and the rigorous justification is beyond the scope of this work.

The aim is to find the expression of a mass flux, $F_{\mathrm{frag}}$, such that
\begin{equation}
\frac{\partial g(x,\tau)}{\partial \tau } + \frac{\partial F_{\mathrm{frag}} \left[ g \right] \left( x,\tau \right)}{\partial x} = 0,
\label{eq:frag_cons_DL}
\end{equation}

where $g(x,\tau) \equiv xf(x,\tau)$ is the mass density of grains per unit mass. The term $F_{\mathrm{frag}}[g] \left( x,\tau \right)$ describes a combined mass flux at time $\tau$ crossing mass $x$ resulting from all collisions involving grain pairs with a total mass greater than $x$ and which fragment to produce grains with mass lower than $x$. A relation for the mass flux follows directly from integrating Eq.~\ref{eq:frag_cons_DL} with respect to x,
\begin{equation}
F_{\mathrm{frag}}[g](x,\tau) = - \int\limits_{0}^{x} \frac{\partial g(x',\tau)}{\partial \tau } \mathrm{d}x',
\end{equation}
where the constant $F_{\mathrm{frag}}[g](0,\tau)$ is equal to zero since the process of grain nucleation from gas phase to small nano-particles is neglected.

Multiplying Eq.~\ref{eq:frag_cont_DL} by $x$ and integrating over $x \in [0,\infty)$,  the mass flux is written as
\begin{equation}
\begin{aligned}
&  F_{\mathrm{frag}}[g](x,\tau) = \\
&  - \frac{1}{2} \int\limits_{0}^{x} \int\limits_{0}^{\infty} \int\limits_{0}^{\infty}   \mathbb{1}_{y+z \geq x'} x' b(x';y,z) \mathcal{K}(y,z) \frac{g(y,\tau) g(z,\tau)}{yz} \mathrm{d}y \mathrm{d}z \mathrm{d}x'  \\
&  +  \int\limits_{0}^{x} \int\limits_0^{\infty} \mathcal{K}(y,z) g(z,\tau) \frac{g(y,\tau)}{y} \mathrm{d}y \mathrm{d}z.
\end{aligned}
\label{eq:flux_1}
\end{equation}

A non-zero mass flux across mass $x$ only has physical meaning if the mass of the colliding grains is greater than $x$. We can make this explicit by introducing the operator $\mathbb{1}_{y+z \geq x}$ to the right-hand side of Eq.~\ref{eq:flux_1},
\begin{equation}
\begin{aligned}
& F_{\mathrm{frag}}[g](x,\tau) = \\
&   -  \frac{1}{2} \int\limits_{0}^{x} \int\limits_{0}^{\infty} \int\limits_{0}^{\infty}   \mathbb{1}_{y+z \geq x} x' b(x';y,z) \mathcal{K}(y,z) \frac{g(y,\tau) g(z,\tau)}{yz} \mathrm{d}y \mathrm{d}z \mathrm{d}x'  \\
&   + \int\limits_{0}^{x} \int\limits_0^{\infty} \mathbb{1}_{y+z \geq x} \mathcal{K}(y,z) g(z,\tau) \frac{g(y,\tau)}{y} \mathrm{d}y \mathrm{d}z, 
\end{aligned}
\label{eq:flux}
\end{equation}
noting that the operator $ \mathbb{1}_{y+z \geq x}$ imposes $ \mathbb{1}_{y+z \geq x'} = 1$, since $x' \in [0,x]$. Thus, the conservative form of Eq.~\ref{eq:frag_cont_DL} is Eq.~\ref{eq:frag_cons_DL} with the flux given in Eq.~\ref{eq:flux}. Inserting this expression for the flux into Eq.~\ref{eq:frag_cons_DL} and applying the Leibniz integral rule correctly reduces to Eq.~\ref{eq:frag_cont_DL}, as we would expect. To our knowledge, this is the first time that $F_{\mathrm{frag}}[g](x,\tau)$ has been derived.

\begin{figure}
\centering
\includegraphics[width=\columnwidth,trim=10 140 10 50, clip]{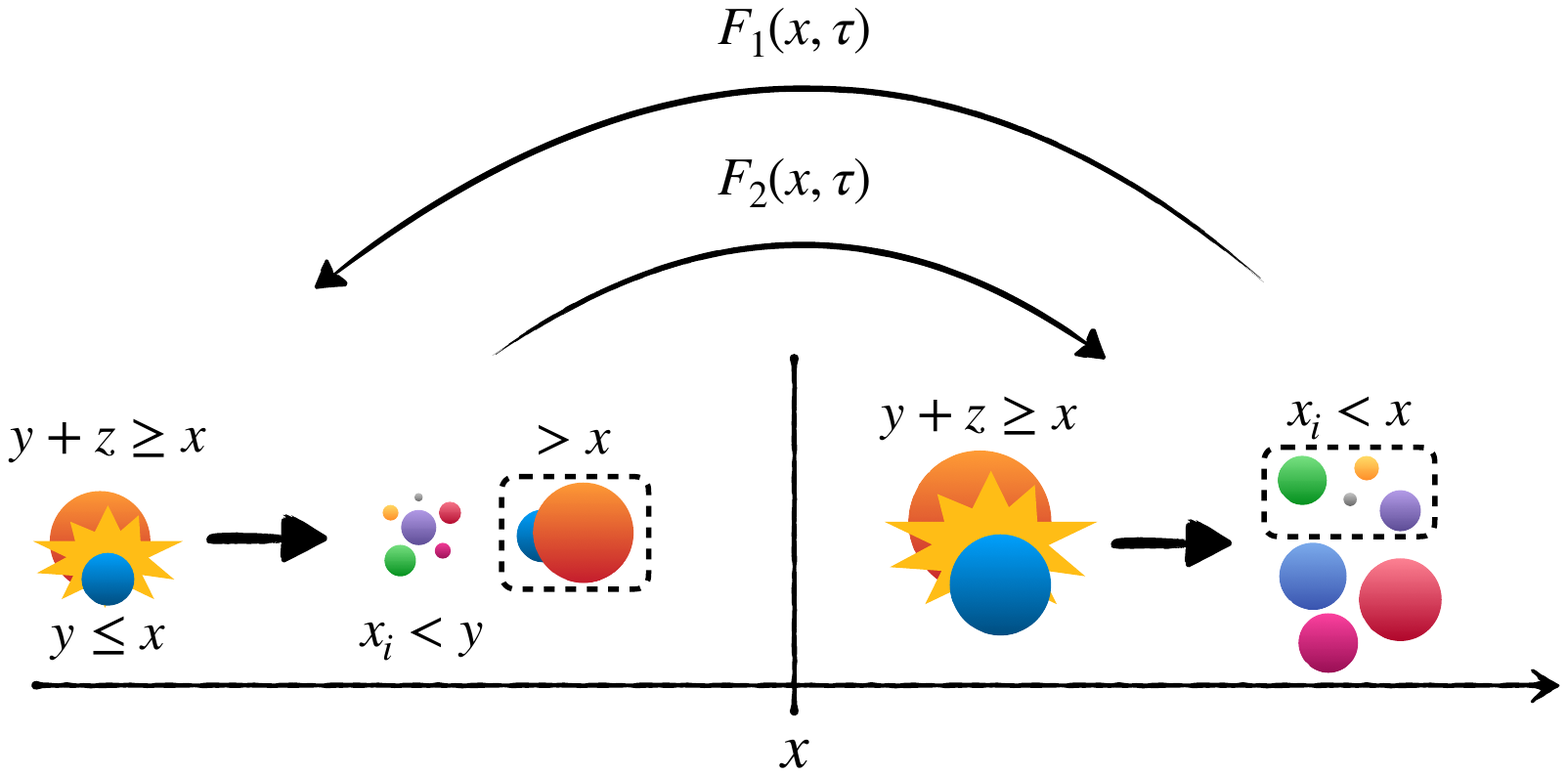}
\caption{Illustration of the fragmentation flux in Eq.~\ref{eq:flux}. The flux is a balance between two terms. The first term describes the production of particles with mass $x_i < x$ from the collision of two particles with total mass $y+z \geq x$. This first term is called $F_1$. The second term, $F_2$, describes the formation of particles with mass greater than $x$ due to the mass transfer process from the collision of two particles with total mass $y+z \geq x$ but with one particle of mass $y \leq x$. The fragmentation flux is a balance between the production of small grains and the production of larger grain due to the mass transfer phenomenon.}
\label{fig:frag_flux}
\end{figure}

The first term on the right-hand side of Eq.~\ref{eq:flux} describes the flux of mass density across the mass $x$ by fragmentation of grains of masses $y$ and $z$ with $y+z \geq x$ producing fragments of mass lower than $x$, as illustrated in Fig.~\ref{fig:frag_flux} with the term $F_1(x,\tau)$. The second term describes the flux of mass through $x$ by fragmentation with mass transfer of grains of masses $y \leq x$ and $z$ with $y+z \geq x$ producing fragments of mass greater than $x$, as shown in Fig.~\ref{fig:frag_flux} with the term $F_2(x,\tau)$. The fragmentation flux is a balance between (i) the negative flux of fragments coming from the breakup of larger grains and (ii) the positive flux of smaller grains gaining mass through mass transfer. In the second term, the gain of mass, due to the mass transfer can be interpreted as a form of coagulation. Indeed the flux of mass in the coagulation process \citep{Tanaka1996,Filbet2004,Liu2019,Lombart2021} is mathematically similar to the second term. The conservative form of the general non-linear fragmentation equation, with the flux of mass Eq.~\ref{eq:flux}, directly highlights the mass transfer phenomenon presented in Sect.~\ref{sec:mt}.

\subsubsection{Alternative equation}
\label{sec:cons_form_alt_eq}
The alternative rate equation in Eq.~\ref{eq:alt_frag_cont_DL} can also be written in a conservative form with an associated mass flux 
\begin{equation}
\left\{
\begin{aligned}
&\frac{\partial g(x,\tau)}{\partial \tau } + \frac{\partial F_{\mathrm{frag,alt}} \left[ g \right] \left( x,\tau \right)}{\partial x} = 0,\\
&  F_{\mathrm{frag,alt}}[g](x,\tau) = \\
& \quad   - \frac{1}{2} \int\limits_{0}^{x} \int\limits_{0}^{\infty} \int\limits_{0}^{\infty}  x' b(x';y,z) \mathbb{1}_{y+z \geq x} \mathcal{K}(y,z) \\
& \qquad \qquad \qquad  \times \frac{g(y,\tau) g(z,\tau)}{yz} \mathrm{d}y \mathrm{d}z \mathrm{d}x' \\
&  \quad +  \int\limits_{0}^{x} \int\limits_0^{\infty} \int\limits_0^{z+y} \mathbb{1}_{z+y \geq x}  x' b(x';z,y)  \mathcal{K}(z,y)  \\
& \qquad \qquad \qquad \qquad  \times \frac{g(z,\tau) g(y,\tau)}{y(z+y)} \mathrm{d}x' \mathrm{d}y \mathrm{d}z.
\end{aligned}
\right.
\label{eq:flux_alt}
\end{equation}
The method to derive the flux $F_{\mathrm{frag,alt}}$ is similar to Sect.~\ref{sec:cons_form_origin_eq}. The alternative rate equation in Eq.~\ref{eq:alt_frag_cont_DL} is only used for benchmarking the discontinuous Galerkin method (Sect.~\ref{sec:num}) since an analytical solution is derived for the constant collision kernel and a specific distribution of fragments which does not strictly satisfy the local mass conservation in Eq.~\ref{eq:b_mass_cons}.

\subsection{Analytical solution of Eq.~\ref{eq:alt_frag_cont_DL}}
\label{sec:analytic}
The only analytical solution to the general non-linear fragmentation equation that we could find in the literature comes from \citet{Feingold1988} in the atmospheric community, who derived an exact solution for Eq.~\ref{eq:frag_cont_DL} assuming a constant kernel, $K_0$, and the following fragment distribution function:
\begin{equation}
\tilde{b}(m;m',m'') = \tilde{\gamma}^2 (m'+m'') e^{-\tilde{\gamma} m},
\label{eq:b_F88}
\end{equation} 
where $K_0$ is a constant with dimension $[\mathrm{length}]^3 [\mathrm{time}]^{-1}$ and $\tilde{\gamma}$ is a positive parameter characterising the physics of the fragmentation process. This distribution of fragments physically represents the case where the number of fragments depends on the mass of the two colliding droplets $m'$ and $m''$ and the shape of the distribution decreases exponentially with $m$. It means that for one collision a large number of small droplets will be generated. The decaying exponential term in Eq.~\ref{eq:b_F88} allows the distribution of fragment to approximatively respect the physical constraint $\forall m > m'+m'',\,\tilde{b}(m;m'm'') \approx 0$. However, Eq.~\ref{eq:b_F88} does not fully satisfy the local mass conservation condition in Eq.~\ref{eq:b_mass_cons}
\begin{equation}
\begin{aligned}
&\int_0^{m'+m''} m \tilde{b}(m;m',m'') \mathrm{d}m \\
&\qquad = (m'+m'') \left[1- \underbrace{e^{-\tilde{\gamma} (m'+m'')} (1+\tilde{\gamma} (m'+m''))}_{\epsilon (\tilde{\gamma},m',m'')} \right] \\
& \qquad < m'+m'',
\end{aligned}
\end{equation}
since here $\epsilon (\tilde{\gamma},m',m'')$ is small but non-zero. \citet{Feingold1988} found that $\epsilon (\tilde{\gamma},m',m'')$ is a few orders of magnitude less than unity whenever the number of fragments is greater than 10. This can easily be shown for a given mass range and number of fragments by evaluating the integral in Eq.~\ref{eq:frag_number}, numerically solving for $\tilde{\gamma}$, and evaluating $\epsilon$. This discussion will be important later for a numerical test that requires a value of $\tilde{\gamma}$  (see Section~\ref{subsec:kconst_tests}).

Let us now show the steps for the derivation of the analytical solution. The dimensionless kernel and distribution of fragments write
\begin{equation}
\mathcal{K}=1,\; b(x;y,z) = \gamma^2 (y+z) e^{- \gamma x},
\label{eq:b_F88_DL}
\end{equation}
where $\gamma \equiv m_0 \tilde{\gamma}$. As the local mass conservation is not exactly respected, we substitute the distribution of fragments and  constant kernel into the alternative rate equation Eq.~\ref{eq:alt_frag_cont_DL}
\begin{equation}
\begin{aligned}
& \frac{\partial f(x,\tau)}{\partial \tau} = \\
& \qquad  \frac{\gamma^2}{2} \int_0^{\infty} \int_0^{\infty}  (y+z) e^{-\gamma x} f(y,\tau) f(z,\tau) \mathrm{d}y \mathrm{d}z  \\
& \qquad \qquad - f(x,\tau) \gamma^2 \int_0^{\infty} \int_0^{\infty} f(y,\tau)  z e^{-\gamma z } \mathrm{d}z \mathrm{d}y,
\end{aligned}
\label{eq:sol_f}
\end{equation}
where the operator $\mathbb{1}_{y+z \geq x}$ is contained in the expression of $b$.  Indeed, the function $b$, approximatively respects the physical constraint $ \forall x > y+z,\,b(x;y,z) \approx 0$.

By continuing the development, we obtain
\begin{equation}
\frac{\partial f(x,\tau)}{\partial \tau} =\gamma^2 e^{-\gamma x} \mathcal{M} \mathcal{N}(\tau)  - f(x,\tau) \mathcal{N}(\tau),
\label{eq:sol_f_step1}
\end{equation}
where 
\begin{equation}
\mathcal{M} \equiv \int_0^{\infty} x f(x,\tau) \mathrm{d}x = 1,\;\mathcal{N}(\tau) \equiv \int_0^{\infty} f(x,\tau) \mathrm{d}x,
\end{equation}
are the dimensionless total mass density and total number of particles. Then, by integrating Eq.~\ref{eq:sol_f_step1} over $x$, we obtain an evolution equation for $\mathcal{N}(\tau)$
\begin{equation}
\frac{\mathrm{d} \mathcal{N}(\tau)}{\mathrm{d}\tau} =\gamma \mathcal{N}(\tau) -  \mathcal{N}(\tau)^2.
\label{eq:eq_diff_N}
\end{equation}
The solution to Eq.~\ref{eq:eq_diff_N} writes
\begin{equation}
\mathcal{N}(\tau) =\frac{e^{\gamma \tau} }{ 1+\frac{e^{\gamma \tau} - 1}{\gamma} },
\label{eq:sol_N}
\end{equation}
where $\mathcal{N}(0) = 1$. We remark that, for a given time, $\mathcal{N}(\tau)$ increases with $\gamma$, a positive integer that characterises the production of fragments. Finally, by injecting Eq.~\ref{eq:sol_N} into Eq.~\ref{eq:sol_f_step1} and solving the differential equation, we obtain the exact solution of Eq.~\ref{eq:sol_f},
\begin{equation}
f(x,\tau) = \frac{f(x,0) + \gamma(e^{\gamma \tau} -1) e^{-\gamma x}}{1+\frac{e^{\gamma \tau}-1}{\gamma}}
\label{eq:exact_sol}
\end{equation}
where $f(x,0)$ is the initial number density. Note that the exact solution only retains a memory of the initial condition for a finite time since, as $t \rightarrow \infty$, the solution converges to the distribution of fragments 
\begin{equation}
f(x,\tau) \underset{\tau \rightarrow \infty}{=} \gamma^2e^{-\gamma x}.
\end{equation}
These characteristics of the exact solution suggest that the solutions to the general non-linear fragmentation equation depend on the initial condition and the evolution is controlled by the distribution of fragments \citep{Feingold1988}. This evolution is also observed for the multiplicative kernel and a power-law distribution of fragments as shown in Sect.~\ref{sec:power_law_distrib}.

%-----------------------------------------------------------------------------------------------------------------
\section{Discontinuous Galerkin method}
\label{sec:DG}
The discontinuous Galerkin (DG) method \citep{Cockburn1989,Zhang2010,Liu2019} is an efficient numerical method to solve the non-linear fragmentation equation \citep{Lombart2022} and the Smoluchowski coagulation equation \citep{Lombart2021} with a reduced number of mass bins. This method is therefore well adapted to treat the coagulation and the fragmentation processes in 3D hydrodynamic simulations. Our objective is to extend the work from \citet{Lombart2022} by applying the DG method to the general non-linear fragmentation equation in Eq.~\ref{eq:frag_cont_DL}. A complete description of the DG method, in the astrophysical context, is given in \citet{Lombart2021}. Here, we outline only the principal steps for the general non-linear fragmentation equation.

\subsection{Summary}
\label{sec:DG_summary}
The unbounded mass interval in the fragmentation equation is reduced to a physical mass range $[x_{\mathrm{min}} >0, x_{\mathrm{max}}< \infty]$ and divided into $N$ bins. Each bin is defined by $I_j = [x_{j-1/2},x_{j+1/2}]$ for $j \in [\![1,N]\!]$. The size and the center position of each bin $j$ are given, respectively, by $h_j = x_{j+1/2}-x_{j-1/2}$ and $x_j = (x_{j+1/2}+x_{j-1/2})/2$. In each bin $j$, the unknown mass density function $g$ is approximated by polynomials of order $k$ defined as a linear combination of Legendre polynomials $\phi_i$ with $i \in [\![0,k]\!]$, 
\begin{equation}
\forall x \in I_j,\; g(x,\tau) \approx g_j(x,\tau) = \sum_{i=0}^k g_j^i(\tau) \phi_i(\xi_j(x)),
\label{eq:approx_g}
\end{equation}
where the function $\xi$ maps the bin interval $ I_j$ into the reference interval $[-1,1]$ where standard Legendre polynomials are defined. Equation~\ref{eq:frag_cons_DL} is multiplied by the Legendre polynomials basis function vector $\bm{\phi}(\xi_j(x)) = [\phi_0(\xi_j(x),...,\phi_k(\xi_j(x)]^{\top}$, then integrated over each bin $I_j$. Therefore, the DG method determines the evolution of the components $g_j^i$ by solving the following equation
\begin{equation}
\begin{aligned}
 \int_{I_j} \frac{\partial  g_j}{\partial \tau} \bm{\phi}(\xi_j(x)) \mathrm{d}x &- \underbrace{ \int_{I_j} F_{\mathrm{frag}}[g](x,\tau) \frac{\partial \bm{\phi}(\xi_j(x))}{\partial x} \mathrm{d}x}_{\text{integral of the flux, see Sect.~\ref{sec:intflux}}} \\
&   + F_{\mathrm{frag}}[g](x_{j+1/2},\tau) \bm{\phi}(\xi_j(x_{j+1/2})) \\
&  - F_{\mathrm{frag}}[g](x_{j-1/2},\tau) \bm{\phi}(\xi_j(x_{j-1/2})) = 0,
\end{aligned}
\label{eq:DG}
\end{equation}
where the term "integral of the flux" is obtained through integration by parts.

\subsubsection{Description of the flux}
The fragmentation flux, given by Eq.~\ref{eq:flux}, is truncated into the physically relevant mass range $[x_{\mathrm{min}},x_{\mathrm{max}}]$ by replacing $0$ and $\infty$ by $x_{\mathrm{min}}$ and $x_{\mathrm{max}}$, respectively,
\begin{equation}
\begin{aligned}
& F_{\mathrm{frag,nc}}[g](x,\tau) = \\
& \quad  - \frac{1}{2} \int\limits_{x_{\mathrm{min}}}^{x} \int\limits_{x_{\mathrm{min}}}^{x_{\mathrm{max}}} \int\limits_{x_{\mathrm{min}}}^{x_{\mathrm{max}}}  x' b(x';y,z) \mathbb{1}_{y+z \geq x} \\
& \qquad \qquad \qquad \qquad  \times \mathcal{K}(y,z) \frac{g(y,\tau) g(z,\tau)}{yz} \mathrm{d}y \mathrm{d}z \mathrm{d}x' \\
& \quad  +  \int\limits_{x_{\mathrm{min}}}^{x} \int\limits_{x_{\mathrm{min}}}^{x_{\mathrm{max}}} \mathbb{1}_{y+z \geq x}  \mathcal{K}(z,y) g(z,\tau) \frac{g(y,\tau)}{y} \mathrm{d}y \mathrm{d}z,
\end{aligned}
\label{eq:flux_trunc_ncons}
\end{equation}
which is a first proposition for the expression of the flux on the physical mass range, but not satisfactory because the flux is not conserved if mass flows out through $x_{\mathrm{max}}$ due to mass transfer, i.e. $F_{\mathrm{frag,nc}}[g](x_{\mathrm{max}},\tau)  \neq 0$. For that reason, this flux is denoted  $ F_{\mathrm{frag,nc}}$  for "non-conservative" flux. Note that no mass flows out through $x_{\mathrm{min}}$, $F_{\mathrm{frag,c}}[g](x_{\mathrm{min}},\tau)  = 0$, meaning that no grains of mass lower than $x_{\mathrm{min}}$ are produced. Therefore, it is necessary to modify the flux in order to conserve the mass for which  $F_{\mathrm{frag,nc}}[g](x,\tau) \big|_{x=x_{\mathrm{min}},x_{\mathrm{max}}} = 0$. The term  $F_{\mathrm{frag,c}}$ stands for the "conservative" flux. To prevent the formation of particles of mass $x \geq x_{\mathrm{max}}$ due to the phenomenon of mass transfer, it is sufficient that the total mass of the two colliding particles is lower than  $x_{\mathrm{max}}+x_{\mathrm{min}}$. For instance, if $ y + z = x_{\rm max} + x_{\rm min} $, the possible values are $ y \leq x_{\rm max} $ and $ z \leq x_{\rm max} $, with the limiting cases being $ y = x_{\rm max} $ and $ z = x_{\rm min} $, or vice versa. Moreover, the operator $\mathbb{1}_{y+z \geq x}$ has to be changed into $\mathbb{1}_{y+z \geq x + x_{\mathrm{min}}}$, since the total mass of the colliding grains is always greater than twice the minimum grain mass. In summary, the condition $x + x_{\mathrm{min}} \leq y+z \leq x_{\mathrm{max}}+ x_{\mathrm{min}}$ must be verified when the range of grain mass is limited to $[x_{\mathrm{min}},x_{\mathrm{max}}]$. By applying these modifications to Eq.~\ref{eq:flux_trunc_ncons}, we obtain the conservative truncation of the flux
\begin{equation}
\begin{aligned}
& F_{\mathrm{frag,c}}[g](x,\tau) = \\
& \quad  - \frac{1}{2} \int\limits_{x_{\mathrm{min}}}^{x} \int\limits_{x_{\mathrm{min}}}^{x_{\mathrm{max}}} \int\limits_{x_{\mathrm{min}}}^{x_{\mathrm{max}}}  x' b(x';y,z) \mathbb{1}_{y+z \geq x + x_{\mathrm{min}}} \mathbb{1}_{x_{\mathrm{max}} + x_{\mathrm{min}}\geq y+z } \\
& \qquad \qquad \qquad \qquad \times \mathcal{K}(y,z) \frac{g(y,\tau) g(z,\tau)}{yz} \mathrm{d}y \mathrm{d}z \mathrm{d}x' \\
& \quad  +  \int\limits_{x_{\mathrm{min}}}^{x} \int\limits_{x_{\mathrm{min}}}^{x_{\mathrm{max}}} \mathbb{1}_{y+z \geq x + x_{\mathrm{min}}}  \mathbb{1}_{x_{\mathrm{max}}+ x_{\mathrm{min}}\geq y+z }  \\
& \qquad \qquad \qquad \qquad \times \mathcal{K}(z,y) g(z,\tau) \frac{g(y,\tau)}{y} \mathrm{d}y \mathrm{d}z.
\end{aligned}
\label{eq:flux_trunc_cons}
\end{equation}
For numerical purpose, it is necessary to extend the operators $\mathbb{1}$ in the limits of integrals. The first term on the right-hand side of Eq.~\ref{eq:flux_trunc_cons} splits into two terms by comparing $z$ to $x$. Then, the operators $\mathbb{1}$ are applied on the variable $y$ for all terms to give
\begin{equation}
\begin{aligned}
& F_{\mathrm{frag,c}}[g](x,\tau) = \\
& \quad - \frac{1}{2} \int\limits_{x_{\mathrm{min}}}^{x} \int\limits_{x_{\mathrm{min}}}^{x} \int\limits_{x-z + x_{\mathrm{min}}}^{x_{\mathrm{max}}-z+x_{\mathrm{min}}}  x' b(x';y,z) \mathcal{K}(y,z) \\
& \qquad \qquad \qquad \qquad \qquad \qquad  \times  \frac{g(y,\tau) g(z,\tau)}{yz} \mathrm{d}y \mathrm{d}z \mathrm{d}x' \\
& \quad - \frac{1}{2} \int\limits_{x_{\mathrm{min}}}^{x} \int\limits_{x}^{x_{\mathrm{max}}} \int\limits_{x_{\mathrm{min}}}^{x_{\mathrm{max}}-z+x_{\mathrm{min}}}  x' b(x';y,z) \mathcal{K}(y,z)  \\
& \qquad \qquad \qquad \qquad \qquad \qquad   \times  \frac{g(y,\tau) g(z,\tau)}{yz} \mathrm{d}y \mathrm{d}z \mathrm{d}x' \\
& \quad  +  \int\limits_{x_{\mathrm{min}}}^{x} \int\limits_{x-z +x_{\mathrm{min}}}^{x_{\mathrm{max}}-z+x_{\mathrm{min}}} \mathcal{K}(z,y) g(z,\tau) \frac{g(y,\tau)}{y} \mathrm{d}y \mathrm{d}z.
\end{aligned}
\label{eq:flux_trunc_cons_2}
\end{equation}
We directly observe that, this new expression of the flux ensures mass conservation of the system evolving in a finite mass range.

\subsubsection{Evaluation of the flux}
\label{sec:eval_flux}
The general non-linear fragmentation equation, Eq.~\ref{eq:frag_cont_DL}, belongs to the family of non-local partial differential equations. The evolution of the number density function $f$ depends on the evaluation of the product of the number density function over all the mass range, similar to the coagulation and non-linear fragmentation equations \citep{Liu2019,Lombart2021,Lombart2022}. In Eq.~\ref{eq:flux_trunc_cons_2}, the evaluation of the flux at the interface $x_{j-1/2}$ depends on the evaluation of $g_j$ in all bins, due to the double integral of the mass density function $g$. The approximation of $g$ is a non continuous function due to the DG method. However, the flux $F_{\mathrm{frag,c}}$, in Eq.~\ref{eq:flux_trunc_cons_2}, is a continuous function of mass across interfaces. This important characteristic of the flux differs from the usual DG solvers applied on local partial differential equations for which the flux is discontinuous and must be reconstructed at interfaces \citep{Cockburn1989,Zhang2010,Guillet2019}.

We assume that the distribution of fragments and the collision kernel are integrable, which is true for the analytical solution in Sect.~\ref{sec:analytic}. The numerical flux, Eq.~\ref{eq:flux_trunc_cons_2}, is analytically integrated over the mass variables by approximating $g$ with $g_j$ in bin $I_j$ (Eq.~\ref{eq:approx_g}). The numerical flux at interface $x_{j-1/2}$ is
\begin{equation}
\begin{aligned}
& F_{\mathrm{frag,c}}[g](x_{j-1/2},\tau) = \\
& \quad  -\frac{1}{2}  \sum_{u=1}^{j-1} \sum_{l'=1}^{j-1}   \int\limits_{I_u} \int\limits_{I_{l'}} \int\limits_{x_{j-1/2}-z + x_{\mathrm{min}}}^{x_{\mathrm{max}}-z+x_{\mathrm{min}}}  x' b(x';y,z) \mathcal{K}(y,z) \\
& \qquad \qquad \qquad \qquad \qquad \qquad  \times  \frac{g(y,\tau) g_{l'}(z,\tau)}{yz} \mathrm{d}y \mathrm{d}z \mathrm{d}x' \\
& \quad   - \frac{1}{2}  \sum_{u=1}^{j-1}  \sum_{l'=j}^N  \int\limits_{I_u} \int\limits_{I_{l'}} \int\limits_{x_{\mathrm{min}}}^{x_{\mathrm{max}}-z+x_{\mathrm{min}}}  x' b(x';y,z) \mathcal{K}(y,z)\\
& \qquad \qquad \qquad \qquad \qquad \qquad  \times  \frac{g(y,\tau) g_{l'}(z,\tau)}{yz} \mathrm{d}y \mathrm{d}z \mathrm{d}x'  \\
& \qquad  + \sum_{l'=1}^{j-1}   \int\limits_{I_{l'}} \int\limits_{x_{j-1/2}-z +x_{\mathrm{min}}}^{x_{\mathrm{max}}-z+x_{\mathrm{min}}} \mathcal{K}(z,y) g_{l'}(z,\tau) \frac{g(y,\tau)}{y} \mathrm{d}y \mathrm{d}z.
\end{aligned}
\label{eq:flux_num_1}
\end{equation}
To be able to analytically calculate the integrals over $y$ , it is necessary to approximate the function $g(y,\tau)$ over the entire mass range, with the following approximation
\begin{equation}
  \begin{aligned}
    &\forall y \in [x_{\mathrm{min}},x_{\mathrm{max}}],\\
    &g\left(y,\tau \right) \approx \\
    & \sum_{l=1}^N \sum_{i=0}^k g_l^i\left(\tau\right) \phi_i(\xi_l(y)) [ H(y-x_{l-1/2}) - H(y-x_{l+1/2})],
  \end{aligned}
  \label{eq:approx_g_global}
\end{equation}
where $H$ is the Heaviside function. Therefore, Eq.~\ref{eq:flux_num_1} writes
\begin{equation}
\begin{aligned}
&  F_{\mathrm{frag,c}}[g](x_{j-1/2},\tau) = \\
&  - \frac{1}{2} \sum_{l'=1}^{j-1} \sum_{l=1}^N \sum_{i'=0}^k \sum_{i=0}^k  g_{l'}^{i'}(\tau) g_l^i(\tau) \\
& \qquad \qquad  \times T_{\mathrm{frag},1}(x_{\mathrm{max}},x_{\mathrm{min}},j,l',l,i',i)  \\
& - \frac{1}{2}  \sum_{l'=j}^N  \sum_{l=1}^N \sum_{i'=0}^k \sum_{i=0}^k g_{l'}^{i'}(\tau) g_l^i(\tau) \\
& \qquad \qquad  \times  T_{\mathrm{frag},2}(x_{\mathrm{max}},x_{\mathrm{min}},j,l',l,i',i)  \\
&  + \sum_{l'=1}^{j-1}  \sum_{l=1}^N  \sum_{i'=0}^k \sum_{i=0}^k g_{l'}^{i'}(\tau) g_l^i(\tau) T_{\mathrm{coag}}(x_{\mathrm{max}},x_{\mathrm{min}},j,l',l,i',i),
\end{aligned}
\label{eq:flux_num_2}
\end{equation}

where the details of $T_{\mathrm{frag},1}$, $T_{\mathrm{frag},2}$ and $T_{\mathrm{coag}}$ are given in appendix \ref{ap:flux}. These terms are calculated analytically with \textsc{Mathematica} before being translated into \texttt{Fortran} and \texttt{C++}. The algorithm is written in \texttt{Fortran}/\texttt{C++} and tested against the \textsc{Mathematica} version for accuracy. $T_{\mathrm{frag},1}$, $T_{\mathrm{frag},2}$ and $T_{\mathrm{coag}}$ are precomputed once at the beginning of the algorithm, since they only depend on the chosen mass grid. This significantly improves the performance of the time solver, similar to \citet{Lombart2021} and \citet{Lombart2022}. In practice, the three terms are stored in arrays with dimensions corresponding to the number of indices they contain. Then, the subarrays for index $j$ are multiplied by $g_{l'}^{i'}(\tau)g_l^i(\tau)$ and summed over all elements to obtain the three terms in the right-hand side of Eq.~\ref{eq:flux_num_2}. The process is repeated for all $j$ to obtain $F_{\mathrm{frag,c}}[g](x_{j-1/2},\tau)$ for all $x_{j-1/2}$.

\begin{figure*}
\centering
\includegraphics[width=\textwidth]{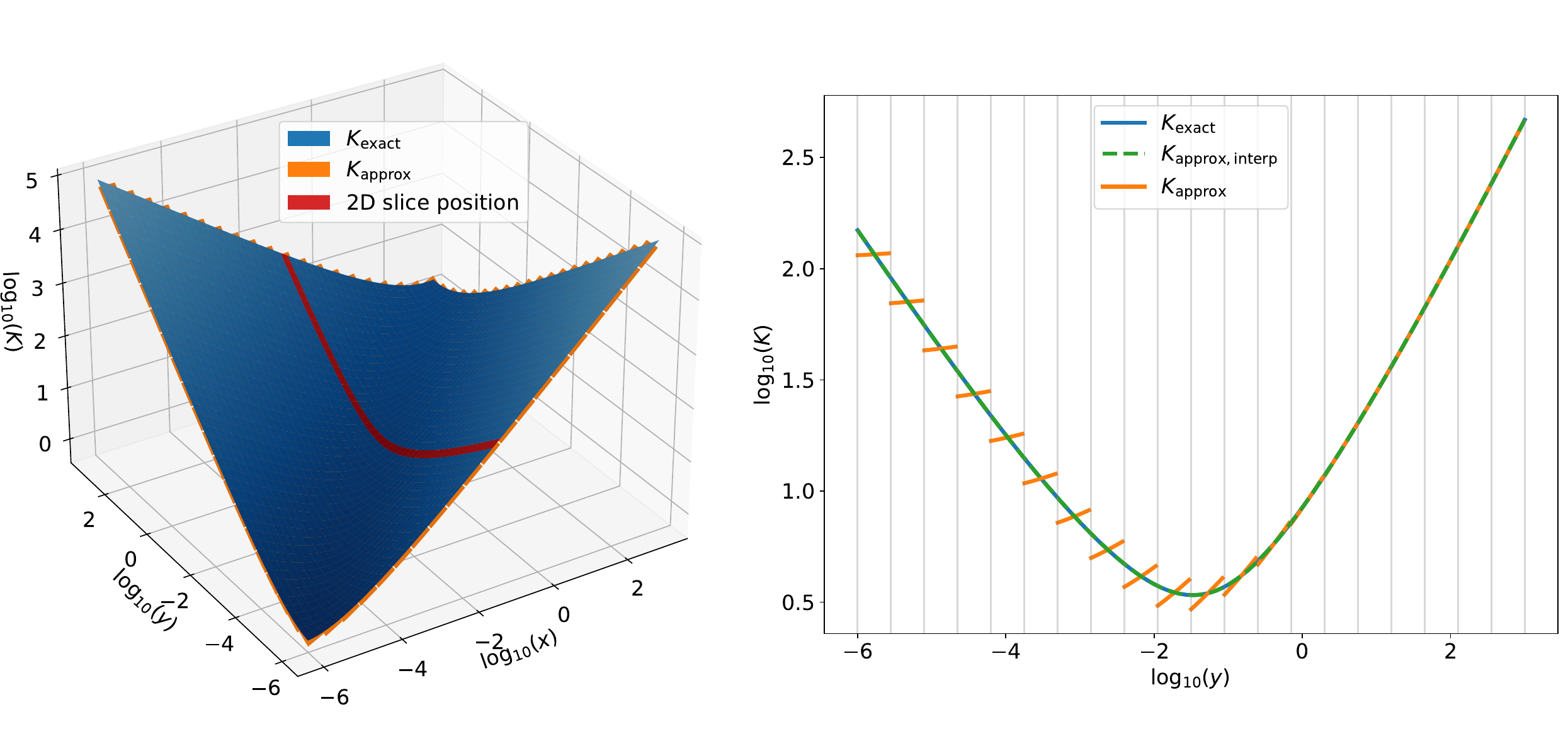}
\caption{Left: Surface plot of the Brownian kernel using the differential velocity relation in Eq.~\ref{eq:dv_br} (blue). The orange patched surface is the approximation of the Brownian kernel using Eq.~\ref{eq:kbr_approx}. The red line indicates the location of the cross-sectional slice found in the right panel. Right: Cross-sectional slice of the Brownian kernel (blue line), with approximations from Eq.~\ref{eq:kbr_approx} (orange line) and a 2D cubic spline interpolation (green dashed line) overplotted. Vertical grey lines represent the boundaries of the bins.}
\label{fig:kbr_approx}
\end{figure*}

The evaluation of the flux assumes that the collision kernel is a 2D continuous function of mass, which is not always the case for physical problems. For instance, the kernel might depend on time through various physical quantities, such as gas temperature. The implementation of the physical collision kernel in Sect.~\ref{sec:kernels} requires the use of the differential velocities between grains given by the 3D hydrodynamic code. The differential velocity term is a 2D piecewise constant function. One approach to couple the DG scheme with the hydrodynamic solver is to compute the integrals with the continuous cross-section, and then multiply by the 2D array for the differential velocity. This approximation of the physical kernel is given by
\begin{equation}
\begin{aligned}
&K(x,y) = \pi \sigma(x,y) \Delta v(x,y) \approx K_{\mathrm{approx}}(x,y), \\
& K_{\mathrm{approx}}(x,y) \equiv \\
& \pi \sigma(x,y) \sum_{l'=1}^N \sum_{l=1}^N \Delta v_{l',l} \mathbb{1}_{x_{l'-1/2} < x < x_{l'+1/2}}   \mathbb{1}_{x_{l-1/2} < y < x_{l+1/2}}.
\end{aligned}
\label{eq:kbr_approx}
\end{equation}
In lieu of working with data obtained from 3D hydrodynamic simulations, we can test the errors from the approximation above by using the well-known differential velocity relation for Brownian motion \citep{Dullemond2005,Brauer2008} , which in dimensionless form is written as
\begin{equation}
\Delta v_{\mathrm{Br}}(x,y) \equiv \sqrt{ \frac{1}{x} + \frac{1}{y}}.
\label{eq:dv_br}
\end{equation}
The Brownian collision kernel is $K_{\mathrm{Br}}(x,y) = \sigma(x,y) \Delta v_{\mathrm{Br}}(x,y)$. To quantify the errors, we define the continuous 2D absolute $L^1$ error
\begin{equation}
\begin{aligned}
&e_{\mathrm{2D,c}} \equiv \frac{\sum_{j,i=1}^N \int_{I_j} \int_{I_i} \left| K_{\mathrm{approx}}(x,y) - K_{\mathrm{Br}}(x,y)  \right| \mathrm{d}x \mathrm{d}y}{\sum_{j,i=1}^N  \int_{I_j} \int_{I_i} K_{\mathrm{Br}}(x,y) \mathrm{d}x \mathrm{d}y}
\end{aligned}
\end{equation}
where the continuous integrals are evaluated with the Gauss-Legendre quadrature using 16 Gauss points.
Figure~\ref{fig:kbr_approx} shows the performance of the kernel approximation in Eq.~\ref{eq:kbr_approx}. The error $e_{\mathrm{2D,c}}$ generated by this approximation is $\sim 3\%$.

While this approximation provides a simple way of coupling to the differential velocities of hydrodynamic solvers, it is important to remember that, in practice, the differential velocities will be time dependent. Consequently, the physical kernel will also be time dependent and the precomputed terms in Eqs.~\ref{eq:flux_Tfrag1}-\ref{eq:flux_Tcoag} have to include the updated array $\Delta v_{i,j}$, e.g. by multiplication with the 2D array $\Delta v_{i,j}$:
\begin{equation}
T_{\mathrm{frag},1} \leftarrow \Delta v_{l',l} \times T_{\mathrm{frag},1}(x_{\mathrm{max}},x_{\mathrm{min}},j,l',l,i',i).
\end{equation}
If we want to enhance the accuracy of the 2D approximation to better preserve the high precision in the DG scheme, we can interpolate the physical collision kernel, as illustrated in the right panel of Figure~\ref{fig:kbr_approx}. The error of the 2D cubic spline interpolation (green dashed line) of the discrete kernel array,
\begin{equation}
\forall (i,j) \in [\![1,N]\!]^2,\; K_{i,j} = \pi \sigma(x_i,x_j) \Delta v_{\mathrm{Br}}(x_i,x_j),
\end{equation}
is only $\sim 0.005 \%$. The 2D interpolation gives this high accuracy only in log-log space where the grid is regular. The inconvenience of using the interpolation for the coupling is that the approximation of the kernel has to be integrated to evaluate the terms in Eqs.~\ref{eq:flux_Tfrag1}-\ref{eq:flux_Tcoag} after each hydrodynamic timestep. This would significantly reduce the global performance of the DG scheme, which relies heavily on precomputed integrals or efficient quadrature methods (see Sect.~\ref{sec:DG_architecture}) to quickly generate the terms in Eqs.~\ref{eq:flux_Tfrag1}-\ref{eq:flux_Tcoag}.

In addition to the differential velocities of grains given by the hydrodynamic solver, sub-grid models for the differential velocities, such as Brownian motion, can be added in the DG scheme like 
\begin{equation}
\Delta v(x_i,x_j) = \sqrt{\Delta v_{\mathrm{hydro}}(x_i,x_j)^2 + \Delta v_{\mathrm{subgrid}}(x_i,x_j)^2}.
\end{equation}
If the sub-grid model is considered independent of time, such as the Brownian motion in Eq.~\ref{eq:dv_br}, the integral terms for the DG scheme need only to be precomputed once.

\subsubsection{Integral of the flux}
\label{sec:intflux}
Now that we have a relation for the conservative flux, $F_{\mathrm{frag,c}}[g](x,\tau)$, we define $\mathcal{F}_{\mathrm{frag},c}$ the term with the integral of the conservative flux, in Eqs.~\ref{eq:DG} and \ref{eq:flux_trunc_cons_2}, which takes the form
\begin{equation}
\begin{aligned}
&  \mathcal{F}_{\mathrm{frag,c}}(j,k',\tau) \equiv  \\
&   - \frac{1}{2} \int_{I_j} \int\limits_{x_{\mathrm{min}}}^{x} \int\limits_{x_{\mathrm{min}}}^{x} \int\limits_{x-z + x_{\mathrm{min}}}^{x_{\mathrm{max}}-z+x_{\mathrm{min}}}  x' b(x';y,z) \mathcal{K}(y,z) \partial_x \phi_{k'}(\xi_j(x)) \\
& \qquad \qquad \qquad \qquad \qquad \qquad  \times  \frac{g(y,\tau) g(z,\tau)}{yz} \mathrm{d}y \mathrm{d}z \mathrm{d}x' \mathrm{d}x \\
&    -  \frac{1}{2} \int_{I_j} \int\limits_{x_{\mathrm{min}}}^{x} \int\limits_{x}^{x_{\mathrm{max}}} \int\limits_{x_{\mathrm{min}}}^{x_{\mathrm{max}}-z+x_{\mathrm{min}}}  x' b(x';y,z) \mathcal{K}(y,z) \partial_x \phi_{k'}(\xi_j(x)) \\
& \qquad \qquad \qquad \qquad \qquad \qquad  \times  \frac{g(y,\tau) g(z,\tau)}{yz} \mathrm{d}y \mathrm{d}z \mathrm{d}x' \mathrm{d}x \\
&   +  \int_{I_j} \int\limits_{x_{\mathrm{min}}}^{x} \int\limits_{x-z +x_{\mathrm{min}}}^{x_{\mathrm{max}}-z+x_{\mathrm{min}}} \mathcal{K}(z,y) \partial_x \phi_{k'}(\xi_j(x)) \\
 & \qquad \qquad \qquad \qquad \qquad \qquad  \times g(z,\tau) \frac{g(y,\tau)}{y} \mathrm{d}y \mathrm{d}z \mathrm{d}x,
\end{aligned}
\label{eq:intflux}
\end{equation}
where $k' \in [\![0,k]\!]$. For each term in the right-hand side of Eq.~\ref{eq:intflux}, the integrals over $z$ are split into two integrals:  $z \in [x_{\mathrm{min}},x_{j-1/2}]$ and $z \in [x_{j-1/2},x]$ for the first term; $z \in [x,x_{j+1/2}]$ and $z \in [x_{j+1/2},x_{\mathrm{max}}]$ for the second term; and $z \in [x_{\mathrm{min}},x_{j-1/2}]$ and $z \in [x_{j-1/2},x]$ for the third term. We apply the same method as in Sect.~\ref{sec:eval_flux} to obtain the numerical integral of the flux 
\begin{equation}
\begin{aligned}
&  \mathcal{F}_{\mathrm{frag,c}}(j,k',\tau) =\\
&  - \frac{1}{2} \sum_{l'=1}^{j-1} \sum_{l=1}^N \sum_{i'=0}^k \sum_{i=0}^k g_{l'}^{i'}(\tau) g_l^i(\tau) \\
& \qquad \qquad  \times \mathcal{T}_{\mathrm{frag},1,A}(x_{\mathrm{max}},x_{\mathrm{min}},j,k',l',l,i',i) \\
&  - \frac{1}{2} \sum_{l=1}^N \sum_{i'=0}^k \sum_{i=0}^k g_{j}^{i'}(\tau) g_l^i(\tau) \\
& \qquad \qquad  \times \mathcal{T}_{\mathrm{frag},1,B}(x_{\mathrm{max}},x_{\mathrm{min}},j,k',l,i',i) \\
&  - \frac{1}{2}  \sum_{l=1}^N \sum_{i'=0}^k \sum_{i=0}^k g_{j}^{i'}(\tau) g_l^i(\tau)\\
& \qquad \qquad  \times \mathcal{T}_{\mathrm{frag},2,A}(x_{\mathrm{max}},x_{\mathrm{min}},j,k',l,i',i)\\
&  - \frac{1}{2}  \sum_{l'=j+1}^N \sum_{l=1}^N \sum_{i'=0}^k \sum_{i=0}^k g_{l'}^{i'}(\tau) g_l^i(\tau)\\
& \qquad \qquad \times \mathcal{T}_{\mathrm{frag},2,B}(x_{\mathrm{max}},x_{\mathrm{min}},j,k',l',l,i',i) \\
&  + \sum_{l'=1}^{j-1} \sum_{l=1}^N \sum_{i'=0}^k \sum_{i=0}^k g_{l'}^{i'}(\tau) g_l^i(\tau)  \\
& \qquad \qquad  \times \mathcal{T}_{\mathrm{coag},A}(x_{\mathrm{max}},x_{\mathrm{min}},j,k',l',l,i',i)  \\
&   +  \sum_{l=1}^N \sum_{i'=0}^k \sum_{i=0}^k g_{j}^{i'}(\tau) g_l^i(\tau) \\
& \qquad \qquad  \times \mathcal{T}_{\mathrm{coag},B}(x_{\mathrm{max}},x_{\mathrm{min}},j,k',l,i',i),
\end{aligned}
\label{eq:intflux_2}
\end{equation}
where the definition of the different terms and the method to compute them are detailed in appendix~\ref{ap:intflux}. Importantly, since the integral terms in $F_{\mathrm{frag,c}}$ and $\mathcal{F}_{\mathrm{frag,c}}$ are evaluated analytically, the accuracy depends only on the order of polynomials to approximate $g$.

\subsection{Scaling limiter}
\label{sec:limiter}
The DG scheme needs the use of a scaling limiter to preserve the positivity of the numerical solutions \citep{Zhang2010,Liu2019,Lombart2022}. The scaling limiter is applied by a reconstruction step based on cell averaging. The reconstructed polynomials in each bin writes \citep{Liu2019,Lombart2022}
\begin{equation}
\left\{
\begin{aligned}
& \forall j \in [\![1,N]\!],\;p_j(x,\tau) \equiv \psi_j(\tau) \left( g_j(x,\tau) - \overline{g}_j(\tau) \right) +  \overline{g}_j(\tau) \,\\
&\psi_j(\tau) \equiv \mathrm{min} \left\{ 1, \left| \frac{\overline{g}_j(\tau)}{ m_j(\tau) - \overline{g}_j(\tau)} \right| \right\},
\end{aligned}
\right.
\end{equation}
where $m_j(\tau) \equiv \underset{x \in I_j}{\mathrm{min}}\; g_j(x,\tau)$ and $\overline{g}_j$ is the average of $g_j$ in bin $I_j$
\begin{equation}
\overline{g}_j(\tau) \equiv \frac{1}{h_j} \int_{I_j} g_j(x,\tau) \mathrm{d}x = g_j^0(\tau).
\end{equation}
Since $g_j$ is approximated by orthogonal polynomials, the scaling limiter coefficient $\psi_j$ is applied to all components of $p_j$ except the average value which writes $p_j^0(\tau) = g_j^0(\tau)$. In \citet{Lombart2021} and \citet{Lombart2022}, there is a mistake in the application of the scaling limiter. We present here the correct implementation of the scaling limiter for the DG scheme by following the description in \citet{Guillet2019}. For the initialisation of the DG scheme, the $L^2$ projection of the initial mass distribution onto the Legendre basis can result in negative values for the numerical mass density. At that step, the use of the scaling limiter is required, therefore, the following replacement is applied on the components of $g_j$
\begin{equation}
\forall j \in [\![1,N]\!],\; \forall i \geq 1,\; g_j^i(\tau) \leftarrow \psi_j g_j^i(\tau).
\label{eq:limiter}
\end{equation}
Then, after each evolution of the component of $g_j$, meaning after each timestep in the SSPRK method (see Sect.~\ref{sec:CFL}), the replacement in Eq.~\ref{eq:limiter} has to be applied to ensure the positivity of the numerical solution.

\subsection{CFL criterion}
\label{sec:CFL}
The DG scheme is associated to the Strong Stability Preserving Runge-Kutta third-order method (SSPRK) to ensure that the high-order accuracy is preserved during time \citep{Liu2019,Lombart2021,Lombart2022}. The SSPRK time solver is stable under a suitable Courant-Friedrichs-Lewy (CFL) condition on the timestep. A precise estimation of the CFL condition has been investigated in several studies on coagulation and fragmentation processes \citep{Filbet2004,Liu2019,Lombart2021,Laibe2022,Lombart2022}, but none so far dedicated to the evaluation of the CFL condition for the general non-linear fragmentation (probably due to the high degree of complexity in the flux given in Eq.~\ref{eq:flux_trunc_cons_2}). The method based on the Laplace transform used in \citet{Laibe2022}  seems difficult to be applied to the general non-linear fragmentation. The method to determine the CFL criterion developed in \citet{Filbet2004,Lombart2022} does not provide an analytical expression to evaluate the CFL criterion because of the non linearity of the general non-linear fragmentation. Therefore, we propose to numerically evaluate the CFL condition. The DG scheme with SSPRK method for order 0 corresponds to the forward Euler discretisation, i.e.
\begin{equation}
\begin{aligned}
& g_j^{0,n+1} = \\
& g_j^{0,n} + \frac{\Delta \tau}{\Delta x_j} \left[ F_{\mathrm{frag}}[g_j^n](x_{j+1/2},\tau) - F_{\mathrm{frag}}[g_j^n](x_{j-1/2},\tau) \right],
\end{aligned}
\end{equation}
for the $n$-th time step. The CFL condition is evaluated to ensure the positivity of the bin average at time step $n+1$, i.e. $ \overline{g}_j^{n+1} = g_j^{0,n+1} \geq 0$ \citep{Liu2019,Lombart2021,Lombart2022}. We obtain the condition
\begin{equation}
\begin{aligned}
& \forall j \in [\![1,N]\!], \\
& g_j^{0,n+1} \geq 0 \\
& \Rightarrow  \Delta \tau_{\mathrm{CFL}} \leq \frac{g_j^{0,n} \Delta x_j}{\left|F_{\mathrm{frag}}[g_j^n](x_{j+1/2},\tau) - F_{\mathrm{frag}}[g_j^n](x_{j-1/2},\tau) \right|}.
\end{aligned}
\label{eq:dt_CFL}
\end{equation}

In practice, we define 
\begin{equation}
\Delta \tau_{\mathrm{CFL}} \equiv \underset{j}{\mathrm{min}} \left[ \frac{g_j^{0,n} \Delta x_j}{\left|F_{\mathrm{frag}}[g_j^n](x_{j+1/2},\tau) - F_{\mathrm{frag}}[g_j^n](x_{j-1/2},\tau) \right|} \right],
\end{equation}
and the time solver is executed with the timestep $\mathrm{d} \tau = C_{\mathrm{CFL}}\Delta \tau_{\mathrm{CFL}}$, where $C_{\mathrm{CFL}} \in [0,1]$ is the CFL coefficient to ensure stability, typically $C_{\mathrm{CFL}} = 0.3$ for SSPRK order 3 \citep{Gottlieb2015}.

%-----------------------------------------------------------------------------------------------------------------
\section{Numerical results}
\label{sec:num}
The DG scheme presented in Sect.~\ref{sec:DG} is tested against the analytical solution outlined in Sect.~\ref{sec:analytic}. Tests are performed with a limited number of mass bins, i.e. $N=20$, in order to reflect the constraints from 3D hydrodynamical simulations. We also performed simulations for a power-law mass distribution of fragments.

\subsection{Evaluation of errors}
\label{sec:errors}
Error measurements are performed to determine the experimental order of convergence (EOC) and the efficiency of the DG algorithm, similar to \citet{Liu2019} and \citet{Lombart2022}.  The continuous $L^1$ norm used to evaluate the errors is
\begin{equation}
\begin{aligned}
e_{\mathrm{c},N}(\tau) &\equiv \sum_{j=1}^N \int\limits_{I_j} \left| g_j(x,\tau) - g_{\mathrm{exact}}(x,\tau) \right| \mathrm{d}x,\\
& \approx \sum_{j=1}^N \frac{h_j}{2} \sum_{\alpha=1}^R \omega_{\alpha} \left| g_j(x_j^{\alpha},\tau) - g_{\mathrm{exact}}(x_j^{\alpha},\tau) \right|,
\end{aligned}
\end{equation}
where the integral is approximated by the Gauss-Legendre quadrature method. The terms $g_{\mathrm{exact}}$ and $g_j$ are respectively the exact and the numerical solutions. $\omega_{\alpha}$ and $x_j^{\alpha}$ are the weights and the points in bin $I_j$ for the Gauss-Legendre quadrature method. We use $R=16$ gaussian points. The discrete $L^1$ error is evaluated at the geometric mean value, which is the mass of the representative grain in each bin. The $L^1$ norm is evaluated in logarithmic scale with the following change of variable
\begin{equation}
\begin{aligned}
||f||_1 \equiv \int\limits_{x_{\mathrm{min}}}^{x_{\mathrm{max}}} |f(x)| \mathrm{d}x &= \sum_{j=1}^N \int_{I_j} |f(x)| \mathrm{d}x \\
&\underset{v \leftarrow \log(x)} {\approx} \sum_{j=1}^N  \int\limits_{\log(x_{j-1/2})}^{\log(x_{j+1/2})}  e^{v} |f(e^v)| \mathrm{d}v,
\end{aligned}
\end{equation}
where $f$ is an arbitrary function. Then, we apply the midpoint rule to evaluate the integral to obtain the discrete $L^1$ error,
\begin{equation}
e_{\mathrm{d},N}(\tau) \equiv \sum_{j=1}^N \log \left(\frac{x_{j+1/2}}{x_{j-1/2}} \right)  \hat{x}_j | g_j(\hat{x}_j,\tau) - g(\hat{x}_j,\tau)|,
\end{equation}
where $\hat{x}_j = \sqrt{x_{j+1/2} x_{j-1/2}}$ is the geometric mean in bin $j$.

The EOC is defined as
\begin{equation}
\mathrm{EOC} \equiv \frac{\ln \left(\frac{e_N(\tau)}{e_{2N}(\tau)} \right)}{\ln(2)},
\end{equation}
where $e_N$ is the error for $N$ bins and $e_{2N}$ for $2N$ bins. The error can be the continuous or discrete $L^1$ error. To avoid time-stepping errors, the errors are calculated after one timestep. The numerical total mass density of the system is the first moment of $g(x,\tau)$ and writes
\begin{equation}
M_{1,N}(\tau) \equiv \int\limits_{x_{\mathrm{min}}}^{x_{\mathrm{max}}} g(x,\tau) \mathrm{d}x = \sum_{j=1}^N h_j g_j^0(\tau).
\end{equation}
Therefore, the mass conservation is analysed with the absolute error of the total mass density given by
\begin{equation}
e_{M_1,N}(\tau) \equiv \frac{| M_{1,N}(\tau) - M_1|}{M_1},
\end{equation}
where $M_1$ is the first moment of the exact solution $g_{\mathrm{exact}}$, which is constant in time.

\subsection{Implementation details}
\label{subsec:benchmark_frag}
Simulations are performed for a mass range $x \in [x_{\mathrm{min}},x_{\mathrm{max}}] $, with $x_{\mathrm{min}} = 10^{-6}$ and $x_{\mathrm{max}} =10^3$, in order to follow the formation of small grains from the fragmentation of larger grains. Numerical solutions of the mass density are shown only for polynomial approximations of order $k \in \{0,1,2,3\}$. Tests are performed with \textsc{Fortran} and the errors are calculated with \textsc{Python}. The initial components $g_j^i$ are evaluated by the $L^2$ projection of the initial condition $g_0(x)$ on the Legendre polynomials basis in each bin (see Eq.17 in \citealt{Lombart2021}) 
\begin{equation}
\left\{
\begin{aligned}
&g_0(x) \equiv xe^{-x}, \\
&\forall j \in [\![1,N]\!],\; g_j^i(0) = \frac{2}{h_j d_i} \int_{I_j}  g_0(x) \phi_i(\xi_j(x)) \mathrm{d}x.
\end{aligned}
\right.
\end{equation}
where $d_i$ is the normalisation coefficient of the Legendre polynomial basis defined as 
\begin{equation}
d_i \equiv \frac{2}{2i+1}.
\end{equation}
The integral is evaluated by a Gauss-Legendre quadrature method with five points. To avoid numerical instabilities, a minimum physical threshold is set to $10^{-20}$. Therefore, during the simulations, any polynomials $g_j(x,\tau)$ with mean value below the threshold are changed into a constant polynomial with value $10^{-20}$. This step is applied after each update of the $g_j^i(\tau)$, i.e. for each sub timestep in the SSPRK solver (Sect.~\ref{sec:CFL} and Eq.38 in \citealt{Lombart2021}). We choose the CFL coefficient $C_{\mathrm{CFL}}=0.3$ and the algorithm is run sequentially on the Apple M1 Max chip. We use the \texttt{gfortran v13.1.0} compiler.

\subsection{Distribution of fragments from \citet{Feingold1988} }
\label{subsec:kconst_tests}
Numerical solutions for the constant kernel with the specific distribution of fragments, Eq.~\ref{eq:b_F88_DL}, are benchmarked against the analytical solution presented in Sect.~\ref{sec:analytic}. Tests are performed in quadruple precision, in order to maintain stability of the DG scheme. 
Simulations are performed from $\tau =0$ to $\tau = 3 \times 10^{-3}$ with $100$ constant timesteps $\Delta \tau = 3 \times 10^{-5}$. The coefficient $\gamma$ in Eq.~\ref{eq:b_F88_DL} is set to $10^4$, meaning that the majority of the grains produced by fragmentation are of mass $10^{-4}$. The evolution of the numerical solutions with $N=20$ bins and polynomials of order $k \in \{0,1,2,3\}$  are shown in Fig.~\ref{fig:kconst_linlog_loglog}. Initially the majority of the mass is represented by large particles of mass $x=1$. Then, the fragmentation process occurs and a large number of fragments with mass $x=10^{-4}$ are produced. At $\tau = 8.10^{-5}$, the number of fragments with mass $x=10^{-4}$ is so large that the majority of the total mass starts to be represented by these fragments. We observe at that time a double-humped curve. The number of the large grains of mass $x=1$ continue to decrease. At the final time, the analytical and numerical solutions converge to the mass distribution of fragments.

\subsubsection{Positivity and mass conservation}
\label{subsubsec:positivity_mass_cons}
\begin{figure*}
\centering
\includegraphics[width=0.9\textwidth]{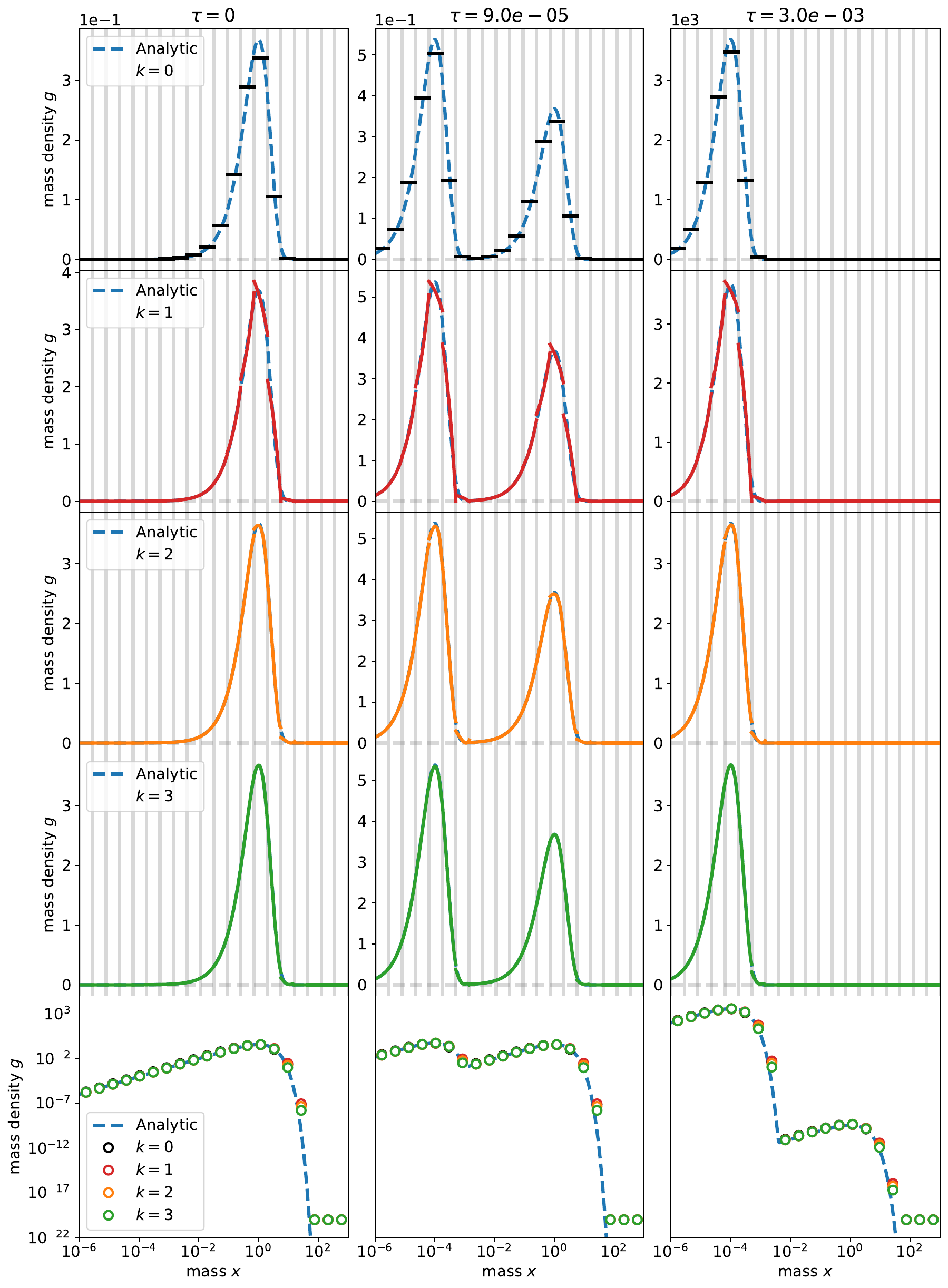}
\caption{Test for the constant kernel and the distribution of fragments given in Eq.~\ref{eq:b_F88_DL}. The first four rows show the numerical solution (solid lines) for $N=20$ bins and the listed value of $k \in \{0,1,2,3\}$ at time $\tau = 0$ (left column), $\tau = 9 \times 10^{-5}$ (middle column) and $\tau = 3 \times 10^{-3}$ (right column). In these rows, the mass density is in linear scale and the mass in logarithmic scale. The exact solution $g(x,\tau)$ is given by the blue dashed line. Vertical grey lines represent the boundaries of the bins. The last row shows the same numerical solutions in log-log scale. The accuracy improves with increasing values of $k$. Order 3 recovers the two peaks at $\tau = 3 \times 10^{-3}$ with an accuracy of order $\sim 0.1\%$ for masses below the cusp, and $\sim 0.01\%$ above.}
\label{fig:kconst_linlog_loglog}
\end{figure*}

\begin{figure*}
\centering
\subfloat[][]{\includegraphics[width=0.9\columnwidth]{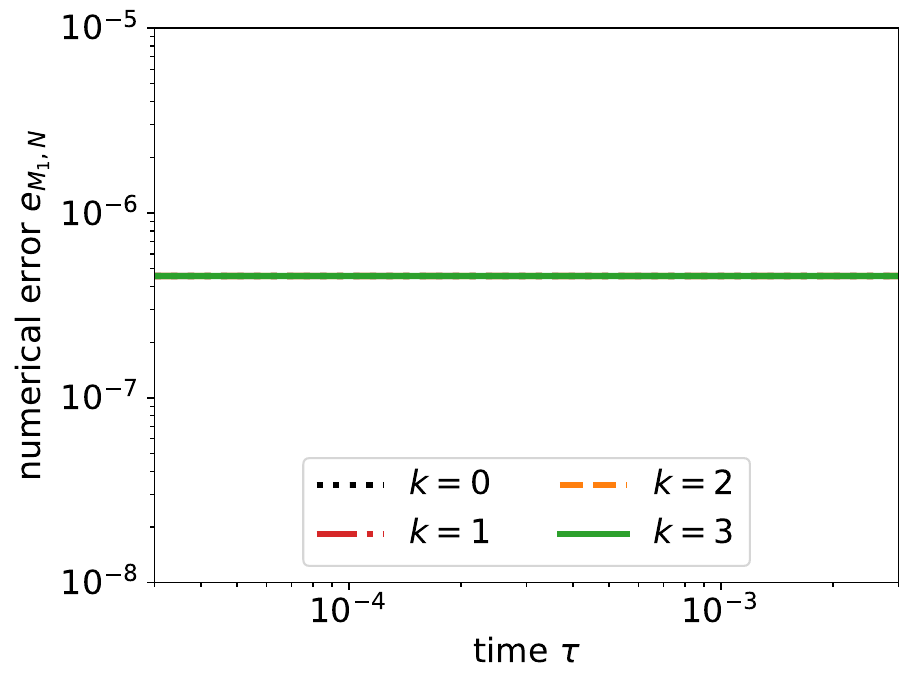}\label{fig:kconst_a}}
\subfloat[][]{\includegraphics[width=0.9\columnwidth]{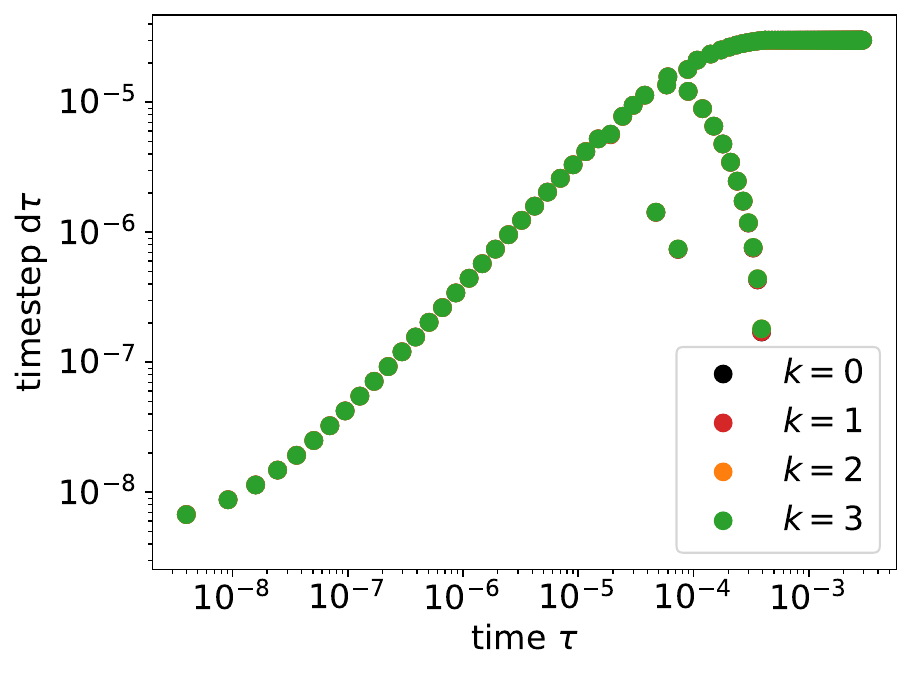}\label{fig:kconst_b}}\\
\subfloat[][]{\includegraphics[width=0.9\columnwidth]{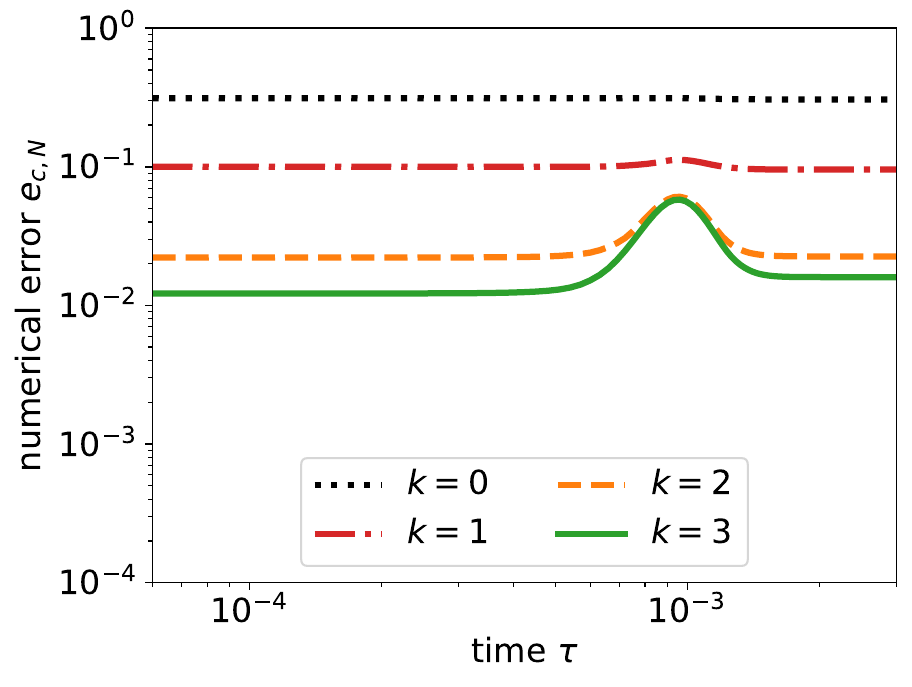}\label{fig:kconst_c}}
\subfloat[][]{\includegraphics[width=0.9\columnwidth]{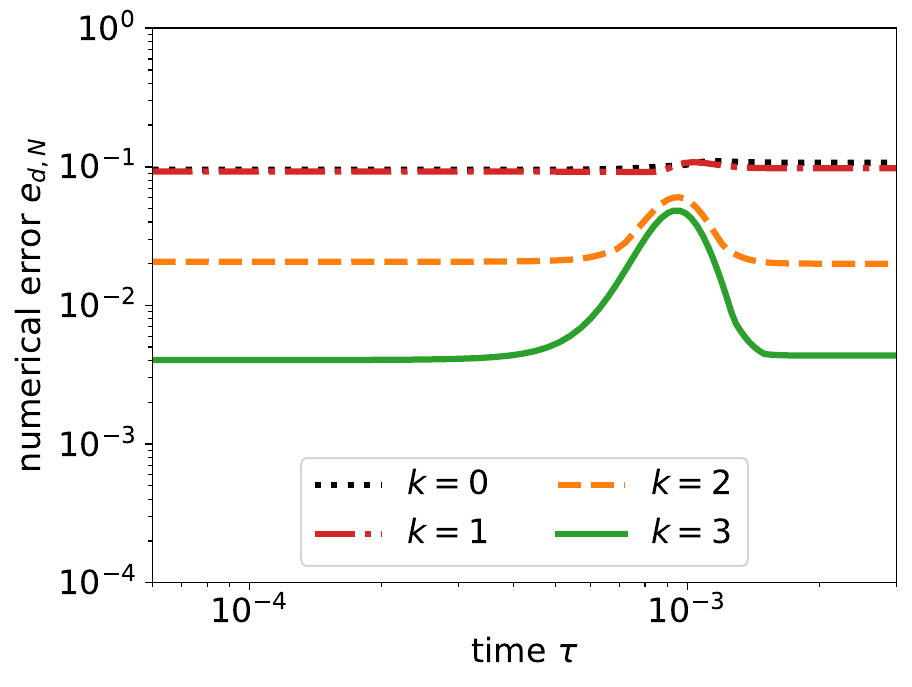}\label{fig:kconst_d}}\\
\subfloat[][]{\includegraphics[width=0.9\columnwidth]{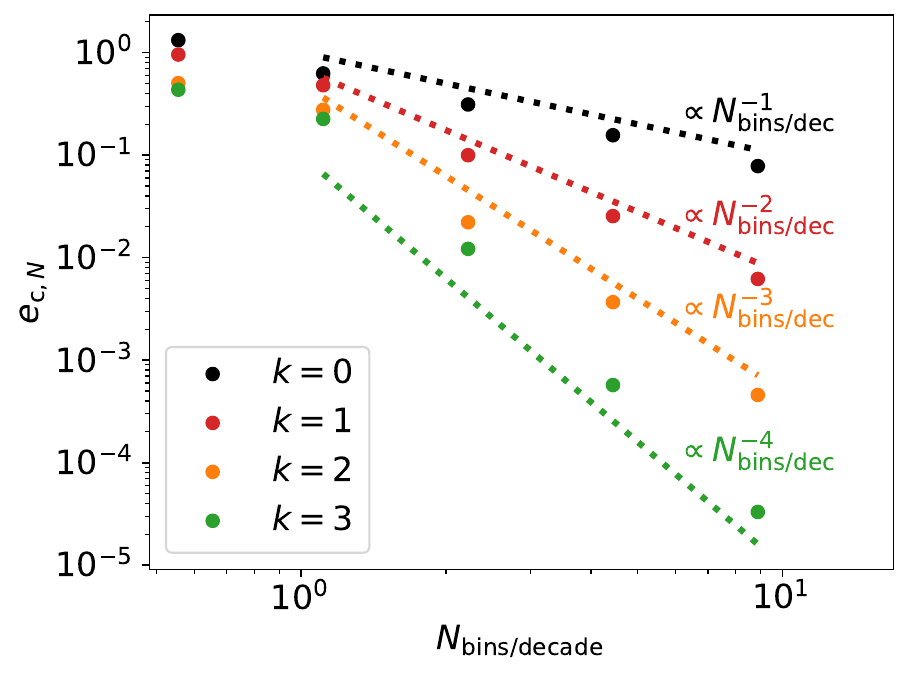}\label{fig:kconst_e}}
\subfloat[][]{\includegraphics[width=0.9\columnwidth]{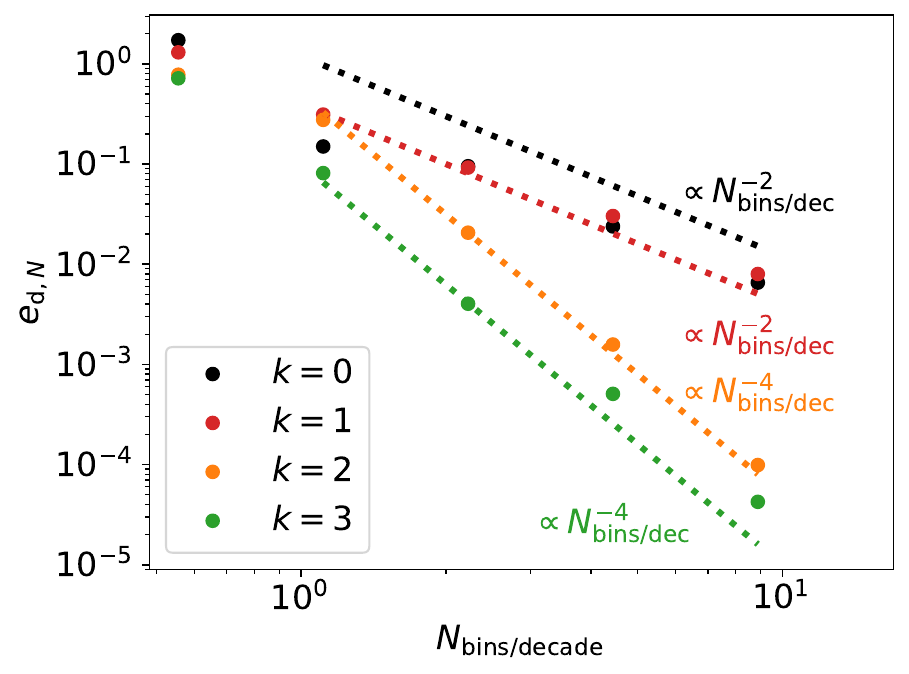}\label{fig:kconst_f}}
\caption{Test for the constant kernel and the distribution of fragments given in Eq.~\ref{eq:b_F88_DL} with $N=20$ bins. Panel~\protect\subref{fig:kconst_a}: evolution of the absolute error $e_{M_1,N}$ of the total mass. The mass is conserved for each order $k$. Panel~\protect\subref{fig:kconst_b}: evolution of the timestep $\mathrm{d} \tau$. The timestep steadily increases with time as more fragments are produced until reaching the global timestep value $\Delta \tau = 3 \times 10^{-5}$. The plateau shows that from $\tau = 10^{-4}$, the CFL condition is greater than $ 3 \times 10^{-5}$. Panels~\protect\subref{fig:kconst_c},\protect\subref{fig:kconst_d}: time evolution of the $L^1$ continuous and discrete errors, showing they remain bounded at large times. The variation of errors observed at $\tau = 10^{-3}$ is explained in Sect.~\ref{subsubsec:stability}. Panels~\protect\subref{fig:kconst_e},\protect\subref{fig:kconst_f}: the errors $e_{c,N}$ and $e_{d,N}$ are plotted versus the number of bins per decade. The experimental order of convergence is $\mathrm{EOC} = k+1$ for $e_{c,N}$ and for $e_{d,N}$, $\mathrm{EOC} = k+1$ for polynomials of odd orders and $\mathrm{EOC}=k+2$ for polynomials of even orders. An accuracy of order $\sim 0.1\%$ is achieved with $\sim 4$ bins/decade or $\sim 1\%$ with  $\sim 2$ bins/decade for $k=3$.}
\label{fig:kconst_analysis}
\end{figure*}

Figure~\ref{fig:kconst_linlog_loglog} shows the numerical results of the mass density (linear scale for the first fourth rows and log scale for the last row) versus the mass $x$ in log scale. For all polynomial orders, the numerical solutions remain positive thanks to the combination of the CFL-limited SSPRK time solver (Sect.~\ref{sec:CFL}) and the scaling limiter (Sect.~\ref{sec:limiter}). Figure~\ref{fig:kconst_analysis},\subref{fig:kconst_a} shows the absolute error $e_{M_{1,N}}$ from $\tau=10^{-5}$ to $\tau=3 \times 10^{-3}$. The mass is conserved for all orders $k$, as expected from the design of the DG scheme with the conservative flux.

\subsubsection{Accuracy}
\label{subsubsec:accuracy}
Figure~\ref{fig:kconst_linlog_loglog} shows the accuracy of the numerical solutions improves with the order of the polynomials. The two plots in log-log scale at $\tau = 8 \times 10^{-5}$ and $\tau = 3 \times 10^{-3}$ show that the numerical diffusion is reduced (up to a factor 10) in the decreasing parts (exponential decays) as the order of polynomials increases. The major part of the total mass is generally localised in the maximum of the curve. However, at $\tau = 8 \times 10^{-5}$, the curve exhibits two maxima, indicating that a significant proportion of the total mass is represented by both small and large grains. Near the peak at lower mass, numerical solutions with order $k=3$ achieve absolute errors of order $\sim 0.1\%$ while errors of order $\sim 1\%$ are obtained with $k \in \{0,1,2\}$. Near the peak at higher mass, numerical solutions with order $k=3$ achieve absolute errors of order $\sim 0.01\%$  while errors of order  $\sim 1\%$ is obtained with $k=0$. 

\subsubsection{Stability of the DG scheme}
\label{subsubsec:stability}
Figures~\ref{fig:kconst_analysis}-\subref{fig:kconst_c},\subref{fig:kconst_d} show the time evolution of the continuous $e_{c,N}$ and discrete $e_{d,N}$ $L^1$ errors. With only $N=20$ bins, both errors for each order remain bounded over the entire time interval. However, notice an unusual behaviour of the curves around $\tau \sim 10^{-3}$. Looking at the error analysis one time step at a time, the increase in the error comes from the fact that the numerical solution near the lower peak is approximately polynomial in shape for $\tau < 10^{-3}$ but transitions to an exponential decay for $\tau > 10^{-3}$. The sudden introduction of a second exponential section in the solution approximated by polynomials leads to a short temporal increase in the error that lasts until the initial size distribution is sufficiently depleted.

\subsubsection{Convergence of the DG scheme}
\label{subsubsec:EOC}
The EOC is determined independently from the mass range by plotting the continuous and discrete $L^1$ errors as a function of the number of bins per decade $N_{\mathrm{bin/dec}}$. Figures~\ref{fig:kconst_analysis}-\subref{fig:kconst_e},\subref{fig:kconst_f} show the numerical errors at time $\tau = 10^{-9}$ for several total bin numbers $N=5,10,20,40,80$. In Fig.~\ref{fig:kconst_analysis}-\subref{fig:kconst_e}, the EOC for the continuous $L^1$ error is of order $k+1$. In Fig.~\ref{fig:kconst_analysis}-\subref{fig:kconst_f}, the EOC for the discrete $L^1$ error is of order $k+2$ for odd polynomials, and $k+1$ for even polynomials. With $e_{d,N}$, an accuracy of order $\sim 0.1\%$ is achieved with $\sim 4$ bins/decade with $k=3$,  $\sim 5$ bins/decade with $k=2$, and more than $10$ bins/decade for $k=0,1$. An accuracy of order $\sim 1\%$ is achieved with $\sim 2$ bins/decade with $k=3$, $\sim 3$ bins/decade with $k=2$, and $\sim 8$ bins/decade with $k \in \{0,1\}$.

\subsubsection{Analysis of the CFL condition}
\label{subsubsec:CFL}
The evolution of the timestep $\mathrm{d}\tau$ under the CFL condition Eq.~\ref{eq:dt_CFL} is shown in Fig.~\ref{fig:kconst_analysis},\subref{fig:kconst_b}. The timestep increases during time until a stable value $\sim 3 \times 10^{-5}$. The green points around $\tau \approx 10^{-4}$ are smaller than the CFL condition to obtain equally-spaced global timesteps $\Delta \tau$. The plateau observed from $\tau = 5 \times 10^{-4}$ shows that the CFL condition is greater than $ 3 \times 10^{-5}$; therefore, the time solver runs at $\mathrm{d}\tau = \Delta \tau$.

\subsection{Power-law distribution of fragments}
\label{sec:power_law_distrib}
\subsubsection{Power-law mass distribution}
\begin{figure*}
\centering
\includegraphics[width=0.8\textwidth]{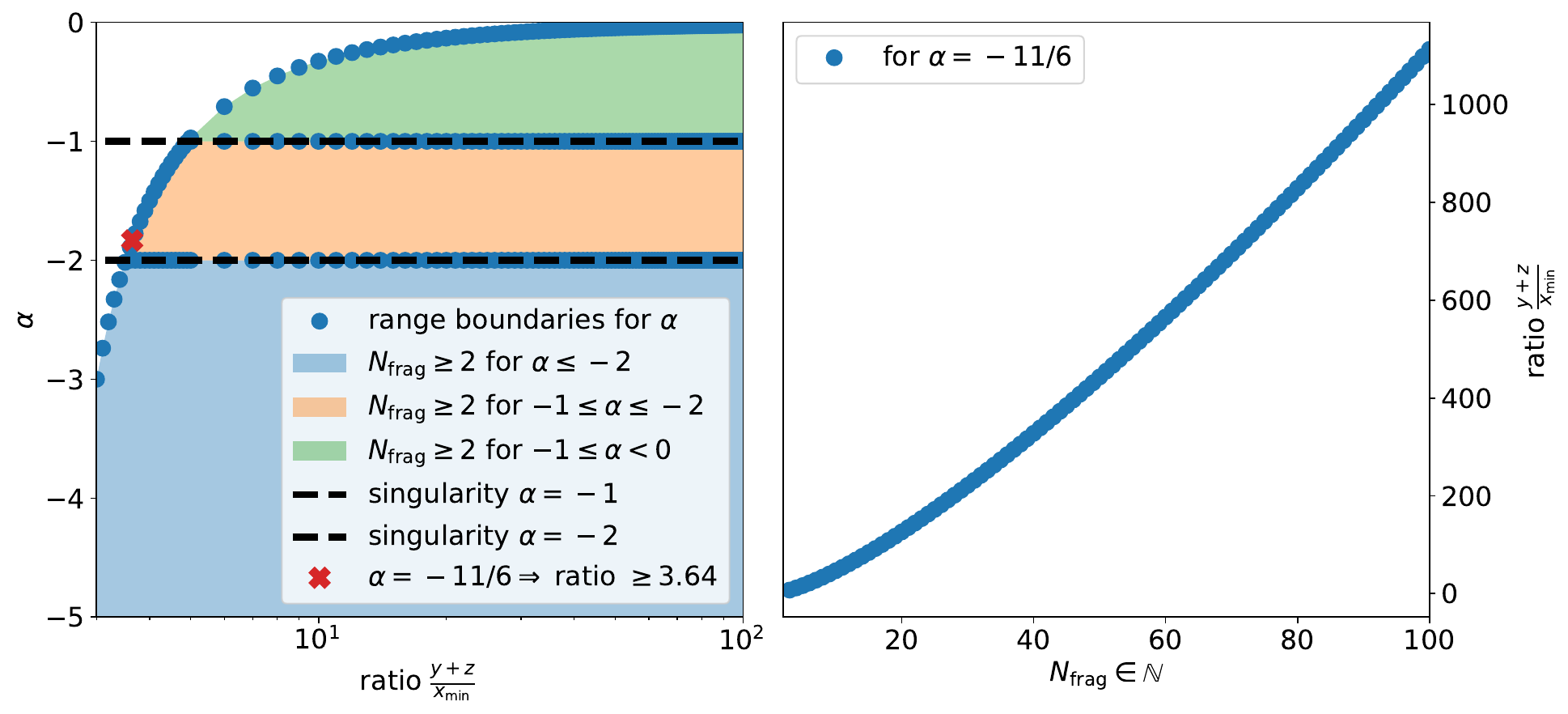}
\caption{Left: the range in values for $\alpha$, respecting the condition $N_{\mathrm{frag}} \geq 2$, is plotted versus mass ratio $(y+z)/x_{\mathrm{min}}$. We observe that the values $\alpha = -2$ and $\alpha = -1$ (dashed black line) correspond to the singularities from Eqs.~\ref{eq:b_power_law} and \ref{eq:b_nb_fragments}. Shaded regions indicate values of $\alpha$ that respect the condition  $N_{\mathrm{frag}} \geq 2$ for a given mass ratio. For $\alpha=-11/6$, the condition is respected only if the mass ratio is greater than $3.64$. Right: For a given number of fragments $N_{\mathrm{frag}} \in \mathbb{N}$, the mass ratio is calculated with  $\alpha=-11/6$. A large number of fragments requires correspondingly high mass ratios (e.g. more than a $1000$ for $N_{\mathrm{frag}} = 100$ when $\alpha = -11/6$).}
\label{fig:b_power_law_nfrag}
\end{figure*}

In astrophysics, the mass distribution of fragments is traditionally described by a power-law 
\begin{equation}
n(m,t)\mathrm{d}m \propto m^{\alpha} \mathrm{d}m,
\end{equation}
where the determination of $\alpha$ remains an open problem. Several theoretical works obtained $\alpha = -11/6 \approx-1.8$  \citep{Dohnayi1969,Williams1994,Jones1996,Tanaka1996}. Experimental studies of grain-grain collision found that $\alpha$ takes value from $-2$ to $-1$ \citep{Blum1993,Guttler2010,Deckers2014,Bukhari_Syed2017}. We present a numerical analysis to constrain the value of $\alpha$. Let us consider a mass distribution of fragments defined as
\begin{equation}
b(x;y,z) \equiv A(y,z) x^{\alpha}.
\label{eq:b_mass_cons_num}
\end{equation}
The normalisation coefficient $A(y,z)$ is obtained by applying local mass conservation to our physical mass domain
\begin{equation}
\int\limits_{x_{\mathrm{min}}}^{y+z} x b(x;y,z) \mathrm{d}x = y+z,
\label{eq:b_mass_cons_coeff_norm}
\end{equation}
giving 
\begin{equation}
A(y,z) = \frac{(\alpha +2) (y+z)}{(y+z)^{\alpha +2} - x_{\mathrm{min}}^{\alpha +2}} = \frac{(\alpha +2) (y+z)}{ x_{\mathrm{min}}^{\alpha +2} \left[ \left( \frac{y+z}{x_{\mathrm{min}}} \right)^{\alpha +2} - 1\right]},
\label{eq:b_power_law}
\end{equation}
where $\alpha \in \mathbb{R}$, $(y+z,x_{\mathrm{min}}) \in \mathbb{R}_{+}^2$ and $y+z >x_{\mathrm{min}}$. The mass distribution of fragments is a positive function if $\alpha \in (-\infty,-2) \cup (-2,\infty)$. A singularity appears for $\alpha = -2$ for the normalisation in Eq.~\ref{eq:b_power_law}. For that specific case, the coefficient $A$ has to be calculated from Eq.~\ref{eq:b_mass_cons_num}. The number of fragments produced after each collision is given by Eq.~\ref{eq:frag_number} with $0 \leftrightarrow x_{\mathrm{min}} $,
\begin{equation}
N_{\mathrm{frag}}(y,z) = \frac{(\alpha+2) (y+z) \left[ \left( \frac{y+z}{x_{\mathrm{min}}} \right)^{\alpha +1} - 1\right]}{(\alpha+1)  x_{\mathrm{min}} \left[ \left( \frac{y+z}{x_{\mathrm{min}}} \right)^{\alpha +2} - 1\right]}.
\label{eq:b_nb_fragments}
\end{equation} 
The number of fragments is strictly positive for $\alpha \in (-\infty,-2) \cup(-2,-1) \cup (-1,\infty) $. Two singularities appear for $\alpha = -2$ and $\alpha = -1$. For the value $\alpha = -1$, Eqs.~\ref{eq:b_power_law}-\ref{eq:b_nb_fragments} have to be calculated from Eq.~\ref{eq:b_mass_cons_num}. The physical condition on $N_{\mathrm{frag}}$ is that at least two fragments are produced per collision,  $N_{\mathrm{frag}} \geq 2$. This inequality is difficult to solve in the general case with the variables $y$, $z$, $x_{\mathrm{min}}$ and $\alpha$. But we can give an estimation of $\alpha$ for given values of $x_{\mathrm{min}}$ and the ratio $(y+z)/x_{\mathrm{min}}$, i.e. the ratio of mass between the total mass of the colliding grains and the mass of the smallest grain considered. Conversely,  by choosing $\alpha$, we can estimate $N_{\mathrm{frag}}$ according to the ratio $(y+z)/x_{\mathrm{min}}$. The left panel in Figure~\ref{fig:b_power_law_nfrag} shows the range $\alpha$ values as a function of mass ratio for which $N_{\mathrm{frag}} \geq 2$.  The physical condition $N_{\mathrm{frag}} \geq 2$ is satisfied over a wide range of values for the ratio $(y+z)/x_{\mathrm{min}}$, when $\alpha \in (-\infty,-2) \cup(-2,-1) \cup (-1,0) $. The important result is that the model can handle the physical value $\alpha = -11/6$ with the condition that $(y+z)/x_{\mathrm{min}} \geq 3.64$.  A large number of fragments require a large mass ratio  (see right panel in Figure ~\ref{fig:b_power_law_nfrag}).

\subsubsection{DG scheme with Gauss quadrature}
\label{sec:DG_GQ}
Applying the method used in Section~\ref{sec:eval_flux} to evaluate the flux with the power-law distribution of fragments from Eq.~\ref{eq:b_mass_cons_num}, the analytic integrals result in Gaussian hypergeometric functions $_2F_1$; however, the correct evaluation of this special function requires high accuracy and is very computationally inefficient. We therefore opt to evaluate the integrals numerically with a Gauss quadrature method. Since the limits in the integral on the variable $y$ in Eqs.~\ref{eq:flux_num_1} and \ref{eq:intflux} depend on the variable $z$, the accurate evaluation of the integrals required a large number of Gauss points. For that reason, we use the Gauss-Kronrod quadrature method. To ensure the stability of the DG scheme, we require 15 Gauss points is $15$ or $31$ Gauss-Kronrod points. We present here an approach that differs from \citet{Liu2019} to evaluate the integral on $y$, for which the limits contain variables, .i.e.
\begin{equation}
\int\limits_{x_{j-1/2}-z+x_{\mathrm{min}}}^{ x_{\mathrm{max}} - z + x_{\mathrm{min}}} g(y,\tau) \mathrm{d}y,
\label{eq:intv}
\end{equation}
where $g$ is approximated by a piecewise polynomial function. The method used in \citet{Liu2019} is to find which bins contain the values  $x_{j-1/2}-z+x_{\mathrm{min}}$ and $x_{\mathrm{max}} - z + x_{\mathrm{min}}$ and then split the integral over the bins to approximate $g$ by the corresponding polynomials. Our approach is to sum the integral of $g_l$ over each bin $l$ with an automatic selection of the bins which intersect the range $[x_{j-1/2}-z+x_{\mathrm{min}},x_{\mathrm{max}} - z + x_{\mathrm{min}}]$. Eq.~\ref{eq:intv} is approximated by
\begin{equation}
\begin{aligned}
&\int\limits_{x_{j-1/2}-z+x_{\mathrm{min}}}^{ x_{\mathrm{max}} - z + x_{\mathrm{min}}} g(y,\tau) \mathrm{d}y \\
&\approx \sum_{l=1}^N \mathbb{1}_{x_{l-1/2}<x_{\mathrm{max}} - z + x_{\mathrm{min}}} \mathbb{1}_{x_{j-1/2}-z+x_{\mathrm{min}} < x_{l+1/2}} \\
& \qquad \times \int\limits_{\mathrm{max} \left(  x_{j-1/2}-z+x_{\mathrm{min}}, x_{l-1/2} \right) }^{\mathrm{min} \left( x_{\mathrm{max}} - z + x_{\mathrm{min}}, x_{l+1/2} \right)} g_l(y,\tau) \mathrm{d}y.
\end{aligned}
\label{eq:approx_gv_GQ}
\end{equation}
The property $[x_{l-1/2},x_{l+1/2}] \cap [x_{j-1/2}-u+x_{\mathrm{min}},x_{\mathrm{max}} - u + x_{\mathrm{min}}]$ is ensured by the use of the operator $\mathbb{1}$. The integral on $y$ is evaluated with the Gauss-Kronrod quadrature with the following formula
\begin{equation}
\int\limits_a^b f(x)\mathrm{d}x \approx \frac{b-a}{2} \sum_{\alpha=1}^Q \omega_{\alpha} f \left(\frac{b+a}{2} + \frac{b-a}{2}s_{\alpha} \right),
\end{equation}
where $Q$ is the number of Gauss-Kronrod points and $\omega_{\alpha}$ and $s_{\alpha}$ are the weights and node coefficients, respectively, of the Gauss-Kronrod quadrature method. Then, the integral on $z$ is evaluated with the Gauss-Kronrod method to obtain the first double integral for the flux in Eq.~\ref{eq:flux_num_1}. The details on the evaluation of the flux $F_{\mathrm{frag,c}}[g](x_{j-1/2},\tau)$ and the integral of the flux $\mathcal{F}_{\mathrm{frag,c}}(j,k',\tau)$ are given in appendix~\ref{ap:DGGKQ}.

\subsubsection{Results}
\begin{figure*}
\centering
\includegraphics[width=0.9\textwidth]{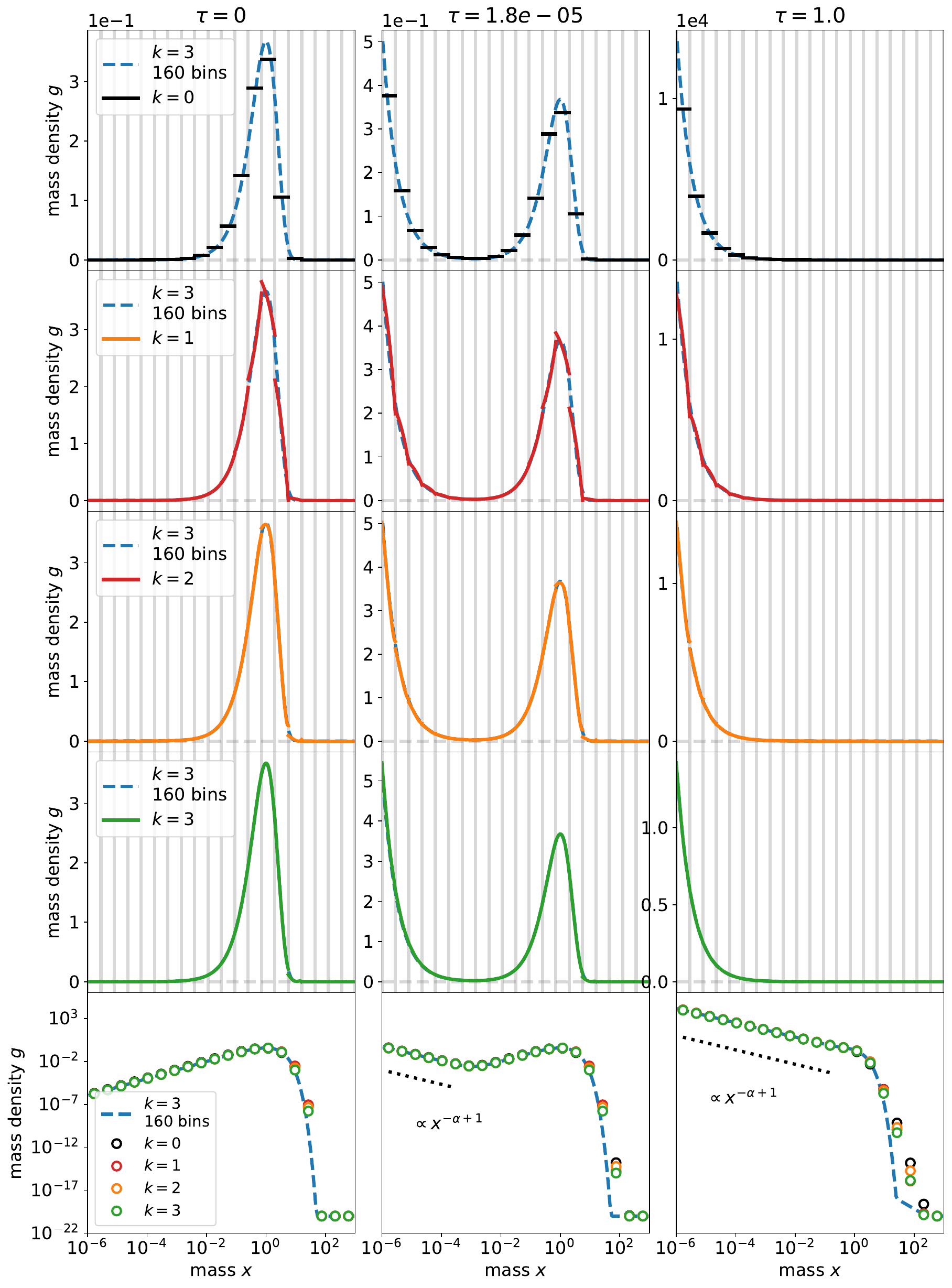}
\caption{Test for the multiplicative kernel and the power-law distribution of fragments in Eq.~\ref{eq:b_power_law} with power-law index $\alpha = -11/6$. The first four rows show the numerical solution (solid lines) for $N=20$ bins and the indicated value $k \in 0,1,2,3$ at time $\tau = 0$ (left column), $\tau = 1.8 \times10^{-5}$ (middle column) and $\tau = 1$ (right column). The mass density is in linear scale and the mass in logarithmic scale. The reference numerical solution $g(x,\tau)$ is given by the blue dashed line. Vertical grey lines represent the boundaries of the bins. The last row shows the same numerical solutions in log-log scale. The accuracy improves with increasing values of $k$.}
\label{fig:kmul_power_law}
\end{figure*}

\begin{figure*}
\centering
\subfloat[][]{\includegraphics[width=0.9\columnwidth]{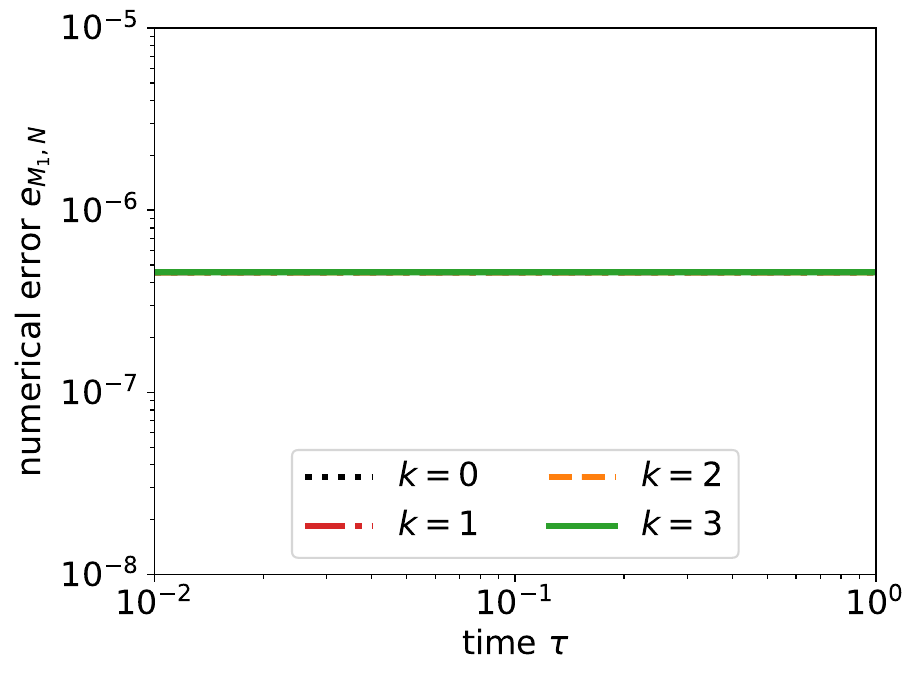}\label{fig:kmul_a}}
\subfloat[][]{\includegraphics[width=0.9\columnwidth]{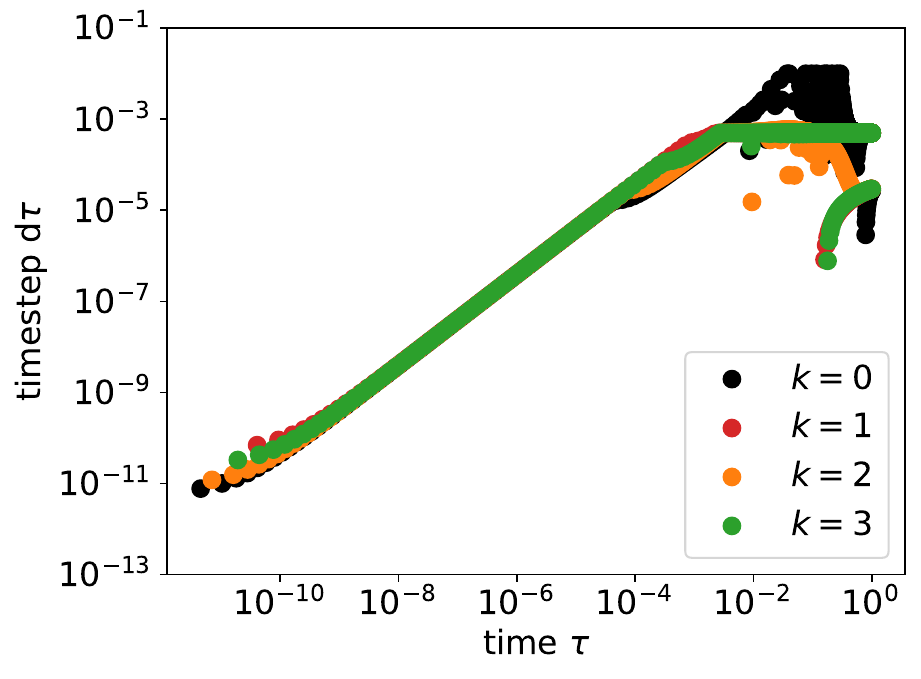}\label{fig:kmul_b}}\\
\subfloat[][]{\includegraphics[width=0.9\columnwidth]{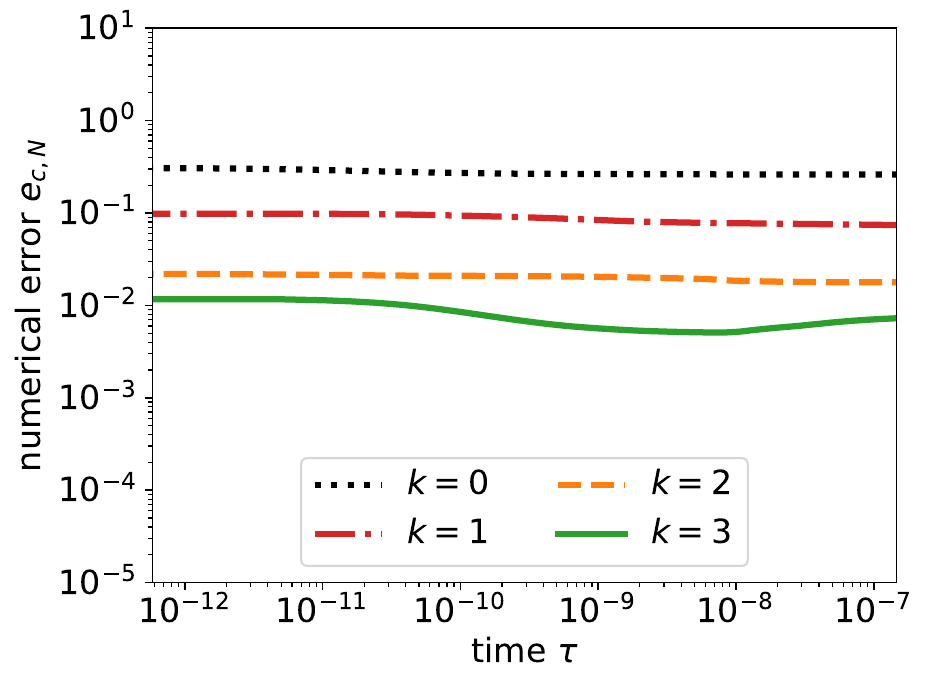}\label{fig:kmul_c}}
\subfloat[][]{\includegraphics[width=0.9\columnwidth]{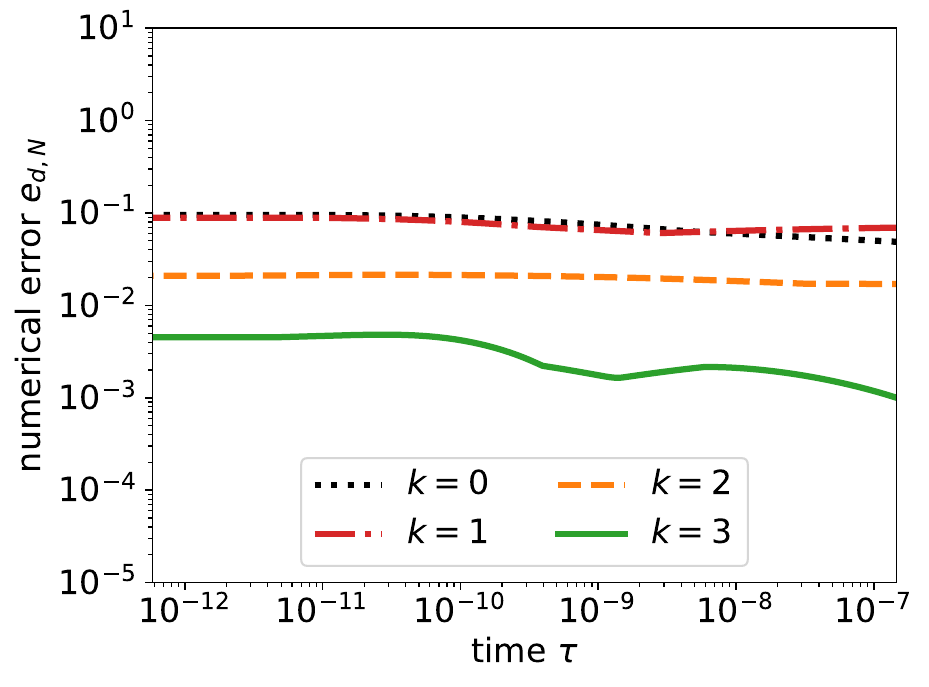}\label{fig:kmul_d}}\\
\caption{Test for the multiplicative kernel and the power-law mass distribution of fragments  in Eq.~\ref{eq:b_power_law} with $N=20$ bins. The power-law coefficient is $\alpha = -11/6$. The results are similar to the constant kernel test in Fig.~\ref{fig:kconst_analysis}. The mass is conserved for each order $k$ in panel~\protect\subref{fig:kmul_a}. Panel~\protect\subref{fig:kmul_b}: evolution of the timestep $d \tau$. For each order k, the timestep globally increases with time as more fragments are produced. The numerical errors decrease with the order of polynomials $k$ in panels~\protect\subref{fig:kmul_c} and \protect\subref{fig:kmul_d}. }
\label{fig:kmul_power_law_analysis}
\end{figure*}

We present here the benchmark results of the the DG scheme with a power-law mass distribution of fragments. We test our numerical algorithm using a multiplicative kernel with $\alpha=-11/6$ and $N=20$. The numerical solutions for $k=0,1,2,3$ are benchmarked with a reference solution obtained by using $N=160$ bins with $k=3$. The simulations are performed from $\tau =0$ to $\tau = 1$ with $100$ timesteps $\Delta \tau$ of length $10^{-2}$. The simulations are performed in double precision. Figure~\ref{fig:kmul_power_law} shows the numerical results of the mass density function versus the mass in linear-log for the first fourth rows and log-log scale for the last row. Analyse of the performance, shown in Fig.~\ref{fig:kmul_power_law_analysis}, shows similar results to those found in Sect.~\ref{subsec:kconst_tests}. High accuracy of the numerical solutions is achieved with high order of polynomials. The experimental order of convergence is not shown since it is similar to the test in Fig.~\ref{fig:kconst_analysis}.

%-----------------------------------------------------------------------------------------------------------------
\section{Discussion}
\label{sec:discussions}
\subsection{Conservative form}
We have derived the first conservative form of the general non-linear fragmentation equation (Eq.~\ref{eq:frag_cont_DL}) as a hyperbolic law equation (Eq.~\ref{eq:frag_cons_DL}) with a fragmentation flux in mass space. The fragmentation flux, Eq.~\ref{eq:flux}, includes a `coagulation' flux that results from potential mass transfer events that take place when one grain breaks against another. Our new expression for the flux combines in a continuous way the gain and loss terms of Eq.~\ref{eq:frag_cont_DL}. The conservative form is of particular interest since it allows the use of robust numerical schemes (e.g. finite-volume or DG schemes) that naturally conserve the mass to machine precision. 

\subsection{Performance}
The high-order DG scheme (see Sect.\ref{sec:DG}) efficiently solves the general non-linear fragmentation equation with only $N=20$ bins. High-order polynomial approximations can drastically reduce the numerical diffusion observed for low-order numerical schemes with the same bin resolution. While this improvement was modest (one order of magnitude) for the analytic test case, the accuracy increased by four orders of magnitude for the power-law test. Moreover, the improved accuracy is occurring near the peak of the mass density curve where the majority of the mass of the system resides. In the test illustrated in Fig.~\ref{fig:kconst_linlog_loglog}, we already achieve an accuracy of approximately $1\%$ with order $k = 2$ and $N = 20$ bins, an accuracy level comparable to those reported by 3D hydrodynamic codes \citep{Teyssier2002,Price2018}. Thus, for the first time, we could treat dust fragmentation in 3D hydrodynamic simulations together with gas and dust dynamics. The details of how to couple the algorithm with a 3D hydrodynamic code will presented in a future work.

\subsection{Limitations}
\subsubsection{DG scheme architecture}
\label{sec:DG_architecture}
\begin{figure}
\centering
\includegraphics[width=\columnwidth]{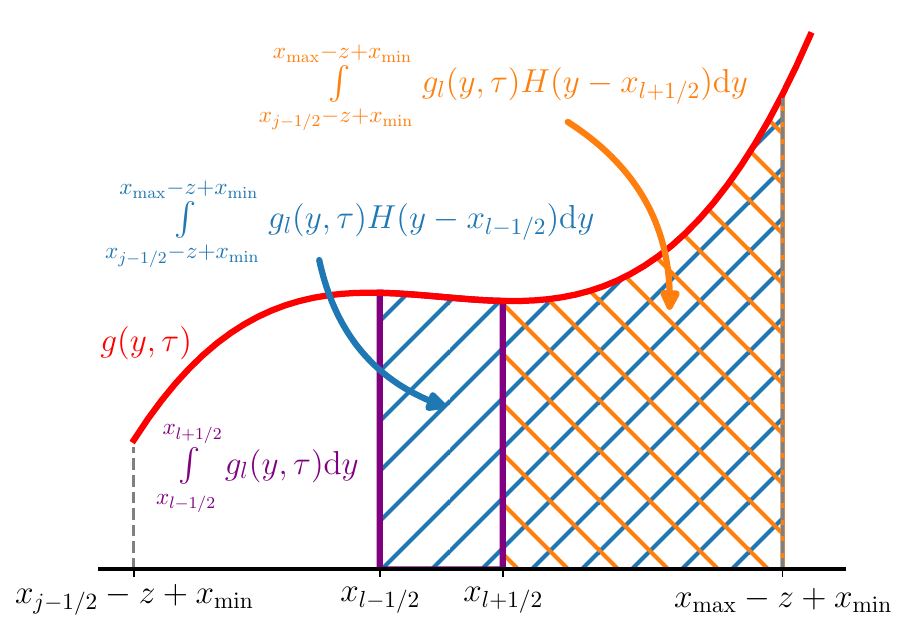}
\caption{Illustration of the method used to analytically evaluate the integral of $g(y,\tau)$ approximated on the entire mass range by $g_l(y,\tau)$ in Eq.~\ref{eq:approx_g_global}. The integral over the range $[x_{j-1/2}-z+x_{\mathrm{min}},x_{\mathrm{max}}-z+x_{\mathrm{min}}]$ is evaluated by summing the integral of $g_l(y,\tau)$ (purple rectangle for a given $l$) over all bins $l \in [\![1,N]\!]$. The integral of $g_l(y,\tau)$ is obtain by the difference of two integrals (blue and orange area). This difference of integrals is propagated all along the derivation to obtain terms in Eqs.~\ref{eq:flux_Tfrag1}-\ref{eq:flux_Tcoag} and Eqs.~\ref{eq:terms_Tfrag1_A_intflux}-\ref{eq:terms_Tcoag_B_intflux}. Therefore, high accuracy, such as quadruple precision, is required to correctly evaluate the terms for the flux and the integral of the flux.}
\label{fig:sketch_heaviside}
\end{figure}

The DG method, with the analytical evaluation of the integrals presented in Sect.~\ref{sec:DG}, requires the use of quadruple precision to handle the arithmetic of large numbers for polynomials of high-order. Thus, in its current form, order $k=3$ is the maximum limit of the algorithm to obtain a good balance between accuracy and computational efficiency. This issue mainly comes from the approximation of $g(y,\tau)$ over the entire mass range with the aid of Heaviside functions (see Eq.~\ref{eq:approx_g_global}). To evaluate the integrals in Eqs.~\ref{eq:flux_Tfrag1}-\ref{eq:flux_Tcoag} and Eqs.~\ref{eq:terms_Tfrag1_A_intflux}-\ref{eq:terms_Tcoag_B_intflux}, the \textit{difference} of Heaviside functions is propagated through the integrals. Therefore, the algorithm needs to evaluate the subtraction of two integrals with potentially large values, as illustrated in Fig.~\ref{fig:sketch_heaviside}. This method of approximation over the entire mass range explains why the accurate evaluation of the terms in Eqs.~\ref{eq:flux_Tfrag1}-\ref{eq:flux_Tcoag} and Eqs.~\ref{eq:terms_Tfrag1_A_intflux}-\ref{eq:terms_Tcoag_B_intflux} requires quadruple precision, which is computationally inefficient. 

One solution to this problem would be to rewrite the approximation of $g$ over the entire mass range as
\begin{equation}
  \begin{aligned}
    &\forall y \in [x_{\mathrm{min}},x_{\mathrm{max}}],\\
    &g\left(y,\tau \right) \approx \\
    & \sum_{l=1}^N \sum_{i=0}^k g_l^i\left(\tau\right) \phi_i(\xi_l(y))H(y-x_{l-1/2}) H(x_{l+1/2}-y)],
  \end{aligned}
  \label{eq:approx_g_global_new}
\end{equation}
which instead uses the \textit{product} of two Heaviside functions. In this way, we avoid the difference of two large numbers. The details of the derivation to obtain the analytical integrals will be presented in a future work. However, as mentioned in Sect.~\ref{sec:DG_GQ}, analytic evaluation of the integrals for the power-law mass distribution of fragments will probably still result in Gaussian hypergeometric functions. We will explore whether there is an efficient way to rewrite or simplify these solutions using \textsc{Mathematica}.

The DG scheme with the Gauss quadrature (DGGQ) method in Sect.~\ref{sec:DG_GQ} is an interesting alternative to the DG scheme with analytical integrals. The DGGQ scheme has several advantages: i) any collision kernel and mass distribution of fragments can be used, ii) only double precision is needed, and iii) it is relatively easy to implement. However, in order to reach high accuracy for the power-law mass distribution test in Sect.~\ref{sec:power_law_distrib}, we found we needed to use the Gauss-Kronrod method with at least $31$ points. This number of points is determined experimentally. For instance, with $21$ points and order 3 polynomials, the simulation does not reach the final time because the timestep under the CFL condition at some point tends to zero. This behaviour can be explained by the fact that the evaluation of the integrals is not sufficiently accurate with $21$ points. A good balance has been found by using $31$ points. This requires a significative amount of time to precompute the integrals used in the evaluation of  $F_{\mathrm{frag,c}}$ and $\mathcal{F}_{\mathrm{frag,c}}$ in Eqs~\ref{eq:flux_num_1} and \ref{eq:intflux} -- particularly for orders $k=2$ and $k=3$. In practice, to reduce the precomputing time, the terms are generated with multi-threading processes by using \texttt{OpenMP}. The required large number of Gauss points suggests that the Gauss quadrature is not well adapted to evaluating integrals with a variable in the limits. It is possible that other integration methods would better suit this application. For example, the double exponential formula or Tanh-Sinh quadrature  \citep{Takahasi1974,Mori2001,Muhammad2005} may be a suitable candidate for the DG fragmentation scheme we present in this paper. Tanh-Sinh quadrature is an efficient quadrature method to evaluate an integral on a finite interval with exponential convergence. With this quadrature method, we are likely to reach better accuracy with less points than the Gauss-Kronrod quadrature method. This too will be explored in a future work.

\subsubsection{Time execution}
\begin{table}
\begin{center}
\begin{tabular}{ccc}
order &analytic test &power-law test \\
  of polynomials  &time (s)&time (s) \\ 
\hline
 $k=0$ & $10^{-3}$ & $10^{-4}$  \\
  $k=1$ & $10^{-2}$ & $8 \times 10^{-4}$\\
  $k=2$ &  $2 \times 10^{-2}$ &$2 \times 10^{-3}$ \\
  $k=3$ &  $6 \times 10^{-2}$ &$8 \times 10^{-3}$\\
\end{tabular}
\caption{Averaged elapsed wall time in seconds for a single timestep for each order of polynomials $k$.}
\label{tab:time}
\end{center}
\end{table}

Table~\ref{tab:time} gives the averaged elapsed wall-clock time for a single timestep for each order of polynomials $k$. We observe that the power-law mass distribution test executes faster than the analytic test since less terms are required to evaluate $F_{\mathrm{frag,c}}$ and $\mathcal{F}_{\mathrm{frag,c}}$ for each timestep (see Appendix~\ref{ap:DGGKQ}). A gain of a factor $10$ is reached with the DGGQ scheme for $k=3$. To couple the DG scheme to multi-fluid 3D hydrodynamic codes, the DG time solver has to be highly efficient. A first test to treat the general non-linear fragmentation equation in a 3D simulation of protoplanetary disc will be performed with the code \texttt{PHANTOM} \citep{Price2018}. A typical gas/dust simulation of disc uses 10 dust sizes (i.e. $N=10$) with $10^6$ SPH particles and $32$ CPUs. In this configuration, one hydrodynamic timestep takes $\sim 1$s. By coupling the DG scheme to \texttt{PHANTOM}, we aim to reach a one-to-one ratio in running time, meaning that the elapsed time for the DG scheme has to be $\sim 1 s$. According to the CFL criterion in Sect.~\ref{sec:CFL}, the fragmentation process will probably impose a sub-cycling compared to the hydrodynamic timestep. Therefore, for the value of $\mathrm{d}\tau = 8 \times 10^{-3}$ for $k=3$, the running time for one hydrodynamic timestep would be at least $10^6 \times 8 \times 10^{-3} = 8 \times 10^3$s. There are several strategies that can be explored in order to reach this one-to-one ratio.

The first strategy is to use the matrix form of the general non-linear fragmentation equation in order to take advantage of the vast computational resources that have been developed for matrix operations. A similar approach was used by \citet{Sandu2006} for the coagulation equation. The details of the matrix form is given in appendix~\ref{ap:matrix_form}. Thus, efficient matrix operation algorithms, included in the library \texttt{BLAS}, can be used to improve the time solver. Moreover, the matrix form in Eq.~\ref{eq:matrix_form} allows for a simple form of the Jacobian that can be paired with an implicit time solver \citep{Sandu2006}, eliminating the need for sub-cycling altogether. Unfortunately, the implicit SSPRK order 3 method can not be made unconditionally stable and has to follow a timestep restriction with an SSP coefficient $C \leq 2$ \citep{Gottlieb2009,Ketcheson2009,Gottlieb2011}. Therefore, considering the time it takes to compute the Jacobian, this implicit method does not provide a significant improvement over the explicit method. However, recently, \citet{Gottlieb2022} derived a high-order unconditionally SSP implicit Runge-Kutta method by considering restrictions on the second derivative of the variable $g$ in the spatial operator $\mathbf{L}$ for the DG scheme (see the definition of $\mathbf{L}$ in \citealt{Lombart2021}). The linearised backward Euler scheme used in \citet{Sandu2006} and the implicit SSP two derivatives Runge-Kutta order 3 \citep{Gottlieb2022} will be applied for the DG scheme and tested in a future work in order to remove the need for any sub-cycling during the evolution of a simulation.

The second strategy is to use GPU parallelisation to run all $10^6$ calls to the fragmentation solver for all SPH particles in $ \sim 1$s. A future work will be concerned with the design of the DG scheme adapted for GPU programming language ( e.g. \texttt{CUDA} and \texttt{SYCL}). Since the general non-linear fragmentation equation is a non-local partial differential equation, the main difficulty will be to precisely manage the memory access on the GPU device in order to obtain the best performance. The implicit solver will be implemented with the GPU version of the  \texttt{BLAS} library. The GPU version of the DG scheme will allow the treatment of  fragmentation in future exascale hydrodynamics codes, such as \texttt{IDEFIX} \citep{Lesur2023a} and \texttt{DYABLO} \citep{Aubert2021}.

\subsection{Physical mass distribution of fragments}
\label{sec:distrib_frag_power_law}
Laboratory experiments and theoretical studies in astrophysics indicate that fragmentation results in small grains with a power-law mass distribution, accompanied by one or two additional remnants with non-power-law distributed masses \citep{Guttler2010,Windmark2012,Blum2018,Hirashita2019,Lebreuilly2022,Hasegawa2023}. To take into account the production of these outlying remnants, the mass distribution of fragments can be modified \citep{Hirashita2021} as 
\begin{equation}
b(x;y,z) \equiv A x^{\alpha} + \delta(x-(y+z-m_{\mathrm{frag}}))
\label{eq:b_power_law_1remnant}
\end{equation} 
where $\delta$ is the Dirac delta function, $m_{\mathrm{frag}}$ the total mass of fragments and $y+z-m_{\mathrm{frag}}$ the mass of the remnant grain (assuming there is only one). In general, $m_{\mathrm{frag}}$ is defined as a percentage of $y+z$, the mass of  the two colliding grains. The normalisation constant $A$ is determined by the following local mass conservation equation
\begin{equation}
\int\limits_{x_{\mathrm{min}}}^{x_{\mathrm{max}}} xAx^{\alpha} \mathrm{d}x = m_{\mathrm{frag}} \Rightarrow A = \frac{(2+\alpha)m_{\mathrm{frag}} }{x_{\mathrm{max}}^{2+\alpha} - x_{\mathrm{min}}^{2+\alpha}}.
\end{equation}
By giving the link between $ x_{\mathrm{min}}$, $ x_{\mathrm{max}}$ and $ m_{\mathrm{frag}}$, the number of fragments per collision can be obtained. For instance, if we take the values from \citet{Hirashita2019},
\begin{equation}
\begin{aligned}
 &x_{\mathrm{min}} = 10^{-6} x_{\mathrm{max}},\,  x_{\mathrm{max}} = 0.02 m_{\mathrm{frag}},\, \alpha = -\frac{11}{6} \\
 &\Rightarrow N_{\mathrm{frag}} \approx 10^6.
 \end{aligned}
\end{equation}
So under these conditions, each collision produces approximately one million fragments and one remnant. To obtain two remnants, like the N-body simulations of collisions in \citet{Hasegawa2023}, we can further modify Eq.~\ref{eq:b_power_law_1remnant} with another Dirac delta function,
\begin{equation}
b(x;y,z) \equiv A x^{\alpha} + \delta(x-m_1) + \delta(x-(y+z-(m_1+m_{\mathrm{frag}})),
\label{eq:b_power_law_2remnants}
\end{equation}
where $m_1$ is the mass of the first remnant and $m_1+m_{\mathrm{frag}}$ is the mass of the second remnant. The application of the DG scheme with the mass distributions of fragments Eqs~\ref{eq:b_power_law_1remnant} and \ref{eq:b_power_law_2remnants} will be investigated in a future work.

\subsection{Coagulation and fragmentation}
The coagulation/fragmentation equation used in astrophysics is the combination of the Smoluchowski coagulation equation and the general non-linear fragmentation equation Eq.~\ref{eq:frag_cont_DL} and writes \citep{Blum2006,Banasiak2019,Barik2020,Giri2021a}
\begin{equation}
\begin{aligned}
& \frac{\partial f (x,\tau)}{\partial t } = \\
&  \frac{1}{2}\int\limits_{0}^{x} \left( 1 - P_{\mathrm{frag}}(x-y,y,\Delta v)  \right) \mathcal{K}(x-y,y) f(x-y,\tau) f(y,\tau) \mathrm{d}y  \\
& + \frac{1}{2}\int\limits_{0}^{\infty} \int\limits_{0}^{\infty} \mathbb{1}_{y+z \geq x} b(x;y,z) P_{\mathrm{frag}}(y,z,\Delta v) \\
& \quad \qquad \qquad \qquad \times \mathcal{K}(y,z) f(y,\tau) f(z,\tau) \mathrm{d}y \mathrm{d}z \\
&   - f(x,\tau) \int\limits_{0}^{\infty} \mathcal{K}(x,y) f(y,\tau) \mathrm{d}y, 
\end{aligned}
\label{eq:coag_frag}
\end{equation}
where $P_{\mathrm{frag}}$ is the probability that two particles fragment. The first term in the right-hand side of Eq.~\ref{eq:coag_frag} describes the formation of particles of mass $x$ due to coagulation. The second term denotes the formation of particles of mass $x$ due to the fragmentation of two colliding particles of masses $y$ and $z$. The third term describes the loss of particles of mass $x$ due to collisions leading to coagulation or fragmentation. Some mathematical papers proved the existence and uniqueness of mass-conserving solutions to Eq.~\ref{eq:coag_frag} for a large class of collision kernels $\mathcal{K}$ but only for fragmentation without mass transfer \citep{Barik2020,Giri2021a}. Moreover, \citet{Feingold1988} derived a steady state solution to the coagulation/fragmentation equation with the alternative fragmentation rate equation in Eq.~\ref{eq:alt_frag_cont_DL} using a constant collision kernel, constant probability of fragmentation and the mass distribution of fragments in Eq.~\ref{eq:b_F88_DL}. In a future work, the DG scheme will be applied to the conservative form of the coagulation/fragmentation equation and benchmarked with the analytical steady-state solution given in \citet{Feingold1988}.

\subsection{Dust aggregates}
The general non-linear fragmentation equation presented in Sect.~\ref{sec:frag} is a Smoluchowski-like equation, meaning that it respects the same assumptions as the Smoluchowski coagulation equation. In particular, grains are considered to be spheres of the same material and, consequently, so are the fragments. However, astrophysical grains are aggregates with a non-spherical shape \citep{Suttner2001,Blum2006,Okuzumi2009,Blum2018}. To account for this, the evolution of the number density of grains would depend on two variables describing the properties of a grain, such as the mass $x$ and the porosity $\Psi$. The evolution of this number density $f(x,\Psi,\tau)$ is governed by the bivariate coagulation/fragmentation equation \citep{Kostoglou2001,Ormel2007}
\begin{equation}
\begin{aligned}
& \frac{\partial f (x,\Psi_x,\tau)}{\partial \tau } = \left. \frac{\partial f (x,\Psi_x,\tau)}{\partial \tau } \right|_{\mathrm{coag}} + \left.\frac{\partial f (x,\Psi_x,\tau)}{\partial \tau }\right|_{\mathrm{frag}},
\end{aligned}
\label{eq:coag_frag_2D}
\end{equation}
with
\begin{equation}
\begin{aligned}
& \left. \frac{\partial f (x,\Psi,\tau)}{\partial \tau } \right|_{\mathrm{coag}} \equiv\\
& \quad   \frac{1}{2} \int\limits_{0}^{x} \int\limits_{\Psi_{\mathrm{min}}}^{\Psi_{\mathrm{max}}} \int\limits_{\Psi_{\mathrm{min}}}^{\Psi_{\mathrm{max}}} \left( 1 - P_{\mathrm{frag}}(x-y,\Psi_{x-y};y,\Psi_y;\Delta v)  \right) \\
& \qquad  \times \mathcal{K}(x-y,\Psi_{x-y};y,\Psi_y) f(y,\Psi_y,\tau)f(x-y,\Psi_{x-y},\tau)  \\
&  \qquad  \times  \delta(\Psi_x - \Gamma(x-y,\Psi_{x-y};y,\Psi_y)) \mathrm{d}\Psi_{x-y} \mathrm{d}\Psi_y   \mathrm{d}y  \\
& \quad - f(x,\Psi_x,\tau) \int\limits_{0}^{\infty}  \int\limits_{\Psi_{\mathrm{min}}}^{\Psi_{\mathrm{max}}} \mathcal{K}(x,\Psi_x;y,\Psi_y) f(y,\Psi_y,\tau)  \mathrm{d}\Psi_y  \mathrm{d}y,
\end{aligned}
\label{eq:coag_2D}
\end{equation}
and
\begin{equation}
\begin{aligned}
&  \left. \frac{\partial f (x,\Psi_x,\tau)}{\partial \tau } \right|_{\mathrm{frag}} \equiv\\
& \quad \frac{1}{2}\int\limits_{0}^{\infty} \int\limits_{0}^{\infty}  \int\limits_{\Psi_{\mathrm{min}}}^{\Psi_{\mathrm{max}}}  \int\limits_{\Psi_{\mathrm{min}}}^{\Psi_{\mathrm{max}}} \mathbb{1}_{y+z \geq x} b(x,\Psi_x|y,\Psi_y;z,\psi_z)  \\
& \qquad \qquad \qquad \qquad \quad \times \mathcal{K}(y,\Psi_y;z,\Psi_z) f(y,\Psi_y,\tau) f(z,\Psi_z,\tau)  \\
& \qquad \qquad \qquad \qquad \quad \times P_{\mathrm{frag}}(y,\Psi_y;z,\Psi_z,\Delta v)  \mathrm{d}\Psi_y \mathrm{d}\Psi_z \mathrm{d}y \mathrm{d}z\\
&   \qquad - f(x,\Psi_x,\tau) \int\limits_{0}^{\infty}  \int\limits_{\Psi_{\mathrm{min}}}^{\Psi_{\mathrm{max}}} \mathcal{K}(x,\Psi_x;y,\Psi_y) f(y,\Psi_y,\tau) \mathrm{d}\Psi_y \mathrm{d}y, 
\end{aligned}
\label{frag_2D}
\end{equation}
where $\Gamma(x,\Psi_x;y,\Psi_y)$ gives the porosity of the resulting aggregate from collision of two grains of mass and porosity $(x,\Psi_x)$ and $(y,\Psi_y)$ \citep{Ormel2007}. The expression of $\Gamma$ depends on the collision algorithm, such as Particle-Cluster Aggregation (PCA) and Cluster-Cluster Aggregation (CCA). The function $ b(x,\Psi_x|y,\Psi_y;z,\psi_z)$ gives the distribution in mass and porosity of the fragments from two colliding grains. The mass of fragments generally follows a power-law, but the distribution in porosity is poorly understood. Some recipes have been used  to determine the porosity of the fragment, for instance in \citet{Hirashita2021}. It may be possible to derive a conservative form of Eq.~\ref{eq:coag_frag_2D} by following \citet{Qamar2007} and \citet{Das2023} and then to apply the high-order discontinuous Galerkin scheme to efficiently solve this 2D coagulation/fragmentation equation with a few number of bins in mass and in porosity.

\subsection{Accounting for shape, porosity, and chemistry}
The general non-linear fragmentation model only considers spherical grains with the same chemical composition. However, recent observations of protoplanetary discs and diffuse ISM require more complex grain models \citep[][and references therein]{Hensley2023,Siebenmorgen2023,Ysard2019b,Ysard2024}, since shape and composition strongly affect the optical properties of the dust. To match the complexity of these observations, we need dust models that can account for different aggregate shapes and/or chemical compositions. For the first time, the DG scheme applied to the conservative form of the coagulation and the fragmentation equations provide a robust numerical framework to handle the complexity of grain models with greater dimensionality (e.g. shape and porosity). For instance, following the evolution of the mass density distribution of dust aggregates requires the 2D coagulation and fragmentation equations with variables mass and porosity \citep{Kostoglou2001,Okuzumi2009,Hirashita2021}. In addition, the chemical composition can be treated by considering that each grain is composed of a fraction of different chemical species \citep{Pilinis1990,Jacobson1994,Sandu2006}. Then, the dust coagulation and fragmentation equation can be written in term of the mass density distribution for each species to obtain a system of coagulation and fragmentation equations which can be solved efficiently by the DG scheme. By combining chemical composition and aggregate shape, the system of multi-dimensional equations can be efficiently solved by the DG scheme since the mass distribution functions are approximated by polynomials with a few number of bins in each dimension, reducing the computational cost. The efficiency will be improved with the optimisations of the DG scheme mentioned in Sect.~\ref{sec:discussions}. In future studies, the multi-dimensional dust evolution model, accounting for the correct optical properties of grains, will be implemented in 3D hydrodynamics simulations to match observations, thanks to this high-order DG scheme.

%-----------------------------------------------------------------------------------------------------------------
\section{Conclusion}
\label{sec:conclusion}
The grain-grain collision outcomes leading to the formation of fragments are important for understanding population levels of small grains population in different astrophysical environments. Several physical processes, such as thermal balance and gas-dust dynamics, are strongly impacted by the evolution of the dust size distribution, highlighting the need to accurately treat dust coagulation and fragmentation in 3D simulations. 

We have presented the derivation of the conservative form of the general non-linear fragmentation equation utilising a mass flux (see Eqs.~\ref{eq:frag_cons_DL} and \ref{eq:flux}). The physical interpretation of the fragmentation equation is enriched by the formulation of the fragmentation flux, which contains a coagulation flux resulting from mass-transfer in sufficiently high-velocity fragmenting collisions that allow some grains to grow in mass \citep{Blum2018,Birnstiel2023,Hasegawa2023}. This conservative form enables the application of robust numerical schemes, such as the finite volume method or the discontinuous Galerkin scheme we presented in this work. The high-order DG scheme accurately solves the general non-linear fragmentation equation on a low-resolution mass grid of only $\sim 20$ bins, thus paving the way to address poly-disperse dust fragmentation in 3D hydrodynamics codes. The DG scheme meets all necessary requirements to be coupled to 3D codes: i) a strictly positive mass density distribution ensured by the SSPRK third-order time solver combined with a slope limiter, ii) conservation of the mass at machine precision, iii) accuracy of $\sim0.1-1\%$ obtained by high-order discretisation in mass and time space, and iv) a fast algorithm facilitated by precomputed analytic integrals or efficient numerical integration. 

Accurately following the evolution of the dust size distribution in 3D simulations due to fragmentation has long been an objective in astrophysics \citep{Haworth2016}. The DG scheme, applied on the conservative form of the fragmentation equation, provides a numerical framework to treat dust coagulation/fragmentation in 3D simulations and allow for dust models that consider more than just grain size (e.g. porosity and chemical composition).

\section*{Acknowledgements}
We thank the referee for providing very useful comments that helped us to improve the manuscript. The authors thank P.-K. Barik for discussion and contribution on the derivation of the conservative form. This work has received funding from the National Science and Technology Council (NSTC 112-2636-M-003-001). MAH acknowledges support from the Excellence Cluster ORIGINS, which is funded by the Deutsche Forschungsgemeinschaft (DFG, German Research Foundation) under Germany Excellence Strategy - EXC-2094 - 390783311 and partial funding by the Deutsche Forschungsgemeinschaft (DFG, German Research Foundation) - 325594231. We used \textsc{Mathematica} \citep{Mathematica}. 

\section*{Data availability}
Data are available on request.

%\label{lastpage}
\bibliography{biblio_frag_gnl}

\appendix

\section{Terms to evaluate the flux}
\label{ap:flux}
The terms  $T_{\mathrm{frag},1}$, $T_{\mathrm{frag},2}$ and $T_{\mathrm{coag}}$ in Eq.~\ref{eq:flux_num_2} take the form
\begin{equation}
\begin{aligned}
&T_{\mathrm{frag},1}(x_{\mathrm{max}},x_{\mathrm{min}},j,l',l,i',i) \equiv  \\
& \sum_{u=1}^{j-1} \int\limits_{I_u} \int\limits_{I_{l'}} \int\limits_{x_{j-1/2}-z + x_{\mathrm{min}}}^{x_{\mathrm{max}}-z+x_{\mathrm{min}}}  x' b(x';y,z) \\
& \qquad \qquad \times \mathcal{K}(y,z)  \frac{\phi_i(\xi_l(y)) \phi_{i'}(\xi_{l'}(z))}{yz} \\
& \qquad \qquad \times \left[ H(y-x_{l-1/2}) - H(y-x_{l+1/2})  \right] \mathrm{d}y \mathrm{d}z \mathrm{d}x',
\end{aligned}
\label{eq:flux_Tfrag1}
\end{equation}
\begin{equation}
\begin{aligned}
& T_{\mathrm{frag},2}(x_{\mathrm{max}},x_{\mathrm{min}},j,l',l,i',i) \equiv \\
&\sum_{u=1}^{j-1}  \int\limits_{I_u} \int\limits_{I_{l'}} \int\limits_{x_{\mathrm{min}}}^{x_{\mathrm{max}}-z+x_{\mathrm{min}}}  x' b(x';y,z) \\
& \qquad \qquad \times \mathcal{K}(y,z) \frac{\phi_i(\xi_l(y)) \phi_{i'}(\xi_{l'}(z))}{yz}\\
& \qquad \qquad \times  \left[ H(y-x_{l-1/2}) - H(y-x_{l+1/2})  \right]  \mathrm{d}y \mathrm{d}z \mathrm{d}x',
\end{aligned}
\label{eq:flux_Tfrag2}
\end{equation}
\begin{equation}
\begin{aligned}
& T_{\mathrm{coag}}(x_{\mathrm{max}},x_{\mathrm{min}},j,l',l,i',i) \equiv \\
& \int\limits_{I_{l'}} \int\limits_{x_{j-1/2}-z +x_{\mathrm{min}}}^{x_{\mathrm{max}}-z+x_{\mathrm{min}}} \mathcal{K}(z,y)\phi_{i'}(\xi_{l'}(z)) \frac{\phi_i(\xi_l(y))}{y} \\
& \qquad \qquad \times \left[ H(y-x_{l-1/2}) - H(y-x_{l+1/2})  \right]  \mathrm{d}y \mathrm{d}z.
\end{aligned}
\label{eq:flux_Tcoag}
\end{equation}

\section{Terms to evaluate the integral of the flux}
\label{ap:intflux}
The terms in Eq.~\ref{eq:intflux_2} are defined as
\begin{equation}
\begin{aligned}
&\mathcal{T}_{\mathrm{frag},1,A}(x_{\mathrm{max}},x_{\mathrm{min}},j,k',l',l,i',i) \equiv \\
&\int_{I_j} \int\limits_{x_{\mathrm{min}}}^{x}  \int_{I_{l'}}   \int\limits_{x-z + x_{\mathrm{min}}}^{x_{\mathrm{max}}-z+x_{\mathrm{min}}}  x' b(x';y,z) \mathcal{K}(y,z) \partial_x \phi_{k'}(\xi_j(x))\\
& \qquad \qquad \times \frac{\phi_i(\xi_l(y)) \phi_{i'}(\xi_{l'}(z))}{yz} \\
& \qquad \qquad \times \left[ H(y-x_{l-1/2}) - H(y-x_{l+1/2})  \right] \mathrm{d}y \mathrm{d}z \mathrm{d}x' \mathrm{d}x,
\end{aligned}
\label{eq:terms_Tfrag1_A_intflux}
\end{equation}
\begin{equation}
\begin{aligned}
&\mathcal{T}_{\mathrm{frag},1,B}(x_{\mathrm{max}},x_{\mathrm{min}},j,k',l,i',i) \equiv \\
& \int_{I_j} \int\limits_{x_{\mathrm{min}}}^{x}  \int\limits_{x_{j-1/2}}^x   \int\limits_{x-z + x_{\mathrm{min}}}^{x_{\mathrm{max}}-z+x_{\mathrm{min}}}  x' b(x';y,z) \mathcal{K}(y,z) \partial_x \phi_{k'}(\xi_j(x)) \\
& \qquad \qquad \times \frac{\phi_i(\xi_l(y)) \phi_{i'}(\xi_{j}(z))}{yz} \\
& \qquad \qquad \times \left[ H(y-x_{l-1/2}) - H(y-x_{l+1/2})  \right]  \mathrm{d}y \mathrm{d}z \mathrm{d}x' \mathrm{d}x,
\end{aligned}
\label{eq:terms_Tfrag1_B_intflux}
\end{equation}
\begin{equation}
\begin{aligned}
& \mathcal{T}_{\mathrm{frag},2,A}(x_{\mathrm{max}},x_{\mathrm{min}},j,k',l,i',i) \equiv \\
& \int_{I_j} \int\limits_{x_{\mathrm{min}}}^{x}  \int\limits_x^{x_{j+1/2}}   \int\limits_{x_{\mathrm{min}}}^{x_{\mathrm{max}}-z+x_{\mathrm{min}}}  x' b(x';y,z) \mathcal{K}(y,z) \partial_x \phi_{k'}(\xi_j(x)) \\
& \qquad \qquad \times \frac{\phi_i(\xi_l(y)) \phi_{i'}(\xi_{j}(z))}{yz} \\
& \qquad \qquad \times \left[ H(y-x_{l-1/2}) - H(y-x_{l+1/2})  \right] \mathrm{d}y \mathrm{d}z \mathrm{d}x' \mathrm{d}x,
\end{aligned}
\label{eq:terms_Tfrag2_A_intflux}
\end{equation}
\begin{equation}
\begin{aligned}
& \mathcal{T}_{\mathrm{frag},2,B}(x_{\mathrm{max}},x_{\mathrm{min}},j,k',l',l,i',i) \equiv \\
&  \int_{I_j} \int\limits_{x_{\mathrm{min}}}^{x}  \int_{I_{l'}}   \int\limits_{x_{\mathrm{min}}}^{x_{\mathrm{max}}-z+x_{\mathrm{min}}}  x' b(x';y,z) \mathcal{K}(y,z) \partial_x \phi_{k'}(\xi_j(x)) \\
& \qquad \qquad \times \frac{\phi_i(\xi_l(y)) \phi_{i'}(\xi_{l'}(z))}{yz} \\
& \qquad \qquad \times \left[ H(y-x_{l-1/2}) - H(y-x_{l+1/2})  \right] \mathrm{d}y \mathrm{d}z \mathrm{d}x' \mathrm{d}x,
\end{aligned}
\label{eq:terms_Tfrag2_B_intflux}
\end{equation}
\begin{equation}
\begin{aligned}
& \mathcal{T}_{\mathrm{coag},A}(x_{\mathrm{max}},x_{\mathrm{min}},j,k',l',l,i',i) \equiv \\
& \int_{I_j} \int_{I_{l'}} \int\limits_{x-z +x_{\mathrm{min}}}^{x_{\mathrm{max}}-z+x_{\mathrm{min}}} \mathcal{K}(z,y) \partial_x \phi_{k'}(\xi_j(x))  \\
& \qquad \qquad \times \phi_{i'}(\xi_{l'}(z)) \frac{\phi_i(\xi_l(y))}{y} \\
& \qquad \qquad \times \left[ H(y-x_{l-1/2}) - H(y-x_{l+1/2})  \right] \mathrm{d}y \mathrm{d}z \mathrm{d}x,
\end{aligned}
\label{eq:terms_Tcoag_A_intflux}
\end{equation}
\begin{equation}
\begin{aligned}
& \mathcal{T}_{\mathrm{coag},B}(x_{\mathrm{max}},x_{\mathrm{min}},j,k',l,i',i) \equiv \\
& \int_{I_j} \int\limits_{x_{j-1/2}}^x \int\limits_{x-z +x_{\mathrm{min}}}^{x_{\mathrm{max}}-z+x_{\mathrm{min}}} \mathcal{K}(z,y) \partial_x \phi_{k'}(\xi_j(x))  \\
& \qquad \qquad \times \phi_{i'}(\xi_{j}(z)) \frac{\phi_i(\xi_l(y))}{y} \\
& \qquad \qquad \times \left[ H(y-x_{l-1/2}) -H(y-x_{l+1/2})  \right] \mathrm{d}y \mathrm{d}z \mathrm{d}x.
\end{aligned}
\label{eq:terms_Tcoag_B_intflux}
\end{equation}

All the terms in Eqs.~\ref{eq:terms_Tfrag1_A_intflux} to \ref{eq:terms_Tcoag_B_intflux} are analytically calculated with \textsc{Mathematica} and translated into \texttt{Fortran}/\texttt{C++}. Then, $\mathcal{F}_{\mathrm{frag,c}}$ is evaluated similarly to the numerical flux in Eq.~\ref{eq:flux_num_2}. $\mathcal{T}_{\mathrm{frag},1,A}$, $\mathcal{T}_{\mathrm{frag},1,B}$, $\mathcal{T}_{\mathrm{frag},2,A}$, $\mathcal{T}_{\mathrm{frag},2,B}$, $\mathcal{T}_{\mathrm{coag},A}$ and $\mathcal{T}_{\mathrm{coag},B}$ are computed once at the beginning of the algorithm. $\mathcal{F}_{\mathrm{frag,c}}$ is obtained by the sum of six terms in Eq.~\ref{eq:intflux_2}. The first term is obtained by computing the product of the subarray for index $(j,k')$  $\mathcal{T}_{\mathrm{frag},1,A}(x_{\mathrm{max}},x_{\mathrm{min}},j,k',l',l,i',i)$  with $g_{l'}^{i'}(\tau) g_l^i(\tau)$ and summing over all elements. The same evaluation is applied to the other terms. The process is repeated for all $(j,k')$ to obtain $\mathcal{F}_{\mathrm{frag,c}}$ on all the mass range for the DG scheme.

\section{DG scheme with the Gauss-Kronrod quadrature}
\label{ap:DGGKQ}
Here we describe the implementation of the Gauss-Kronrod quadrature method to evaluate the integrals in the term $F_{\mathrm{frag,c}}[g](x_{j-1/2},\tau)$ in Eq.~\ref{eq:flux_trunc_cons_2} and the term $\mathcal{F}_{\mathrm{frag,c}}(j,k',\tau)$ in Eq.~\ref{eq:intflux}.

\subsection{Gauss-Kronrod quadrature for the flux}
By using the evaluation of the integral on $y$ in Eq.~\ref{eq:approx_gv_GQ}, the numerical flux in Eq.~\ref{eq:flux_trunc_cons_2} writes
\begin{equation}
\begin{aligned}
& F_{\mathrm{frag,c}}[g](x_{j-1/2},\tau) =\\
& - \frac{1}{2}   \sum_{l'=1}^{j-1} \sum_{l=1}^N \sum_{i'=0}^k \sum_{i=0}^k  g_{l'}^{i'}(\tau) g_l^i(\tau) \\
& \qquad \qquad \times T_{\mathrm{GK,frag},1}(x_{\mathrm{max}},x_{\mathrm{min}},j,l',l,i',i) \\
& - \frac{1}{2}    \sum_{l'=j}^N  \sum_{l=1}^N \sum_{i'=0}^k \sum_{i=0}^k g_{l'}^{i'}(\tau) g_l^i(\tau) \\
& \qquad \qquad \times  T_{\mathrm{GK,frag},2}(x_{\mathrm{max}},x_{\mathrm{min}},l',l,i',i)   \\
& + \sum_{l'=1}^{j-1}  \sum_{l=1}^N  \sum_{i'=0}^k \sum_{i=0}^k g_{l'}^{i'}(\tau) g_l^i(\tau) T_{\mathrm{GK,coag}}(x_{\mathrm{max}},x_{\mathrm{min}},j,l',l,i',i),
\end{aligned}
\label{eq:flux_num_GK}
\end{equation}
where GK stands for Gauss-Kronrod and the terms write
\begin{equation}
\begin{aligned}
&T_{\mathrm{GK,frag},1}(x_{\mathrm{max}},x_{\mathrm{min}},j,l',l,i',i) \equiv  \\
& \sum_{u=1}^{j-1} \int\limits_{I_u} \int\limits_{I_{l'}} \mathbb{1}_{x_{l-1/2}<x_{\mathrm{max}} - z + x_{\mathrm{min}}} \mathbb{1}_{x_{j-1/2}-z+x_{\mathrm{min}} < x_{l+1/2}}  \\
& \qquad \times \int\limits_{\mathrm{max} \left(x_{j-1/2}-z + x_{\mathrm{min}}, x_{l-1/2}\right)}^{\mathrm{min} \left(x_{\mathrm{max}}-z+x_{\mathrm{min}}, x_{l+1/2} \right)} x' b(x';y,z) \mathcal{K}(y,z) \\
& \qquad  \times \frac{\phi_i(\xi_l(y)) \phi_{i'}(\xi_{l'}(z))}{yz}  \mathrm{d}y \mathrm{d}z \mathrm{d}x', \\
\end{aligned}
\label{eq:terms_Tfrag1_flux_GK}
\end{equation}
\begin{equation}
\begin{aligned}
& T_{\mathrm{GK,frag},2}(x_{\mathrm{max}},x_{\mathrm{min}},l',l,i',i) \equiv \\
& \sum_{u=1}^{j-1} \int\limits_{I_u} \int\limits_{I_{l'}} \mathbb{1}_{x_{l-1/2}<x_{\mathrm{max}} - z + x_{\mathrm{min}}} \\
& \qquad \times \int\limits_{  \mathrm{max} \left( x_{\mathrm{min}}, x_{l-1/2}  \right)}^{\mathrm{min} \left(x_{\mathrm{max}}-z+x_{\mathrm{min}}, x_{l+1/2} \right)}  x' b(x';y,z) \mathcal{K}(y,z) \\
& \qquad  \times \frac{\phi_i(\xi_l(y)) \phi_{i'}(\xi_{l'}(z))}{yz}  \mathrm{d}y \mathrm{d}z \mathrm{d}x',
\end{aligned}
\label{eq:terms_Tfrag1_flux_GK}
\end{equation}
\begin{equation}
\begin{aligned}
& T_{\mathrm{GK,coag}}(x_{\mathrm{max}},x_{\mathrm{min}},j,l',l,i',i) \equiv \\
& \int\limits_{I_{l'}} \mathbb{1}_{x_{l-1/2}<x_{\mathrm{max}} - z + x_{\mathrm{min}}} \mathbb{1}_{x_{j-1/2}-z+x_{\mathrm{min}} < x_{l+1/2}}  \\
& \qquad \times \int\limits_{\mathrm{max} \left(x_{j-1/2}-z + x_{\mathrm{min}}, x_{l-1/2}\right)}^{\mathrm{min} \left(x_{\mathrm{max}}-z+x_{\mathrm{min}}, x_{l+1/2} \right)} \mathcal{K}(y,z) \\
& \qquad \times \phi_{i'}(\xi_{l'}(z)) \frac{\phi_i(\xi_l(y))}{y}  \mathrm{d}y \mathrm{d}z.
\end{aligned}
\label{eq:terms_Tcoag_flux_GK}
\end{equation}
The Gauss-Kronrod quadrature is then applied to evaluate the integrals. The first term writes
\begin{equation}
\begin{aligned}
& T_{\mathrm{GK,frag},1}(x_{\mathrm{max}},x_{\mathrm{min}},j,l',l,i',i) = \\
& \sum_{u=1}^{j-1} \sum_{\gamma = 1}^Q \sum_{\alpha=1}^Q \sum_{\beta=1}^Q \frac{h_u h_{l'} h_{\mathrm{f}_1,j,l',l,\alpha}}{8} \\
& \qquad \times \mathbb{1}_{x_{l-1/2}<x_{\mathrm{max}} - \hat{x}_{l'}^{\alpha} + x_{\mathrm{min}}} \mathbb{1}_{x_{j-1/2}-\hat{x}_{l'}^{\alpha}+x_{\mathrm{min}} < x_{l+1/2}} \\
& \qquad \times \omega_{\gamma} \omega_{\alpha} \omega_{\beta} \hat{x}_u^{\gamma} \;b \left(\hat{x}_u^{\gamma};\hat{x}_{\mathrm{f}_1,j,l',l}^{\alpha,\beta},\hat{x}_{l'}^{\alpha} \right) \mathcal{K}\left( \hat{x}_{\mathrm{f}_1,j,l',l}^{\alpha,\beta},\hat{x}_{l'}^{\alpha}\right) \\
& \qquad \times \frac{\phi_i \left(\frac{2}{h_l}\left(  \hat{x}_{\mathrm{f}_1,j,l',l}^{\alpha,\beta} - x_l \right) \right) \phi_{i'}(s_{\alpha})}{\hat{x}_{\mathrm{f}_1,j,l',l}^{\alpha,\beta} \hat{x}_{l'}^{\alpha}},
\end{aligned}
\end{equation}
where $Q$ is the number of Gauss-Kronrod points, $s$ and $\omega$ are the node and the weight coefficients, respectively, and we define the terms $\hat{x}_{l'}^{\alpha}$, $h_{\mathrm{f}_1,j,l',l,\alpha}$ and $\hat{x}_{\mathrm{f}_1,j,l',l}^{\alpha,\beta}$ as
\begin{equation}
\begin{aligned}
& \hat{x}_{l'}^{\alpha} \equiv x_{l'} + \frac{h_{l'}}{2}s_{\alpha}, \\
& h_{\mathrm{f}_1,j,l',l,\alpha} \equiv \mathrm{min} \left(x_{\mathrm{max}}-\hat{x}_{l'}^{\alpha}+x_{\mathrm{min}}, x_{l+1/2} \right) \\
& \qquad \qquad \qquad  - \mathrm{max} \left(x_{j-1/2}-\hat{x}_{l'}^{\alpha} + x_{\mathrm{min}}, x_{l-1/2}\right),\\
& \hat{x}_{\mathrm{f}_1,j,l',l}^{\alpha,\beta} \equiv \frac{1}{2} \left[ \mathrm{min} \left(x_{\mathrm{max}}-\hat{x}_{l'}^{\alpha}+x_{\mathrm{min}}, x_{l+1/2} \right) \right. \\
& \qquad \qquad \qquad \left. + \mathrm{max} \left(x_{j-1/2}-\hat{x}_{l'}^{\alpha} + x_{\mathrm{min}} \right) \right] + \frac{h_{\mathrm{f}_1,j,l',l,\alpha}}{2} s_{\beta}.
\end{aligned}
\end{equation}
The second term $T_{\mathrm{GK,frag},2}$ writes 
\begin{equation}
\begin{aligned}
& T_{\mathrm{GK,frag},2}(x_{\mathrm{max}},x_{\mathrm{min}},l',l,i',i) = \\
& \sum_{u=1}^{j-1} \sum_{\gamma = 1}^Q \sum_{\alpha=1}^Q \sum_{\beta=1}^Q \frac{h_u h_{l'} h_{\mathrm{f}_2,l',l,\alpha}}{8} \\
& \qquad \times \mathbb{1}_{x_{l-1/2}<x_{\mathrm{max}} - \hat{x}_{l'}^{\alpha} + x_{\mathrm{min}}} \\
& \qquad \times \omega_{\gamma} \omega_{\alpha} \omega_{\beta} \hat{x}_u^{\gamma} b \left(\hat{x}_u^{\gamma};\hat{x}_{\mathrm{f}_2,l',l}^{\alpha,\beta},\hat{x}_{l'}^{\alpha} \right) \mathcal{K}\left( \hat{x}_{\mathrm{f}_2l',l}^{\alpha,\beta},\hat{x}_{l'}^{\alpha}\right) \\
& \qquad \times \frac{\phi_i \left(\frac{2}{h_l}\left(  \hat{x}_{\mathrm{f}_2l',l}^{\alpha,\beta} - x_l \right) \right) \phi_{i'}(s_{\alpha})}{\hat{x}_{\mathrm{f}_2l',l}^{\alpha,\beta} \hat{x}_{l'}^{\alpha}},
\end{aligned}
\end{equation}
where 
\begin{equation}
\begin{aligned}
& h_{\mathrm{f}_2,l',l,\alpha} \equiv \mathrm{min} \left(x_{\mathrm{max}}-\hat{x}_{l'}^{\alpha}+x_{\mathrm{min}}, x_{l+1/2} \right) - x_{l-1/2}, \\
& \hat{x}_{\mathrm{f}_2,l',l}^{\alpha,\beta} \equiv \frac{1}{2} \left[ \mathrm{min} \left(x_{\mathrm{max}}-\hat{x}_{l'}^{\alpha}+x_{\mathrm{min}}, x_{l+1/2} \right) + x_{l-1/2}  \right] \\
& \qquad \qquad + \frac{h_{\mathrm{f}_2,l',l,\alpha}}{2} s_{\beta}.
\end{aligned}
\end{equation}
The last term $T_{\mathrm{GK,coag}}$ writes
\begin{equation}
\begin{aligned}
& T_{\mathrm{GK,coag}}(x_{\mathrm{max}},x_{\mathrm{min}},j,l',l,i',i) = \\
& \sum_{\alpha=1}^Q \sum_{\beta=1}^Q \frac{ h_{l'} h_{\mathrm{c},j,l',l,\alpha}}{4} \\
& \qquad \times \mathbb{1}_{x_{l-1/2}<x_{\mathrm{max}} - \hat{x}_{l'}^{\alpha} + x_{\mathrm{min}}} \mathbb{1}_{x_{j-1/2}-\hat{x}_{l'}^{\alpha}+x_{\mathrm{min}} < x_{l+1/2}} \\
& \qquad \times \omega_{\alpha} \omega_{\beta}  \mathcal{K}\left( \hat{x}_{\mathrm{c},l',l}^{\alpha,\beta},\hat{x}_{l'}^{\alpha}\right) \\
& \qquad \times \frac{\phi_i \left(\frac{2}{h_l}\left(  \hat{x}_{\mathrm{c},j,l',l}^{\alpha,\beta} - x_l \right) \right) \phi_{i'}(s_{\alpha})}{\hat{x}_{\mathrm{c},j,l',l}^{\alpha,\beta}},
\end{aligned}
\end{equation}
where $h_{\mathrm{c},j,l',l,\alpha} = h_{\mathrm{f}_1,j,l',l,\alpha}$ and $\hat{x}_{\mathrm{c},j,l',l}^{\alpha,\beta} = \hat{x}_{\mathrm{f}_1,j,l',l}^{\alpha,\beta}$.

\subsection{Gauss-Kronrod quadrature for the term integral of the flux}
By using Eq.~\ref{eq:approx_gv_GQ}, the term $\mathcal{F}_{\mathrm{frag},c}$ in Eq.~\ref{eq:intflux} writes
\begin{equation}
\begin{aligned}
& \mathcal{F}_{\mathrm{frag,c}}(j,k',\tau) =\\
& - \frac{1}{2} \sum_{l'=1}^{j} \sum_{l=1}^N \sum_{i'=0}^k \sum_{i=0}^k g_{l'}^{i'}(\tau) g_l^i(\tau) \\
& \qquad \qquad \times \mathcal{T}_{\mathrm{GK,frag},1}(x_{\mathrm{max}},x_{\mathrm{min}},j,k',l',l,i',i) \\
& - \frac{1}{2}  \sum_{l'=j}^N \sum_{l=1}^N \sum_{i'=0}^k \sum_{i=0}^k g_{l'}^{i'}(\tau) g_l^i(\tau) \\
& \qquad \qquad \times \mathcal{T}_{\mathrm{GK,frag},2}(x_{\mathrm{max}},x_{\mathrm{min}},j,k',l',l,i',i) \\
&  + \sum_{l'=1}^{j} \sum_{l=1}^N \sum_{i'=0}^k \sum_{i=0}^k g_{l'}^{i'}(\tau) g_l^i(\tau) \\
& \qquad \qquad \times \mathcal{T}_{\mathrm{GK,coag}}(x_{\mathrm{max}},x_{\mathrm{min}},j,k',l',l,i',i) \\
\end{aligned}
\end{equation}
where the terms write
\begin{equation}
\begin{aligned}
& \mathcal{T}_{\mathrm{GK,frag},1}(x_{\mathrm{max}},x_{\mathrm{min}},j,k',l',l,i',i) \equiv \\
& \int_{I_j} \sum_{u=1}^j \mathbb{1}_{x_{u-1/2}<x} \int\limits_{\mathrm{max}\left( x_{\mathrm{min}}, x_{u-1/2} \right)}^{\mathrm{min} \left( x, x_{u+1/2} \right)} \mathbb{1}_{x_{l'-1/2} < x} \\
&  \times \int\limits_{\mathrm{max}\left( x_{\mathrm{min}}, x_{l'-1/2} \right)}^{\mathrm{min} \left( x, x_{l'+1/2} \right)} \mathbb{1}_{x_{l-1/2}<x_{\mathrm{max}} - z + x_{\mathrm{min}}} \\
& \times \mathbb{1}_{x-z+x_{\mathrm{min}} < x_{l+1/2}} \\
& \times  \int\limits_{ \mathrm{max} \left( x-z + x_{\mathrm{min}}, x_{l-1/2} \right)}^{ \mathrm{min} \left( x_{\mathrm{max}}-z+x_{\mathrm{min}}, x_{l+1/2} \right)}  x' b(x';y,z) \mathcal{K}(y,z) \partial_x \phi_{k'}(\xi_j(x))\\
&  \times  \frac{\phi_i(\xi_l(y)) \phi_{i'}(\xi_{l'}(z))}{yz} \mathrm{d}y \mathrm{d}z \mathrm{d}x' \mathrm{d}x,
\end{aligned}
\end{equation}
\begin{equation}
\begin{aligned}
& \mathcal{T}_{\mathrm{GK,frag},2}(x_{\mathrm{max}},x_{\mathrm{min}},j,k',l',l,i',i) \equiv \\
& \int_{I_j} \sum_{u=1}^j \mathbb{1}_{x_{u-1/2}<x} \int\limits_{\mathrm{max}\left( x_{\mathrm{min}}, x_{u-1/2} \right)}^{\mathrm{min} \left( x, x_{u+1/2} \right)} \mathbb{1}_{ x < x_{l'+1/2}} \\
&  \times \int\limits_{\mathrm{max}\left( x, x_{l'-1/2} \right)}^{\mathrm{min} \left( x_{\mathrm{max}}, x_{l'+1/2} \right)} \mathbb{1}_{x_{l-1/2}<x_{\mathrm{max}} - z + x_{\mathrm{min}}} \\
& \times  \int\limits_{ \mathrm{max} \left( x_{\mathrm{min}}, x_{l-1/2} \right)}^{ \mathrm{min} \left( x_{\mathrm{max}}-z+x_{\mathrm{min}}, x_{l+1/2} \right)}  x' b(x';y,z) \mathcal{K}(y,z) \partial_x \phi_{k'}(\xi_j(x))\\
&  \times  \frac{\phi_i(\xi_l(y)) \phi_{i'}(\xi_{l'}(z))}{yz} \mathrm{d}y \mathrm{d}z \mathrm{d}x' \mathrm{d}x,
\end{aligned}
\end{equation}
\begin{equation}
\begin{aligned}
& \mathcal{T}_{\mathrm{GK,coag}}(x_{\mathrm{max}},x_{\mathrm{min}},j,k',l',l,i',i) \equiv \\
& \int_{I_j} \mathbb{1}_{x_{l'-1/2} < x}  \int\limits_{\mathrm{max}\left( x_{\mathrm{min}}, x_{l'-1/2} \right)}^{\mathrm{min} \left( x, x_{l'+1/2} \right)} \mathbb{1}_{x_{l-1/2}<x_{\mathrm{max}} - z + x_{\mathrm{min}}} \\
& \times \mathbb{1}_{x-z+x_{\mathrm{min}} < x_{l+1/2}} \\
& \times  \int\limits_{ \mathrm{max} \left( x-z + x_{\mathrm{min}}, x_{l-1/2} \right)}^{ \mathrm{min} \left( x_{\mathrm{max}}-z+x_{\mathrm{min}}, x_{l+1/2} \right)}   \mathcal{K}(y,z) \partial_x \phi_{k'}(\xi_j(x))\\
&  \times  \frac{\phi_i(\xi_l(y)) \phi_{i'}(\xi_{l'}(z))}{y} \mathrm{d}y \mathrm{d}z \mathrm{d}x' \mathrm{d}x, \\
\end{aligned}
\end{equation}
We then apply the Gauss-Kronrod quadrature method to evaluate the integrals. The first term writes
\begin{equation}
\begin{aligned}
&\mathcal{T}_{\mathrm{GK,frag},1}(x_{\mathrm{max}},x_{\mathrm{min}},j,k',l',l,i',i) =\\
&  \sum_{\lambda = 1}^Q \sum_{u=1}^{j} \sum_{\gamma = 1}^Q \sum_{\alpha=1}^Q \sum_{\beta=1}^Q \frac{h_j h_{u,\lambda} h_{l',\lambda} h_{\mathrm{f}_1,j,l',l,\lambda,\alpha}}{16} \\
&\quad \times \mathbb{1}_{x_{u-1/2} < \hat{x}_j^{\lambda}} \mathbb{1}_{x_{l'-1/2} < \hat{x}_j^{\lambda}} \\
& \quad \times  \mathbb{1}_{x_{l-1/2}<x_{\mathrm{max}} - \hat{x}_{l'}^{\alpha} + x_{\mathrm{min}}} \mathbb{1}_{\hat{x}_j^{\lambda}-\hat{x}_{l'}^{\alpha}+x_{\mathrm{min}} < x_{l+1/2}} \\
& \quad \times \omega_{\lambda}  \omega_{\gamma} \omega_{\alpha} \omega_{\beta}  \hat{x}_u^{\gamma} b \left(\hat{x}_u^{\gamma};\hat{x}_{\mathrm{f}_1,l',l}^{\lambda,\alpha,\beta},\hat{x}_{l'}^{\alpha} \right) \mathcal{K}\left( \hat{x}_{\mathrm{f}_1,l',l}^{\lambda,\alpha,\beta},\hat{x}_{l'}^{\alpha}\right) \\
& \quad \times \frac{\phi_i \left(\frac{2}{h_l}\left(  \hat{x}_{\mathrm{f}_1,l',l}^{\alpha,\beta} - x_l \right) \right) \phi_{i'}(s_{\alpha})}{\hat{x}_{\mathrm{f}_1,l',l}^{\alpha,\beta} \hat{x}_{l'}^{\alpha}}\\
& \quad \times \left. \partial_x \phi_{k'}(\xi_j(x))\right|_{x=\hat{x}_j^{\lambda}},
\end{aligned}
\end{equation}
where we define the terms 
\begin{equation}
\begin{aligned}
& \hat{x}_j^{\lambda} \equiv x_j + \frac{h_j}{2} s_{\lambda},\\
& \hat{x}_u^{\gamma} \equiv x_u + \frac{h_u}{2} s_{\gamma},\\
& \hat{x}_{l'}^{\alpha} \equiv x_{l'} + \frac{h_{l'}}{2} s_{\alpha},\\
&h_{u,\lambda} \equiv \mathrm{min} \left( \hat{x}_j^{\lambda},x_{u+1/2}\right) - x_{u-1/2},\\
&h_{l',\lambda} \equiv \mathrm{min} \left(  \hat{x}_j^{\lambda},x_{l'+1/2}\right) - x_{l'-1/2},\\
&h_{\mathrm{f}_1,j,l',l,\lambda,\alpha} \equiv  \mathrm{min} \left( x_{\mathrm{max}} -\hat{x}_{l'}^{\alpha} + x_{\mathrm{min}} , x_{l+1/2} \right)  \\
& \qquad \qquad \qquad \qquad - \mathrm{max} \left( \hat{x}_j^{\lambda} - \hat{x}_{l'}^{\alpha} + x_{\mathrm{min}}, x_{l-1/2} \right), \\
& \hat{x}_{\mathrm{f}_1,l',l}^{\lambda,\alpha,\beta} \equiv \frac{1}{2} \left[  \mathrm{min} \left( x_{\mathrm{max}} -\hat{x}_{l'}^{\alpha} + x_{\mathrm{min}} , x_{l+1/2} \right)  \right. \\
& \qquad \qquad \qquad  + \left.  \mathrm{max} \left( \hat{x}_j^{\lambda} - \hat{x}_{l'}^{\alpha} + x_{\mathrm{min}}, x_{l-1/2} \right) \right] \\
& \qquad \qquad + \frac{h_{\mathrm{f}_1,j,l',l,\lambda,\alpha}}{2} s_{\beta}.
\end{aligned}
\end{equation}
The second term $\mathcal{T}_{\mathrm{GK,frag},2}$ writes 
\begin{equation}
\begin{aligned}
&\mathcal{T}_{\mathrm{GK,frag},1}(x_{\mathrm{max}},x_{\mathrm{min}},j,k',l',l,i',i) =\\
&  \sum_{\lambda = 1}^Q  \sum_{u=1}^{j} \sum_{\gamma = 1}^Q \sum_{\alpha=1}^Q \sum_{\beta=1}^Q \frac{h_j h_{u,\lambda} h_{l',\lambda} h_{\mathrm{f}_2,j,l',l,\alpha}}{16} \\
& \quad \times  \mathbb{1}_{x_{u-1/2} < \hat{x}_j^{\lambda}} \mathbb{1}_{ \hat{x}_j^{\lambda} < x_{l'+1/2}}  \mathbb{1}_{x_{l-1/2}<x_{\mathrm{max}} - \hat{x}_{l'}^{\alpha} + x_{\mathrm{min}}}\\
& \quad \times \omega_{\lambda}  \omega_{\gamma} \omega_{\alpha} \omega_{\beta}  \hat{x}_u^{\gamma} b \left(\hat{x}_u^{\gamma};\hat{x}_{\mathrm{f}_1,l',l}^{\alpha,\beta},\hat{x}_{l'}^{\alpha} \right) \mathcal{K}\left( \hat{x}_{\mathrm{f}_2,l',l}^{\alpha,\beta},\hat{x}_{l'}^{\alpha}\right) \\
& \quad \times \frac{\phi_i \left(\frac{2}{h_l}\left(  \hat{x}_{\mathrm{f}_2,l',l}^{\alpha,\beta} - x_l \right) \right) \phi_{i'}(s_{\alpha})}{\hat{x}_{\mathrm{f}_2,l',l}^{\alpha,\beta} \hat{x}_{l'}^{\alpha}}\\
& \quad \times \left. \partial_x \phi_{k'}(\xi_j(x))\right|_{x=\hat{x}_j^{\lambda}},
\end{aligned}
\end{equation}
where we define the terms 
\begin{equation}
\begin{aligned}
&h_{l',\lambda} \equiv \mathrm{min} \left( x_{\mathrm{max}},x_{l'+1/2}\right) - \mathrm{min} \left( \hat{x}_j^{\lambda} ,x_{l'-1/2} \right),\\
&h_{\mathrm{f}_2,j,l',l,\alpha} \equiv  \mathrm{min} \left( x_{\mathrm{max}} -\hat{x}_{l'}^{\alpha} + x_{\mathrm{min}} , x_{l+1/2} \right)  \\
& \qquad \qquad \qquad \qquad - \mathrm{max} \left( x_{\mathrm{min}}, x_{l-1/2} \right), \\
& \hat{x}_{\mathrm{f}_1,l',l}^{\alpha,\beta} \equiv \frac{1}{2} \left[  \mathrm{min} \left( x_{\mathrm{max}} -\hat{x}_{l'}^{\alpha} + x_{\mathrm{min}} , x_{l+1/2} \right)  \right. \\
& \qquad \qquad \qquad  + \left.  \mathrm{max} \left( x_{\mathrm{min}}, x_{l-1/2} \right) \right] + \frac{h_{\mathrm{f}_1,j,l',l,\lambda,\alpha}}{2} s_{\beta}.
\end{aligned}
\end{equation}
The last term $\mathcal{T}_{\mathrm{GK,coag}}$ writes 
\begin{equation}
\begin{aligned}
&\mathcal{T}_{\mathrm{GK,coag}}(x_{\mathrm{max}},x_{\mathrm{min}},j,k',l',l,i',i) =\\
&  \sum_{\lambda = 1}^Q  \sum_{\alpha=1}^Q \sum_{\beta=1}^Q \frac{h_j  h_{l',\lambda} h_{\mathrm{c},j,l',l,\lambda,\alpha}}{8}  \\
& \quad \times \mathbb{1}_{x_{l'-1/2} < \hat{x}_j^{\lambda}} \mathbb{1}_{x_{l-1/2}<x_{\mathrm{max}} - \hat{x}_{l'}^{\alpha} + x_{\mathrm{min}}}  \mathbb{1}_{\hat{x}_j^{\lambda}-\hat{x}_{l'}^{\alpha}+x_{\mathrm{min}} < x_{l+1/2}} \\
& \quad \times \omega_{\lambda} \omega_{\alpha} \omega_{\beta}  \mathcal{K}\left( \hat{x}_{\mathrm{c},l',l}^{\lambda,\alpha,\beta},\hat{x}_{l'}^{\alpha}\right) \\
& \quad \times \frac{\phi_i \left(\frac{2}{h_l}\left(  \hat{x}_{\mathrm{c},l',l}^{\lambda,\alpha,\beta} - x_l \right) \right) \phi_{i'}(s_{\alpha})}{\hat{x}_{\mathrm{c},l',l}^{\lambda,\alpha,\beta}}\\
& \quad \times \left. \partial_x \phi_{k'}(\xi_j(x))\right|_{x=\hat{x}_j^{\lambda}},
\end{aligned}
\end{equation}
where $ h_{\mathrm{c},j,l',l,\lambda,\alpha} =  h_{\mathrm{f}_1,j,l',l,\lambda,\alpha}$ and $\hat{x}_{\mathrm{c},l',l}^{\lambda,\alpha,\beta}$.

\section{Matrix form}
\label{ap:matrix_form}
Originally used by \citet{Sandu2006} for the coagulation equation, the matrix form of Eq.~\ref{eq:DG} writes
\begin{equation}
 A \frac{\mathrm{d} \mathbf{g} (\tau)}{\mathrm{d} \tau} = 
 \begin{bmatrix} 
 \mathbf{g}^{T}(\tau) \cdot B^1 \cdot \mathbf{g}(\tau) \\
 \vdots  \\
 \mathbf{g}^{T}(\tau) \cdot B^{N\times(k+1)}  \cdot \mathbf{g}(\tau)
 \end{bmatrix},
\label{eq:matrix_form}
\end{equation}
where $\mathbf{g} = \left[g_m \right]_{ 0 \leq m \leq N \times (k+1)}$ with the index $m=i + j \times (k+1)$ for which $i \in [\![0,k]\!]$ and $j \in [\![1,N]\!]$. The matrix $A$ and the 3-tensor $B$ write
\begin{equation}
\begin{aligned}
&A = \mathrm{diag} \left[A_m \right]_{ 0 \leq m \leq N \times (k+1)},\\
&B = \left[ B_{m,n}^{s} \right]_{0 \leq m \leq N \times (k+1), 0 \leq n \leq N \times (k+1),0 \leq s \leq N \times (k+1) },
\end{aligned}
\end{equation}
where $\forall m \in [\![1,N \times (k+1)]\!],\;  i = m/N,\; j=m(\mathrm{mod} N),\; A_m = \frac{2}{h_j(2i+1)}$ and  $B_{m,n}^{s}$ gathers all the integrals for the flux and the integral of the flux terms. The 3-tensor B is a sparse tensor with elements
\begin{equation}
\begin{aligned}
&\forall (m,n,s) \in [\![0,N \times (k+1)]\!]^3,\\
&B_{m,n}^{s} \equiv \\
&  \frac{1}{2}  \left( \mathbb{1}_{l' \leq j} \mathcal{T}_{\mathrm{GK,frag},1}(x_{\mathrm{max}},x_{\mathrm{min}},j,k',l',l,i',i) \right)\\
& +   \frac{1}{2} \left( \mathbb{1}_{j \leq l'} \mathcal{T}_{\mathrm{GK,frag},2}(x_{\mathrm{max}},x_{\mathrm{min}},j,k',l',l,i',i) \right)\\
& -  \left( \mathbb{1}_{l' \leq j} \mathcal{T}_{\mathrm{GK,coag}}(x_{\mathrm{max}},x_{\mathrm{min}},j,k',l',l,i',i) \right. \\
& - \phi_k(1) \mathbb{1}_{j \leq N-1} \\
& \times \left(  \left[   \mathbb{1}_{l' \leq j} T_{\mathrm{frag},1}(x_{\mathrm{max}},x_{\mathrm{min}},j+1,l',l,i',i) \right. \right.\\
& \qquad \qquad \left. + \mathbb{1}_{j+1 \leq l'} T_{\mathrm{frag},2}(x_{\mathrm{max}},x_{\mathrm{min}},j+1,l',l,i',i) \right] \\
&  \left. \qquad -  \mathbb{1}_{l' \leq j} T_{\mathrm{coag}}(x_{\mathrm{max}},x_{\mathrm{min}},j+1,l',l,i',i) \right), \\
& + \phi_k(-1) \\
& \times \left(  \left[ \mathbb{1}_{l' \leq j-1} T_{\mathrm{frag},1}(x_{\mathrm{max}},x_{\mathrm{min}},j,l',l,i',i) \right. \right.\\
& \qquad \qquad \left. + \mathbb{1}_{j \leq l'} T_{\mathrm{frag},2}(x_{\mathrm{max}},x_{\mathrm{min}},j,l',l,i',i) \right] \\
&  \left. \qquad - \mathbb{1}_{l' \leq j-1} T_{\mathrm{coag}}(x_{\mathrm{max}},x_{\mathrm{min}},j,l',l,i',i) \right), \\
& k' = s/N,\; j = s\; (\mathrm{mod} N), \\
& i' = m/N,\; l' = m\; (\mathrm{mod} N), \\
& i = n/N,\; l = n\; (\mathrm{mod} N).
\end{aligned} 
\end{equation}

The presence of the operator $\mathbb{1}$ in $B_{m,n}^{s}$ implies that the components of the 3-tensor $B$ are sparse matrices.

\end{document}